\documentclass[zpreprint,zbstepj]{./LaTeX/zeus/zeus_paper}
\usepackage[english]{babel}

\newcommand{\ZcoosysB}{%
The ZEUS coordinate system is a right-handed Cartesian system, with the $Z$
axis pointing in the proton beam direction, referred to as the ``forward
direction'', and the $X$ axis pointing left towards the centre of HERA.
The coordinate origin is at the nominal interaction point.\xspace}

\newcommand{\ZcoosysfnB}{\footnote{\ZcoosysB}}

\chardef\usc=95
\chardef\til=126
\catcode`\@=11 
\DeclareRobustCommand\xdotspace{\futurelet\@let@token\@xdotspace}
\def\@xdotspace{%
  \ifx\@let@token.\else
  \ifx\@let@token\bgroup.\else
  \ifx\@let@token\egroup.\else
  \ifx\@let@token\/.\else
  \ifx\@let@token\ .\else
  \ifx\@let@token~.\else
  \ifx\@let@token!.\else
  \ifx\@let@token,.\else
  \ifx\@let@token:.\else
  \ifx\@let@token;.\else
  \ifx\@let@token?.\else
  \ifx\@let@token/.\else
  \ifx\@let@token'.\else
  \ifx\@let@token).\else
  \ifx\@let@token-.\else
  \ifx\@let@token\@xobeysp.\else
  \ifx\@let@token\space.\else
  \ifx\@let@token\@sptoken.\else
   .\space
   \fi\fi\fi\fi\fi\fi\fi\fi\fi\fi\fi\fi\fi\fi\fi\fi\fi\fi}
\catcode`\@=12 

\newcommand{\stru}[2]{%
   \relax\ifmmode\hbox{\vrule height#1 depth#2 width0pt}%
   \else\vrule height#1 depth#2 width0pt\fi}

\newcommand{\Ronum}[1]{\uppercase\expandafter{\romannumeral#1}}
\newcommand{\ronum}[1]{\expandafter{\romannumeral#1}}
\DeclareRobustCommand{\LaTeXZ}{%
  \LaTeX\kern-.05em4\kern-.1em
  {\raisebox{-0.2ex}{$\scriptstyle\text{ZEUS}$}}\xspace}

\newcommand{\fig}[1]{Fig.~\ref{fig-#1}}
\newcommand{\Fig}[1]{Figure~\ref{fig-#1}}

\DeclareMathAlphabet{\mathbf}{OT1}{cmr}{bx}{sl}
\newcommand{\eVdist}{\kern-0.06667em}

\newcommand{\Gev}{{\text{Ge}\eVdist\text{V\/}}}

\newcommand{\gev}{{\,\text{Ge}\eVdist\text{V\/}}}

\newcommand{\pb}{\,\text{pb}}

\newcommand{\pbi}{\,\text{pb}^{-1}}

\newcommand{\cm}{\,\text{cm}}

\newcommand{\mrad}{\,\text{mrad}}

\newcommand{\Tesla}{\,\text{T}}

\newcommand{\slashfrac}[2]{%
  \raisebox{0.5ex}{\ensuremath #1}\kern-0.12em/\kern-0.08em
  \raisebox{-.8ex}{\ensuremath #2}}

\newcommand{\sqr}[3]{%
    {\vcenter{\hrule height.#3ex\hbox{\vrule width.#2ex height#1ex
     \kern#1ex\vrule width.#3ex}\hrule height.#2ex}}}

\catcode`\@=11 
\newcommand{\parenbar}{\mathpalette\p@renb@r}
\def\p@renb@r#1#2{\vbox{%
  \ifx#1\scriptscriptstyle \dimen@.7em\dimen@ii.2em\else
  \ifx#1\scriptstyle \dimen@.8em\dimen@ii.25em\else
  \dimen@1em\dimen@ii.4em\fi\fi \offinterlineskip
  \ialign{\hfill##\hfill\cr
    \vbox{\hrule width\dimen@ii}\cr
    \noalign{\vskip-.3ex}%
    \hbox to\dimen@{$\mathchar300\hfil\mathchar301$}\cr
    \noalign{\vskip-.3ex}%
    $#1#2$\cr}}}
\catcode`\@=12 

\newcommand{\sitil}{{\tilde\sigma}}

\newcommand{\DA}{{\rm DA}}

\newcommand{\IP}{{\rm I$\kern-0.01667em$P}\xspace}
\newcommand{\JB}{{\rm JB}}

\mathchardef\qsm=63
\mathchardef\pls=43
\mathchardef\mns=512
\mathchardef\plm=518
\mathchardef\eql=61
\mathchardef\smallleft=300
\mathchardef\smallright=301
\mathchardef\les=316
\mathchardef\gre=318
\mathchardef\leq=532
\mathchardef\grq=533

\catcode`\@=11 
\newcounter{pict@width}
\newcounter{pict@height}
\newlength{\pict@scale}
\newcommand{\psfigadd}[4]{%
\setcounter{pict@width}{1*\ratio{#2+\pict@scale/2}{\pict@scale}}
\setcounter{pict@height}{1*\ratio{#3+\pict@scale/2}{\pict@scale}}
\setlength{\unitlength}{\pict@scale}
\hbox to #2{\hspace{-\fill}\begin{picture}(\thepict@width,\thepict@height)
\put(0,0){\psfig{figure=#1,width=#2,height=#3,clip=}}
\SetScale{0.283466457}
\SetWidth{1.763889}
{#4}
\end{picture}}
}
\newcounter{pict@widthfst}
\newcounter{pict@widthscd}
\newcounter{pict@widthtot}
\newcommand{\psfigaddtwo}[7]{%
\setcounter{pict@widthfst}{1*\ratio{#2+\pict@scale/2}{\pict@scale}}
\setcounter{pict@widthscd}{1*\ratio{#2+#4+\pict@scale/2}{\pict@scale}}
\setcounter{pict@widthtot}{1*\ratio{#2+#4+#6+\pict@scale/2}{\pict@scale}}
\setcounter{pict@height}{1*\ratio{#3+\pict@scale/2}{\pict@scale}}
\setlength{\unitlength}{\pict@scale}
\hbox{\hspace{-\fill}\begin{picture}(\thepict@widthtot,\thepict@height)
\put(0,0){\psfig{figure=#1,width=#2,height=#3,clip=}}
\put(\thepict@widthscd,0){\psfig{figure=#5,width=#6,height=#3,clip=}}
\SetScale{0.283466457}
\SetWidth{1.763889}
{#7}
\end{picture}}
}
\newcommand{\psfigror}[4]{%
\setcounter{pict@width}{1*\ratio{#2+\pict@scale/2}{\pict@scale}}
\setcounter{pict@height}{1*\ratio{#3+\pict@scale/2}{\pict@scale}}
\setlength{\unitlength}{\pict@scale}
\hbox{\begin{picture}(\thepict@width,\thepict@height)
\put(0,\thepict@height){\psfig{figure=#1,width=#3,height=#2,clip=,angle=270}}
\SetScale{0.283466457}
\SetWidth{1.763889}
{#4}
\end{picture}}
}
\newcommand{\psfigrol}[4]{%
\setcounter{pict@width}{1*\ratio{#2+\pict@scale/2}{\pict@scale}}
\setcounter{pict@height}{1*\ratio{#3+\pict@scale/2}{\pict@scale}}
\setlength{\unitlength}{\pict@scale}
\hbox{\begin{picture}(\thepict@width,\thepict@height)
\put(0,0){\psfig{figure=#1,width=#3,height=#2,clip=,angle=90}}
\SetScale{0.283466457}
\SetWidth{1.763889}
{#4}
\end{picture}}
}
\catcode`\@=12 
\newlength\listtextwidth

\catcode`\@=11 
\newlength{\@tabfninsert}
\newlength{\@tabfnwidth}
\newcommand{\tabfootnote}[2]{%
  \setlength{\@tabfninsert}{0.8em}
  \setlength{\@tabfnwidth}{\textwidth}
  \addtolength{\@tabfnwidth}{-\@tabfninsert}
  \addtolength{\@tabfnwidth}{-0.4em}
  \noindent\makebox[\@tabfninsert][r]{\footnotesize$^{#1}$\hfil}\hfill%
  \parbox[t]{\@tabfnwidth}{\footnotesize #2\hfill}}
\catcode`\@=12 

\newcommand{\be}{\begin{equation}}
\newcommand{\ee}{\end{equation}}
\newcommand{\bea}{\begin{eqnarray}}
\newcommand{\eea}{\end{eqnarray}}

\newcommand{\ddif}[3]{\frac{d^{2}#1}{d#2 d#3}}

\def\citeCTD{{\cite{%
nim:a279:290,*npps:b32:181,*nim:a338:254%
}}\xspace}
\def\citeCAL{{\cite{%
nim:a309:77,*nim:a309:101,*nim:a321:356,*nim:a336:23%
}}\xspace}

\begin{document}
\title{
Measurement of high-$\bold{Q^{2}}$ neutral current\\
deep inelastic $\bold{e^- p}$ scattering cross sections\\
with a longitudinally polarised electron beam at HERA \\
}

\author{ZEUS Collaboration}
\prepnum{DESY-08-202}
\date{}

\abstract{
Measurements of the neutral current cross sections for deep inelastic
scattering in $e^{-}p$ collisions at HERA
with a longitudinally polarised electron beam are presented.
The single-differential cross-sections $d\sigma/dQ^{2}$,
$d\sigma/dx$ and $d\sigma/dy$
and the double-differential cross sections in $Q^{2}$ and $x$ are
measured in the kinematic region $y < 0.9$ and $Q^2 > 185 \gev^2$
for both positively and negatively polarised electron beams
and for each polarisation state separately.
The measurements are based on an integrated luminosity of $169.9\pbi$
taken with the ZEUS detector in 2005 and 2006
at a centre-of-mass energy of $318 \gev$.
The structure functions $x\tilde{F_3}$ and $xF_3^{\gamma Z}$
are determined by combining the $e^- p$ results presented in this paper
with previously measured $e^ + p$ neutral current data.
The asymmetry parameter $A^-$ is used to demonstrate the parity violating
effects of electroweak interactions at large spacelike photon virtuality.
The measurements agree well with the predictions of the Standard Model.
}

\makezeustitle

\def\3{\ss}                                                                                        
\pagenumbering{Roman}                                                                              
                                                   %
\begin{center}                                                                                     
{                      \Large  The ZEUS Collaboration              }                               
\end{center}                                                                                       
  S.~Chekanov,                                                                                     
  M.~Derrick,                                                                                      
  S.~Magill,                                                                                       
  B.~Musgrave,                                                                                     
  D.~Nicholass$^{   1}$,                                                                           
  \mbox{J.~Repond},                                                                                
  R.~Yoshida\\                                                                                     
 {\it Argonne National Laboratory, Argonne, Illinois 60439-4815, USA}~$^{n}$                       
\par \filbreak                                                                                     
  M.C.K.~Mattingly \\                                                                              
 {\it Andrews University, Berrien Springs, Michigan 49104-0380, USA}                               
\par \filbreak                                                                                     
  P.~Antonioli,                                                                                    
  G.~Bari,                                                                                         
  L.~Bellagamba,                                                                                   
  D.~Boscherini,                                                                                   
  A.~Bruni,                                                                                        
  G.~Bruni,                                                                                        
  F.~Cindolo,                                                                                      
  M.~Corradi,                                                                                      
\mbox{G.~Iacobucci},                                                                               
  A.~Margotti,                                                                                     
  R.~Nania,                                                                                        
  A.~Polini\\                                                                                      
  {\it INFN Bologna, Bologna, Italy}~$^{e}$
% ...
\newpage                                                                                           
                                                           %
$^{\    1}$ also affiliated with University College London,                                        
United Kingdom\\                                                                                   
% ...
$^{\    3}$ also working at Max Planck Institute, Munich, Germany \\                               
% ...
%
\newpage   
                                                           %
                                                           %
\begin{tabular}[h]{rp{14cm}}                                                                       
$^{a}$ &  supported by the Natural Sciences and Engineering Research Council of Canada (NSERC) \\  
$^{b}$ &  supported by the German Federal Ministry for Education and Research (BMBF), under        
          contract numbers 05 HZ6PDA, 05 HZ6GUA, 05 HZ6VFA and 05 HZ4KHA\\                         
$^{c}$ &  supported in part by the MINERVA Gesellschaft f\"ur Forschung GmbH, the Israel Science   
          Foundation (grant no. 293/02-11.2) and the U.S.-Israel Binational Science Foundation \\  
$^{d}$ &  supported by the Israel Science Foundation\\                                             
$^{e}$ &  supported by the Italian National Institute for Nuclear Physics (INFN) \\                
$^{f}$ &  supported by the Japanese Ministry of Education, Culture, Sports, Science and Technology 
          (MEXT) and its grants for Scientific Research\\                                          
$^{g}$ &  supported by the Korean Ministry of Education and Korea Science and Engineering          
          Foundation\\                                                                             
$^{h}$ &  supported by the Netherlands Foundation for Research on Matter (FOM)\\                   
$^{i}$ &  supported by the Polish State Committee for Scientific Research, project no.             
          DESY/256/2006 - 154/DES/2006/03\\                                                        
$^{j}$ &  partially supported by the German Federal Ministry for Education and Research (BMBF)\\   
$^{k}$ &  supported by RF Presidential grant N 1456.2008.2 for the leading                         
          scientific schools and by the Russian Ministry of Education and Science through its      
          grant for Scientific Research on High Energy Physics\\                                   
$^{l}$ &  supported by the Spanish Ministry of Education and Science through funds provided by     
          CICYT\\                                                                                  
$^{m}$ &  supported by the Science and Technology Facilities Council, UK\\                         
$^{n}$ &  supported by the US Department of Energy\\                                               
$^{o}$ &  supported by the US National Science Foundation. Any opinion,                            
findings and conclusions or recommendations expressed in this material                             
are those of the authors and do not necessarily reflect the views of the                           
National Science Foundation.\\                                                                     
$^{p}$ &  supported by the Polish Ministry of Science and Higher Education                         
as a scientific project (2006-2008)\\                                                              
$^{q}$ &  supported by FNRS and its associated funds (IISN and FRIA) and by an Inter-University    
          Attraction Poles Programme subsidised by the Belgian Federal Science Policy Office\\     
$^{r}$ &  supported by an FRGS grant from the Malaysian government\\                               
\end{tabular}                                                                                      
                                                           %
                                                           %

\pagenumbering{arabic} 
\pagestyle{plain}

\section{Introduction}
\label{sec-int}

The study of deep inelastic scattering (DIS) of leptons off
nucleons has been instrumental in establishing not only the structure
of nucleons but also many other aspects of the Standard
Model~(SM). Recently, the HERA $e p$ collider with a centre-of-mass energy of
$318\,\gev$, has expanded the accessible kinematic region for DIS
measurements allowing for direct observation of the effects of the weak
interaction at high values of the negative four-momentum transfer squared,
$Q^2$. In particular, the structure function $x\tilde{F}_3$ can be obtained from the
difference of cross sections in $e^+ p$ and $e^- p$ scattering.
At HERA $x\tilde{F}_3$ is dominated
by the interference of photon and $Z$-exchange and can be extracted
from data on a pure proton target with no complications due to target
mass or higher twist effects. This furnishes not only a precise test
of the electroweak sector of the standard model but also provides
direct information on the valence quark distributions in the nucleon.

The data samples collected from 1992--2000 by the H1 and ZEUS
collaborations were used for determinations of the neutral current
(NC) cross sections up to values of $Q^2\approx 30\,000\,\gev^2$
~\mcite{epj:c11:427,epj:c21:443,epj:c28:175,pr:d70:052001,np:b470:3,
        *np:b497:3,*epj:c13:609,*epj:c21:33,*epj:c19:269,*epj:c30:1}.
A first extraction of $x\tilde{F}_3$ clearly demonstrated the effect of
$Z$-exchange. A measurement of $e^+ p$ NC DIS cross sections for a
longitudinally polarised positron beam using a limited sample of data
collected during 2004 has also been published by the ZEUS
collaboration~\cite{pl:b637:210}.

In this paper, measurements of the cross sections and the asymmetry
parameter, the difference in the behaviour of negatively and
positively polarised electrons, are presented. The measurement is
made using data collected during 2005 and 2006 when HERA collided both
positively and negatively polarised electron beams of $27.5\,\gev$
with protons of $920\,\gev$. The integrated luminosity
amounts to $169.9\,\pb^{-1}$ with mean luminosity-weighted
polarisations of +0.29 and -0.27. This is a ten-fold increase over the
previously available $e^- p$ sample. This allows a detailed and direct probe of
electro-weak effects at high $Q^2$ and a more precise measurement of the structure
function $x\tilde{F}_3$.

\section{Standard Model predictions}
\label{sec-kin}

Inclusive deep inelastic lepton-proton scattering can be described in
terms of the kinematic variables $x$, $y$, and $Q^2$.
The variable $Q^2$ is defined as $Q^2 = -q^2 = -(k-k')^2$,
where $k$ and $k'$ are the four-momenta of the incoming and scattered lepton,
respectively.
Bjorken $x$ is defined as $x=Q^2/2P \cdot q$, where $P$ is
the four-momentum of the incoming proton.
The fraction of the lepton energy transferred to the proton in its rest frame
is given by $y = P \cdot q / P \cdot k$.
The variables $x$, $y$ and $Q^2$ are related by $Q^2=sxy$,
where $s$, the centre-of-mass energy is approximately given
by $s=4E_e E_p$, and $E_{e}$ and $E_{p}$
are the initial energies of the electron and proton, respectively.

The electroweak Born-level cross section for the $e^ \pm p$ NC
interaction is given by~\cite{devenish:2003:dis, zfp:c24:151}
\begin{equation}
\ddif{\sigma(e^{\pm}p)}{x}{Q^{2}} =
\frac{2 \pi \alpha^{2} }{xQ^{4}}
[Y_{+} \tilde{F_{2}}(x,Q^{2})
\mp Y_{-} x\tilde{F_{3}}(x,Q^{2})
- y^{2}\tilde {F_{L}}(x,Q^{2})],
\label{eqn:unpol_xsec}
\end{equation}
where $\alpha$ is the fine-structure constant,
$Y_{\pm} = 1 \pm (1 - y)^{2}$, and
$\tilde{F_{2}}(x,Q^{2})$, $\tilde{F_{3}}(x,Q^{2})$ and
$\tilde{F_{L}}(x,Q^{2})$
are generalised structure functions. Next-to-leading order (NLO)
QCD calculations predict that the contribution of the longitudinal
structure function $\tilde {F_L}$ to $d^2\sigma /dx dQ^2$ is approximately
$1.5\%$, averaged over the kinematic range considered,
and therefore neglected in the discussion in this section.
However, this term is included in SM calculations which are compared to
the measurements presented in this paper.

The generalised structure functions depend on the longitudinal polarisation
of the lepton beam which is defined as

\begin{equation}
P_{e}=\frac{N_{R}-N_{L}}{N_{R}+N_{L}} \ ,
\nonumber
\label{eqn:pol}
\end{equation}
where $N_{R}$ and $N_{L}$ are the numbers of right- and left-handed leptons
in the beam\footnote{At HERA beam energies the mass of the incoming leptons
may be neglected, and therefore the difference between handedness and
helicity may also be neglected.}.

Photon exchange dominates the cross section at low $Q^{2}$
and is described by $\tilde{F_{2}}$.
It is only at $Q^{2}$ values comparable to $M_Z^2$ that
the $\gamma / Z$ interference and pure $Z$ exchange terms become important
and the $\tilde{F_{3}}$ term contributes significantly to the cross section.
The sign of the $\tilde{F_3}$ term in Eq.~(\ref{eqn:unpol_xsec}) shows that
electroweak effects increase (decrease) the $e^- p$ ($e^+ p$) cross sections.

Reduced cross sections, $\tilde{\sigma}$, for $e^-p$ and $e^+p$ scattering are defined as

\begin{equation}
\tilde{\sigma}^{e^{\pm} p}
=
\frac {xQ^{4}} {2 \pi \alpha^{2} }
\frac {1} {Y_{+}}
\ddif{\sigma(e^{\pm}p)}{x}{Q^{2}}
=
\tilde{F_{2}}(x,Q^{2}) \mp \frac {Y_{-}} {Y_{+}} x \tilde{F_{3}}(x,Q^{2}).
\label{eqn:red}
\end{equation}

The difference in the $e^{-} p$ and $e^{+} p$ reduced cross sections yields

\begin{equation}
x\tilde{F_3}
= \frac {Y_{+}} {2Y_{-}}( \tilde{\sigma}^{e^{-} p} -
\tilde{\sigma}^{e^{+} p} ).
\label{eqn:xf3}
\end{equation}

The generalised structure functions can be split into terms
depending $\gamma$ exchange ($F_2^{\gamma}$),
$Z$ exchange ($F_2^Z$, $xF_3^Z$)
and $\gamma/Z$ interference ($F_2^{\gamma Z}$, $xF_3^{\gamma Z}$) as

\begin{equation}
\tilde{F_2} = F_2^{\gamma} - (v_e - P_e a_e) \chi_{Z} F_2^{\gamma Z} +
(v_e^2 + a_e^2 - 2 P_e v_e a_e) {\chi_{Z}^{2}} F_2^{Z} ,
\label{eqn:gen_f2}
\end{equation}
\begin{equation}
x\tilde{F_3} = - (a_e - P_e v_e) \chi_{Z} xF_3^{\gamma Z} + (2 v_e a_e
- P_e(v_e^2 + a_e^2)) {\chi_{Z}^{2}} xF_3^{Z}.
\label{eqn:gen_xf3}
\end{equation}

In these equations, the respective vector and axial couplings
of the electron to the $Z$ boson in the SM
are given by $v_{e} = -1/2 + 2\sin^2\theta_W$ and $a_{e} = -1/2$,
where $\theta_W$ is the Weinberg angle.
The relative contribution of $Z$ and $\gamma$ exchange is given by
$\chi_{Z}=\frac{1}{\sin^2{2\theta_W}} \frac{Q^{2}}{M_{Z}^{2}+Q^{2}}$ and
varies between 0.2 and 1.1 over the range $1\,500 < Q^2 < 30\,000~\gev^2$.
For the unpolarised case ($P_{e} = 0$),
the interference structure function $xF_3^{\gamma Z}$
is the dominant term in $x\tilde{F_3}$
as $v_e$ is small ($\approx -0.04$) and thus terms containing $v_e$
in Eq.~(\ref{eqn:gen_xf3}) can be ignored, so that

\begin{equation}
x\tilde{F_3} \simeq - a_e \chi_{Z} xF_3^{\gamma Z}.
\label{eqn:gen_xf3gz}
\end{equation}

In this paper, measurements of $x\tilde{F_3}$ and $xF_3^{\gamma Z}$
are presented using the full $e^- p$ dataset collected during 2005 and 2006.

The structure functions can be written in terms of the sums and differences
of the quark and anti-quark momentum distributions.
In leading order (LO) QCD

\begin{equation}
[F_2^{\gamma},F_2^{\gamma Z},F_2^{Z}] =
\sum _q [e_{q}^{2}, 2e_{q}v_{q},v_{q}^{2}+a_{q}^{2}]
x (q + \bar{q}),
\label{eqn:struc1}
\end{equation}
\begin{equation}
[xF_3^{\gamma Z},xF_3^{Z}] =
\sum _q [e_{q}a_{q},v_{q}a_{q}]
2x (q - \bar{q}),
\label{eqn:struc2}
\end{equation}
where $v_{q}$ and $a_{q}$ are the vector and axial couplings
of the quark $q$ to the $Z$ boson, and $e_{q}$ is the electric charge of the quark.
The densities of the quarks and anti-quarks
are given by parton distribution functions (PDFs)
$q$ and $\bar{q}$, respectively.
The sums run over all quark flavours except the top quark.

The sensitivity of $xF_3^{\gamma Z}$ to $u_v$ and $d_v$,
the valence quark momentum distributions,
is demonstrated in LO QCD through

\begin{equation}
xF_3^{\gamma Z}
= 2x [e_u a_u u_v + e_d a_d d_v] = \frac{x}{3}(2u_{v} + d_{v}).
\label{eqn:xf3gz_simple}
\end{equation}

In addition, the integral of $xF^{\gamma Z}_{3}$
should obey the sum rule~\cite{epjd:cn2:1} :

\be
\int^{1}_{0}xF^{\gamma Z}_{3}\frac{dx}{x} =
\frac{1}{3}\int^{1}_{0} (2u_{v} + d_{v})dx = \frac{5}{3} \ .
\label{eqn:xf3gzSum}
\ee

The charge-dependent polarisation asymmetry, $A^-$,
is defined in terms of pure right-handed ($P_e$ = +1)
and left-handed ($P_e$ = -1) electron beams as

\begin{equation}
A^- \equiv \frac{\sigma^{-}(P_e = +1) -\sigma^{-}(P_e = -1)}
                {\sigma^{-}(P_e = +1) +\sigma^{-}(P_e = -1)} \ ,
\label{eqn:asymDefn}
\end{equation}

where $\sigma^{-}(P_e=+1)$ and $\sigma^{-}(P_e=-1)$
are the cross sections at $P_e$ values of $+1$ and $-1$, respectively.
When the beam polarisation is not unity $A^-$ is given by

\begin{equation}
A^{-} = \frac{\sigma^-(P_{e,+}) - \sigma^-(P_{e,-})}
{P_{e,+}\sigma^-(P_{e,-}) - P_{e,-}\sigma^-(P_{e,+})} \ ,
\label{eqn:asymMeas}
\end{equation}

where $\sigma^{-}(P_{e,+})$ and $\sigma^{-}(P_{e,-})$ are
the cross sections evaluated at positive and negative electron
polarisation values.
To a good approximation the asymmetry is the ratio of
the $F_2^{\gamma Z}$ and $F_2^{\gamma}$ structure functions,
and is proportional to the combination $a_e$$v_q$ :

\begin{equation}
A^- \simeq \chi_{Z} a_e \frac{F_2^{\gamma Z}}{F_2^{\gamma}}
= 2 a_e v_q e_q/e_q^2 \propto a_e v_q.
\label{fgf}
\end{equation}

Thus a measurement of $A^{-}$ can give direct evidence of parity violation
with minimal sensitivity to the proton PDFs,
and a comparison to SM predictions provides
a test of the electroweak sector of the SM.

\section{Experimental apparatus}

\label{sec-ncdet}

A detailed description of the ZEUS detector can be found
elsewhere~\cite{zeus:1993:bluebook}.
A brief outline of the components most relevant for this analysis 
is given below.

Charged particles were tracked in the central tracking detector (CTD)~\citeCTD,
which operated in a magnetic field of $1.43\Tesla$
provided by a thin superconducting solenoid.
The CTD consisted of 72~cylindrical drift chamber layers,
organised in nine superlayers covering
the polar-angle\ZcoosysfnB~region \mbox{$15^\circ<\theta<164^\circ$}.
A silicon microvertex detector (MVD)~\cite{nim:581:656}
was installed between the beampipe and the inner radius of the CTD.
The MVD was organised into a barrel with 3 cylindrical layers and a forward section
with four planar layers perpendicular to the HERA beam direction.
Charged-particle tracks were reconstructed using information from the CTD and MVD.

The high-resolution uranium--scintillator calorimeter (CAL)~\citeCAL
consisted of three parts: the forward (FCAL), the barrel (BCAL)
and the rear (RCAL) calorimeters, covering 99.7\% of the solid angle
around the nominal interaction point.
Each part was subdivided transversely into towers and longitudinally
into one electromagnetic section (EMC) and either one (RCAL)
or two (BCAL and FCAL) hadronic sections (HAC).
The smallest subdivision of the calorimeter was called a cell.
The CAL relative energy resolutions,
as measured under test-beam conditions, were $\sigma(E)/E=0.18/\sqrt{E}$
for electrons and $\sigma(E)/E=0.35/\sqrt{E}$ for hadrons,
with $E$ in $\gev$.
The timing resolution of the CAL was better than 1~ns
for energy deposits exceeding 4.5~\gev.

An iron structure that surrounded the CAL was instrumented
as a backing calorimeter (BAC)~\cite{nim:a313:126}
to measure energy leakage from the CAL.
Muon chambers in the forward, barrel and rear~\cite{nim:a333:342}
regions were used in this analysis to veto background events
induced by cosmic-ray or beam-halo muons.

The luminosity was measured
using the Bethe-Heitler reaction $ep \rightarrow e \gamma p$
with the luminosity detector which consisted of two independent systems,
a photon calorimeter and a magnetic spectrometer.

The lepton beam in HERA became naturally transversely polarised
through the Sokolov-Ternov effect~\cite{sovpdo:8:1203}.
The characteristic build-up time in HERA
was approximately 40 minutes.
Spin rotators on either side of the ZEUS detector
changed the transverse polarisation of the beam
into longitudinal polarisation.
The electron beam polarisation was measured
using two independent polarimeters,
the transverse polarimeter (TPOL)~\cite {sovjnp:238:1969,nim:a329:79} and
the longitudinal polarimeter (LPOL)~\cite {nim:a479:334}.
Both devices exploited the spin-dependent cross section
for Compton scattering of circularly polarised photons off electrons
to measure the beam polarisation.
The luminosity and polarisation measurements were made over times
that were much shorter than the polarisation build-up time.

\section{Monte Carlo simulation}
\label{sec-mc}

Monte Carlo (MC) simulated events were used
to determine the efficiency for selecting events,
the accuracy of kinematic reconstruction,
to estimate the background rate
and to correct for detector acceptance.

NC DIS events were simulated including radiative effects,
using the {\sc Heracles}~\cite{cpc:69:155} program
with the {\sc Djangoh} 1.6~\cite{proc:hera:1991:1419,*spi:www:djangoh16}
interface to the hadronisation programs
and using CTEQ5D \cite{epj:c12:375} PDFs.
The hadronic final state was simulated using the colour-dipole model
in {\sc Ariadne} 4.10~\cite{cpc:71:15}.
To investigate systematic uncertainties,
the {\sc meps} model of {\sc lepto}6.5~\cite{cpc:101:108} was also used.
The Lund string model of {\sc Jetset}7.4~\cite{cpc:39:347,*cpc:43:367,*cpc:82:74}
was used for the hadronisation.
Diffractive NC events were generated using the {\sc Rapgap}~~2.08/06~\cite{cpc:86:147}
program and mixed with the non-diffractive MC events
to simulate the observed hadronic final states.
Background from photoproduction events
was simulated using {\sc Herwig}~5.9~\cite{cpc:67:465}.

The simulated samples were at least five times larger than
the corresponding data samples.
They were normalised to the integrated luminosity of the data.

The ZEUS detector response was simulated using a program
based on {\sc Geant} 3.21~\cite{tech:cern-dd-ee-84-1}.
The generated events were passed through the detector simulation,
subjected to the same trigger requirements as the data
and processed by the same reconstruction programs.

\section{Event reconstruction}
\label{sec-ncrecon}
Neutral Current events at high $Q^2$ are characterised
by the presence of an isolated high-energy electron in the final state.
The transverse momentum of the scattered electron
balances that of the hadronic final state,
and therefore the measured net transverse momentum, $p_T$, should be small.
The measured $p_T$ and the net transverse energy, $E_T$, are defined as

\begin{alignat}{2}
p_T^2 & = & p_X^2 + p_Y^2 = & \left( \sum\limits_{i} E_i \sin \theta_i \cos
\phi_i \right)^2+ \left( \sum\limits_{i} E_i \sin \theta_i \sin \phi_i
\right)^2,
\label{eq-PT2}\\
E_T & = & \sum\limits_{i} E_i \sin \theta_i, \nonumber
\end{alignat}
where the sum runs over all calorimeter energy deposits, $E_i$.
The polar and azimuthal angles, $\theta_i$ and $\phi_i$,
of the calorimeter energy deposits are measured in a coordinate system
with the event vertex as origin.
The variable $\delta$, also used in the event selection, is defined as

\begin{equation}
\delta \equiv \sum\limits_{i} (E-p_Z)_{i} = \sum\limits_{i} ( E_i - E_i \cos
\theta_{i} ).
\label{eq-Delta}
\end{equation}
Conservation of energy and longitudinal momentum
requires $\delta = 2E_e= 55 \gev$ if all final-state particles are
detected and perfectly measured.
Undetected particles that escape through the forward beam-pipe
have a negligible effect on $\delta$.
However, particles lost through the rear beam-pipe
can lead to a substantial reduction in $\delta$.

Backsplash of low energy particles
originating from secondary interactions and deposited at large angles
were suppressed by removing low energy deposits
with a polar angle greater than $\gamma_{\rm max}$,
which was calculated on an event-by-event bases.
The specific value of $\gamma_{\rm max}$ was tuned
using both data and MC samples
containing a reduced amount of backsplash.
Studies~\cite{thesis:sun:2007} have shown that
this procedure depends on the $Q^2$ threshold of the sample.
This effect is accounted for in the study of systematic effects.

Furthermore, CAL energy deposits are separated
into those associated with the scattered electron,
and all other energy deposits.
The sum of the latter is called the hadronic energy.
In the naive quark-parton model, the hadronic polar angle, $\gamma_h$,
defined as

\begin{equation}
\cos\gamma_h = \frac{p_{T,h}^2 - \delta_h^2}{p_{T,h}^2 + \delta_h^2} \ ,
\label{eqn:gamma_h}
\end{equation}
is the scattering angle of the struck quark
where the quantities $p_{T,h}$ and $\delta_h$ are derived from
Eqs.~(\ref{eq-PT2}$-$\ref{eq-Delta}) using only the hadronic energy.

The double-angle (DA) method~\cite{proc:hera:1991:23,*proc:hera:1991:43}
is used for reconstructing the kinematic variables.
It makes use of the polar angle of the scattered electron, $\theta_e$, and $\gamma_h$
to reconstruct the kinematic variables $x_\DA$, $y_\DA$, and $Q^2_\DA$.
The DA method is insensitive to uncertainties
in the overall energy scale of the calorimeter.
However, it is sensitive to initial-state QED radiation
and an accurate simulation of the detector response is necessary.
The variable $y$ is also reconstructed using the electron method, $y_e$,
and the Jacquet-Blondel method, $y_\JB$ \cite{proc:epfacility:1979:391}.
These estimators for $y$
are used only for the event selection.

\section{Neutral Current event selection}
\label{sec-evsel}

ZEUS operated a three-level trigger
system~\cite{zeus:1993:bluebook,uproc:chep:1992:222,nim:allfrey:2007}.
At the first level,
only coarse calorimeter and tracking information were available.
Events were selected using criteria based on an energy deposit in the CAL
consistent with an isolated electron.
In addition, events with high $E_{T}$ in coincidence with a CTD track
were accepted.
At the second level, a requirement on $\delta$
was used to select NC DIS events.
Timing information from the calorimeter
was used to reject events inconsistent with the bunch-crossing time.
At the third level, events were fully reconstructed.
The requirements were similar to, but looser than,
the offline cuts described below.

Scattered electrons were identified using an algorithm
that combined information from the energy deposits in the calorimeter
with tracks measured in the central tracking detectors~\cite{epj:c11:427}.
To ensure high electron finding efficiency and to reject backgrounds,
the identified electron was required to have an energy of at least $10\gev$.
A track matched to the energy deposit in the calorimeter
was required for events in which an electron was found
within the acceptance of the tracking detectors.
This was done by requiring the distance of closest approach (DCA)
between the track extrapolated to the calorimeter surface
and the energy cluster position to be less than 10\,cm
and the electron track momentum, $p_e^{\rm trk}$, to be larger than $3\,\gev$.
A matched track was not required
if the electron emerged at a polar angle
outside the acceptance of the tracking detectors
and had an energy greater than $30\,\gev$ in the FCAL.
An isolation requirement was imposed such that
the energy not associated with the electron
in an $\eta-\phi$ cone of radius 0.8 centred on the electron
was less than $5 \gev$.

In photoproduction events
where the electron emerges at very small scattering angles,
$\delta$ is substantially smaller than $55 \gev$,
and in beam-gas events overlaid on NC events,
$\delta$ is substantially larger than $55 \gev$.
A requirement $38 < \delta < 65 \gev$ was imposed
to remove these backgrounds.
To further reduce background from photoproduction events,
$y_{e}$ was required to be less than $0.95$.
The net transverse momentum was expected to be small for balanced NC events,
so to remove cosmic-ray events and beam-related background events
the quantity $p_{T}/\sqrt{E_{T}}$ was required to be less than $4\sqrt{\gev}$
and the quantity $p_{T}/E_{T}$ was required to be less than 0.7.

In order to reject events where most of the hadronic final state
was lost in the forward beam-pipe, the projection of $\gamma_h$
onto the face of FCAL was required to be outside
a radius of $20~{\mathrm{cm}}$ centred on the beam-pipe axis.
The $Z$ coordinate of the $ep$ interaction vertex, reconstructed using
tracks in the CTD and the MVD, was required to satisfy
$| Z_{\rm vtx} | < 50$~cm.
The final event sample was defined by requiring
$Q^{2}_{\rm DA}>185 \gev^2$ and $y_{\rm DA} < 0.9$.

A total of 360\,437 candidate events passed the selection criteria.
The background is dominated by photoproduction which was estimated
to contribute about 0.3\% on average to the event sample.
Other backgrounds were negligible.

\begin{figure}[p]
\vfill
\begin{center}
\includegraphics[width=6in]{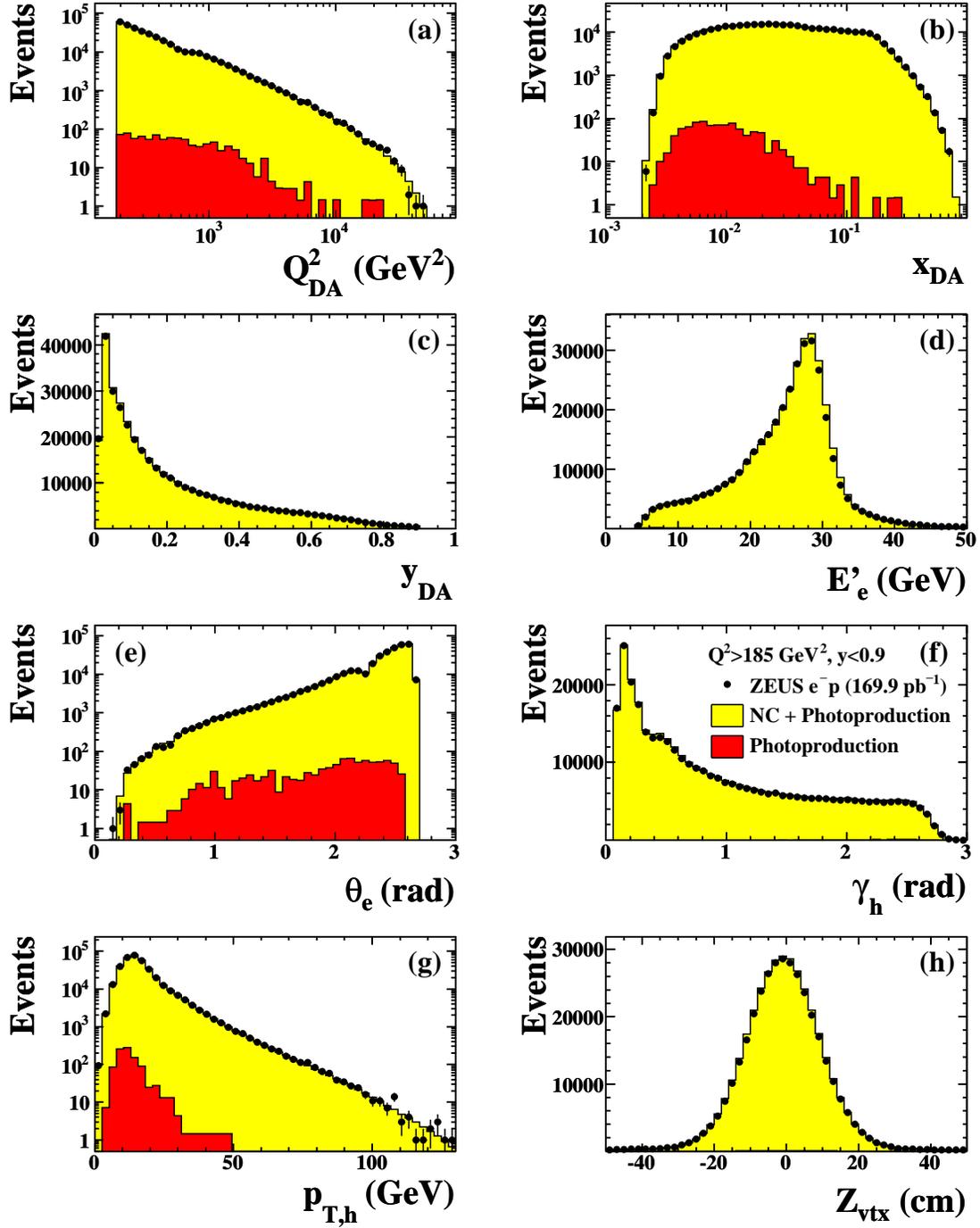}
\end{center}
\caption{
  Comparison of the final $e^- p$ NC data sample with the expectations of
  the MC simulation described in the text.
  The distributions of (a) $Q^{2}_{\rm DA}$, (b) $x_{\rm DA}$, (c)
  $y_{\rm DA}$, (d) the energy of the scattered electron,
  $E_{\rm e}^{\prime}$, (e) the angle of the scattered electron, $\theta_{\rm e}$,
  (f) the hadronic angle, $\gamma_{h}$, (g) the transverse momentum of the
  hadronic system, $p_{T,h}$, and (h) the $Z$ coordinate of the event vertex,
  $Z_{\rm vtx}$, are shown. 
}
\label{fig-con}
\vfill
\end{figure}

A comparison between data and MC distributions
is shown in Fig. \ref{fig-con}
for the variables $Q^{2}_{\rm DA}$, $x_{\rm DA}$, $y_{\rm DA}$,
energy $E_e^{\prime}$ and $\theta_e$ of the scattered electron,
$\gamma_h$ and $p_{T,h}$ of the final hadronic system,
and $Z_{\rm vtx}$ for the event.
The distributions from the data and MC (NC $+$ photoproduction) agree well.

\section{Cross section determination}

The single-differential cross-sections
$d\sigma/dQ^2$, $d\sigma/dx$ and $d\sigma/dy$
for $Q^2 > 185\gev^2$ and $y < 0.9$,
and $d\sigma/dx$ and $d\sigma/dy$
for $Q^2 > 3\,000\gev^2$ and $y < 0.9$,
and the double-differential cross-section $d^2\sigma/dxdQ^2$
were measured.
The cross sections in a particular bin
($d^2\sigma/dx dQ^2$ is shown as an example) was determined according to

\begin{equation}
  \frac{d^2 \sigma}{dx dQ^2} = \frac{N_{\rm data}-N_{\rm bg}}{N_{\rm MC}}
  \cdot \frac{d^2 \sigma^{\rm SM}_{\rm Born}}{dx dQ^2} \nonumber \ ,
\label{eq-xsect}
\end{equation}
where $N_{\rm data}$ is the number of data events in the bin,
$N_{\rm bg}$ is the number of background events
predicted from the photoproduction MC, and
$N_{\rm MC}$ is the number of signal MC events
normalised to the luminosity of the data.
The SM prediction for the Born-level cross section,
$d^2 \sigma^{\rm SM}_{\rm Born}/dx dQ^2$, was evaluated
using CTEQ5D PDFs~\cite{epj:c12:375} and using the PDG ~\cite{epj:c15:1} values 
for the fine-structure constant, the mass of the $Z$ boson,
and the weak mixing angle.
This procedure implicitly takes into account the acceptance,
bin-centering, and radiative corrections from the MC simulation.
The bin sizes used for the determination of the single- and double-differential
cross sections were chosen to be commensurate with the detector resolutions.
The statistical uncertainties on the cross sections
were calculated from the numbers of events observed in the bins,
taking into account the statistical uncertainty of the MC simulation
(signal and background).
Poisson statistics were used for all bins.

\section{Systematic uncertainties}
\label{sec-sys}
\begin{figure}[p]
\vfill
\begin{center}
\includegraphics[width=6in]{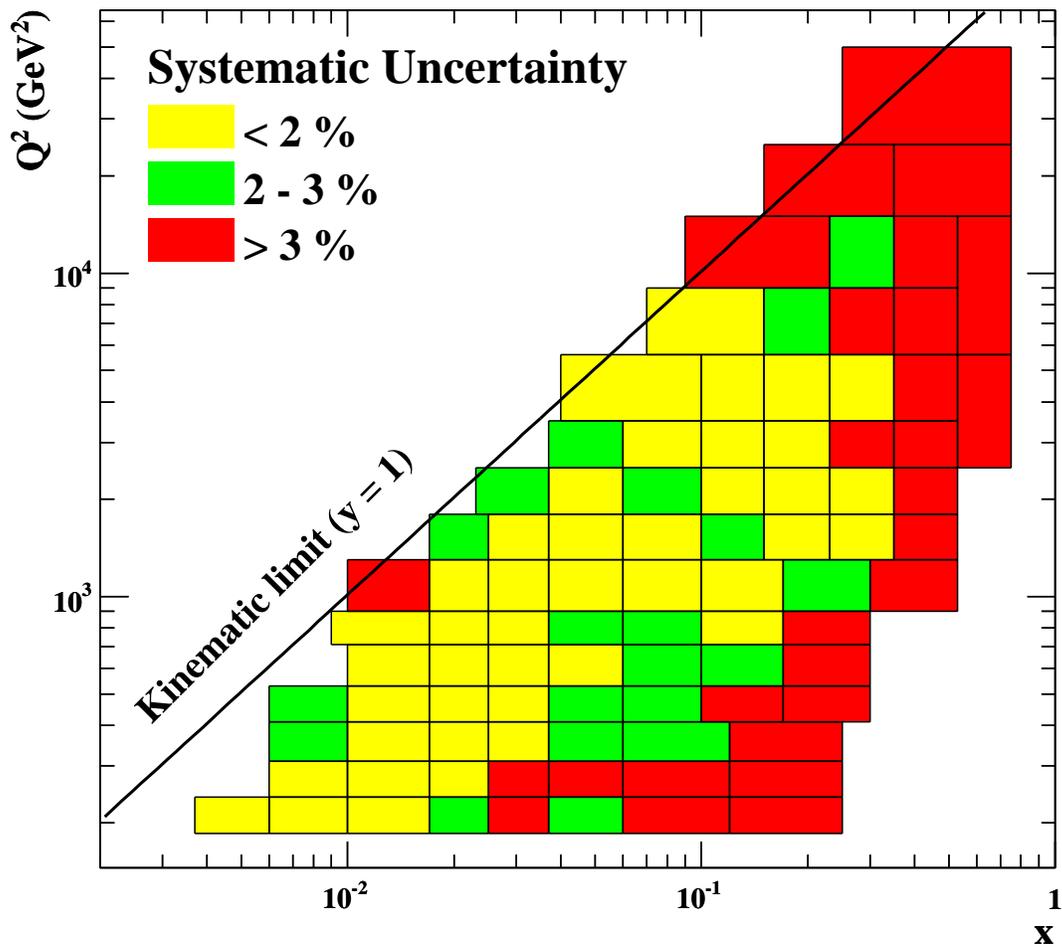}
\end{center}
\vspace{-1cm}
\caption{
  The total systematic uncertainty for the bins
  used in the reduced cross section measurements.
}
\label{fig-sys}
\vfill
\end{figure}

Systematic uncertainties were estimated by re-calculating the cross
sections after modifying the analysis to account for known uncertainties.
The positive and negative deviations from the nominal cross-section values
were added in quadrature separately
to obtain the total positive and negative systematic uncertainty.
The total systematic uncertainties for the bins
used in the reduced cross section measurements are shown in \fig{sys}.
The description of each systematic uncertainty follows.

The following systematic uncertainties were treated
as correlated between bins :
\begin{itemize}
\item $\{\delta_1\}$ to estimate the systematic uncertainty
associated with the electron finder,
an alternative electron-finding algorithm~\cite{nim:a365:508}
was used and the results were compared to those using the nominal algorithm.
In addition, to evaluate the systematic uncertainty of electron finding
in an environment of densely packed energy deposits,
the electron isolation requirement was varied by $\pm 2 \gev$.
These two checks were combined to give the systematic uncertainty
from electron finding which was less than 1\% for the bulk of the phase space.
In the double-differential cross-section bins at high $Q^{2}$ and high $y$,
the uncertainty was about 4\%, and increased to 18\% in the high-$y$ bins
of $d\sigma / dy$ for $Q^{2} > 3\,000 \gev^{2}$;

\item $\{\delta_2\}$ the variation of the electron energy scale
by $\pm 2 \%$ in the MC resulted in changes less than $1\%$
in the cross sections over most of the kinematic region
due to the use of the DA reconstruction method.
The effect was at most  $5\%$ at high $y$ in $d\sigma/dy$;

\item $\{\delta_3\}$ the nominal procedure to calculate $\gamma_{\rm max}$
used the low-$Q^2$ sample.
To account for the $Q^2$ dependence of $\gamma_{\rm max}$
in the backsplash removal procedure,
it was also derived using a high-$Q^2$ sample
and the results were compared.
The effect on the cross sections was generally less than $1\%$,
but increased to typically $5\%$ in the high-$x$ bins;

\item $\{\delta_4\}$ the systematic uncertainty in the parton-shower scheme
was evaluated by using the {\sc meps} model of {\sc lepto}
to calculate the acceptance instead of {\sc ariadne}
\footnote{Since the simulation of parton-shower scheme
also changes the description of the electron isolation,
the comparison was made without the electron isolation requirement
to prevent double counting of systematic errors.}.
The uncertainty was typically within $\pm 2\%$ but reached 5\%
in some bins of the double-differential cross sections;

\item $\{\delta_5\}$ the cut of $20 \cm$ on the projected radius
of the hadronic angle onto the FCAL was varied by $\pm 3 \cm$.
The cross sections typically changed less than $\pm 1\%$.
The effect increased up to $\pm 6\%$ for the highest $x$ bins
of both $d\sigma/dx$ and the double-differential cross section;

\item $\{\delta_6\}$ the uncertainty due to ``overlay'' events,
in which a normal DIS event coincided with additional energy deposits
in the RCAL from some other interaction, and photoproduction contamination
was estimated by narrowing or widening the $38 < \delta < 65\gev$ interval 
symmetrically by $\pm 4\gev$. The effect on the cross sections was typically
below 2\%. In a few high-$Q^2$ bins the uncertainty was
as large as 4\%;

\item $\{\delta_7\}$ systematic uncertainties
arising from the normalisation of the photoproduction background
were estimated by changing the background normalisation
by $\pm 40\%$.
In addition, systematic uncertainties arising from the estimation
of the photoproduction background were also estimated
by reducing the cut on $y_e$ to $y_e < 0.9$.
The resulting changes in the cross sections were typically below $\pm 1\%$,
and at most $2\%$ in the high-$Q^2$ bins of the double-differential cross-section.
\end{itemize}

The following systematic uncertainties are either small
or not correlated between bins :

\begin{itemize}
\item $\{\delta_8\}$ the energy resolution used in the MC
for the scattered electron was varied by $\pm 1\%$,
and the effect was less than $0.5 \%$ over the full kinematic range;

\item $\{\delta_9\}$ to reflect uncertainties in the alignment of the CAL
with respect to the CTD,
the electron scattering angle was varied by $\pm 1 \mrad$.
Typically, the deviations were within $\pm 1\%$ over the full kinematic range;

\item $\{\delta_{10}\}$ to account for differences
in the description of the $p_e^{\rm trk}$
distribution between data and MC,
the $p_e^{\rm trk}$ requirement was varied by $\pm 1 \gev$,
resulting in a variation of the cross section by $\pm 1\%$
over most of the kinematic range,
and up to $2\%$ in a few double-differential cross-section bins;

\item $\{\delta_{11}\}$ the uncertainty resulting from the hadronic energy scale
was evaluated by varying the hadronic energy in the MC
by $\pm 1\%$.
This caused changes of less than $\pm 1\%$ over the full kinematic range
in the MC;

\item $\{\delta_{12}\}$ the DCA requirement was changed to $8 \cm$
to estimate the uncertainty in the background contamination
due to falsely identified electrons.
The uncertainties in the cross sections associated with
this variation were below $\pm 1\%$ over the full kinematic range;

\item $\{\delta_{13}\}$ the systematic uncertainty associated
with cosmic-ray rejection was evaluated
by varying the $p_T/\sqrt{E_T}$ cut by $\pm 1 \sqrt{\Gev}$.
The cross-section uncertainties were below $\pm 1 \%$
over the full kinematic range.
\end{itemize}

The relative uncertainty in the measured polarisation was
3.6\% using the LPOL and 4.2\% using the TPOL.
The choice of polarimeter was made run-by-run
to maximise the available luminosity
and minimise the uncertainty in the measured polarisation.
The measured luminosity was assigned a relative uncertainty of 2.6\%.
The uncertainties in the luminosity and polarisation measurements were 
not included in the total systematic uncertainty shown in the final results.

\section{Results}
\label{sec-res}

\subsection{Unpolarised cross sections} 
\begin{figure}[p]
\vfill
\begin{center}
\includegraphics[width=6in]{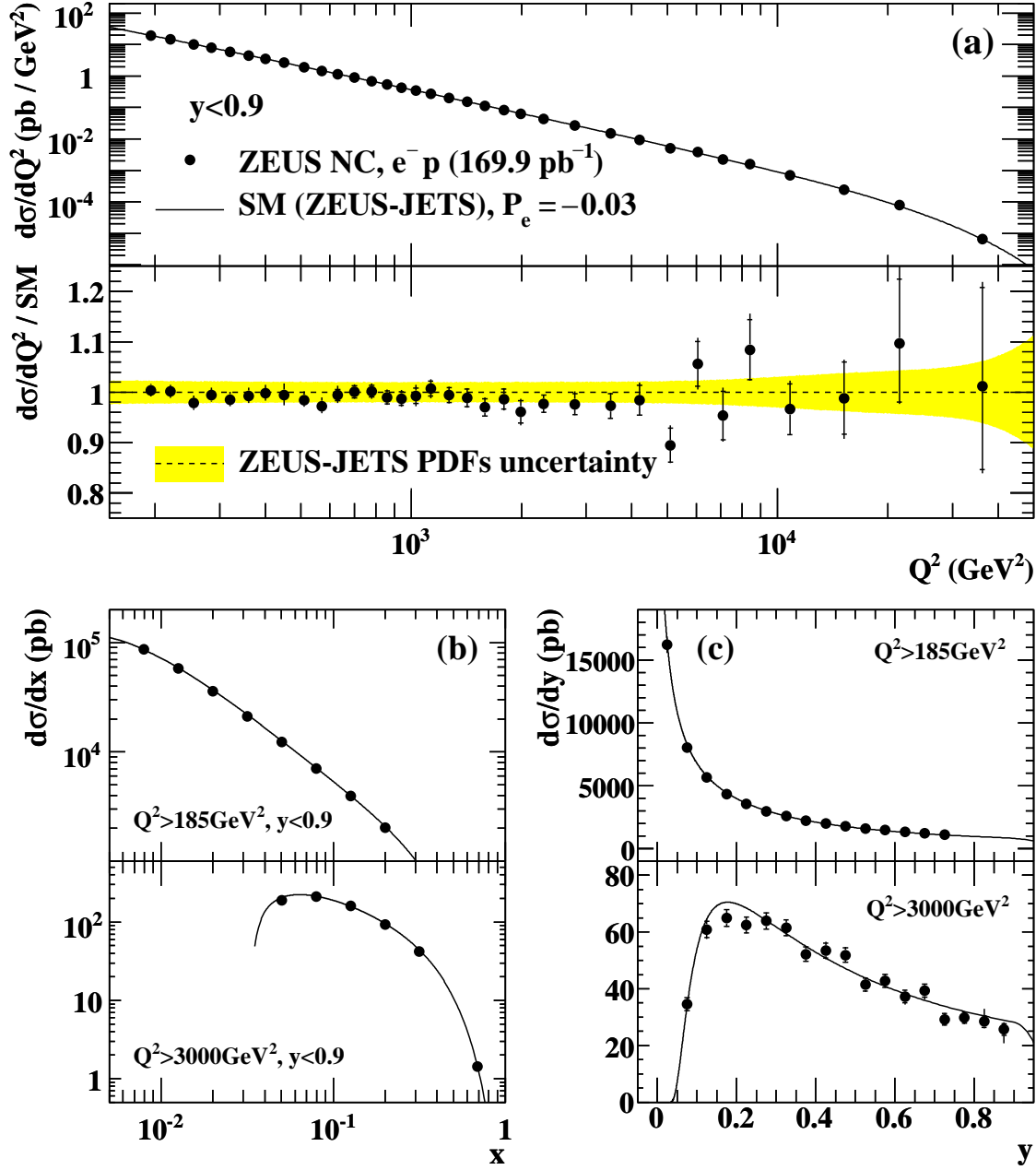}
\end{center}
\caption{
  The $e^- p$ NC DIS cross sections:
  (a) $d\sigma/dQ^{2}$ for $y < 0.9$
  and the ratio to the SM prediction,
  (b) $d\sigma/dx$ for $Q^2 > 185 \gev^2$ and $Q^2 > 3000 \gev^2$
  for $y < 0.9$ and
  (c) $d\sigma/dy$ for $Q^2 > 185 \gev^2$ and $Q^2 > 3000 \gev^2$.
  The closed circles represent data points in which
  the inner error bars show the statistical uncertainty
  while the outer bars show the statistical and systematic
  uncertainties added in quadrature.
  The curves show the predictions of the SM evaluated using
  the ZEUS-JETS PDFs at a polarisation corresponding to the residual
  polarisation in the data
  and the shaded band shows the uncertainties from the ZEUS-JETS PDFs.
}
\label{fig-allsing}
\vfill
\end{figure}

The single-differential cross-sections
with respect to $Q^{2}$, $x$ and $y$,
tabulated in Tables~\ref{tab:dsdq2Total}, \ref{tab:dsdq2TotalSys},
\ref{tab:dsdxTotal}, \ref{tab:dsdxTotalSys},
\ref{tab:dsdyTotal} and \ref{tab:dsdyTotalSys},
are shown in \fig{allsing}
for $Q^{2} > 185\gev^2$ and $y < 0.9$
for the combined positive and negative polarisation samples
having a residual polarisation of $-0.03$. 
The cross-sections $d\sigma/dx$ and
$d\sigma/dy$ for $Q^2 > 3\,000 \gev^2$ and $y < 0.9$
are also presented in \fig{allsing}.
The measured cross sections demonstrate the precision of this measurement.
The measured cross sections are well described by the SM
prediction evaluated using the ZEUS-JETS PDFs\cite{epj:c42:1}.
The measurement of $d\sigma/dQ^{2}$ spans
two orders of magnitude in $Q^{2}$,
and at $Q^2 \sim 40\,000 \gev^2$,
the spatial resolution reaches $\sim 10^{-18}$ m.

\begin{figure}[p]
\vfill
\begin{center}
\includegraphics[width=6in]{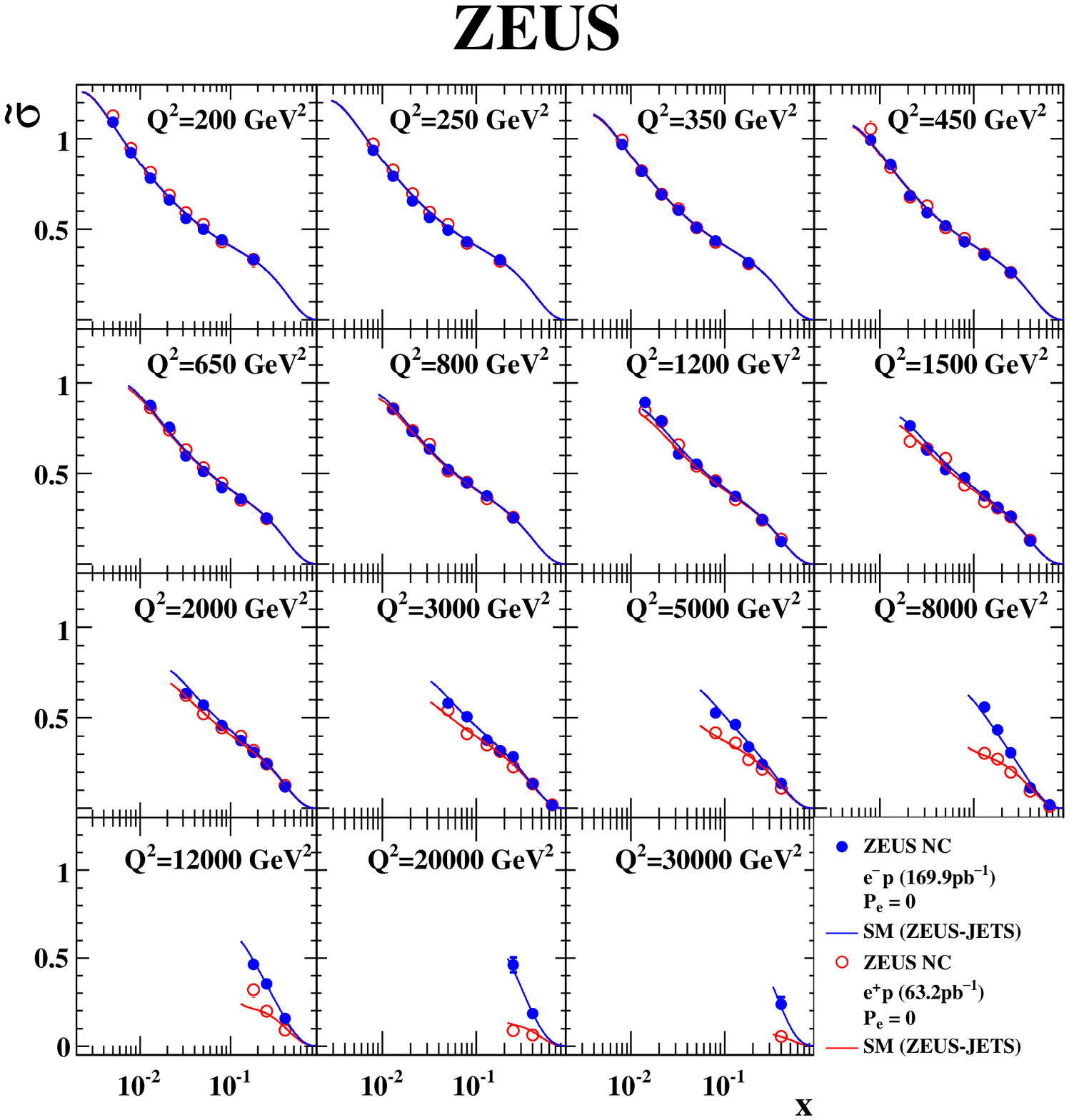}
\end{center}
\caption{
  The $e^\pm p$ unpolarised NC DIS reduced cross section $\sitil$ 
  plotted as a function of $x$ at fixed $Q^2$.
  The closed (open) circles represent data points
  for $e^- p$ ($e^+ p$) collisions in which
  the inner error bars show the statistical uncertainty
  while the outer bars show the statistical and systematic uncertainties
  added in quadrature,
  although errors are too small to be seen in most cases.
  The curves show the predictions of the SM
  evaluated using the ZEUS-JETS PDFs.
}
\label{fig-red_unpol}
\vfill
\end{figure}

The reduced cross sections of unpolarised $e^ - p$ NC DIS,
tabulated in Tables~\ref{tab:ds2dxdq2Total_1} and \ref{tab:ds2dxdq2TotalSys_1},
are presented in \fig{red_unpol}
with the residual polarisation of -0.03 corrected using theoretical predictions.
The correction factors were at most 2\% in the highest-$Q^{2}$ bins.
The SM predictions are in good agreement with the measurements
over the full kinematic range. 
Also shown are unpolarised $e^ + p$ NC DIS measurements
with an integrated luminosity of $63.2 \pb^{-1}$
collected in 1999 and 2000\cite{pr:d70:052001}.
As discussed earlier, a significant difference
between the $e^- p$ and $e^+ p$ unpolarised reduced cross sections
is seen at high-$Q^2$ values due to the contribution of $x\tilde{F_3}$.

\begin{figure}[p]
\vfill
\begin{center}
\includegraphics[width=6in]{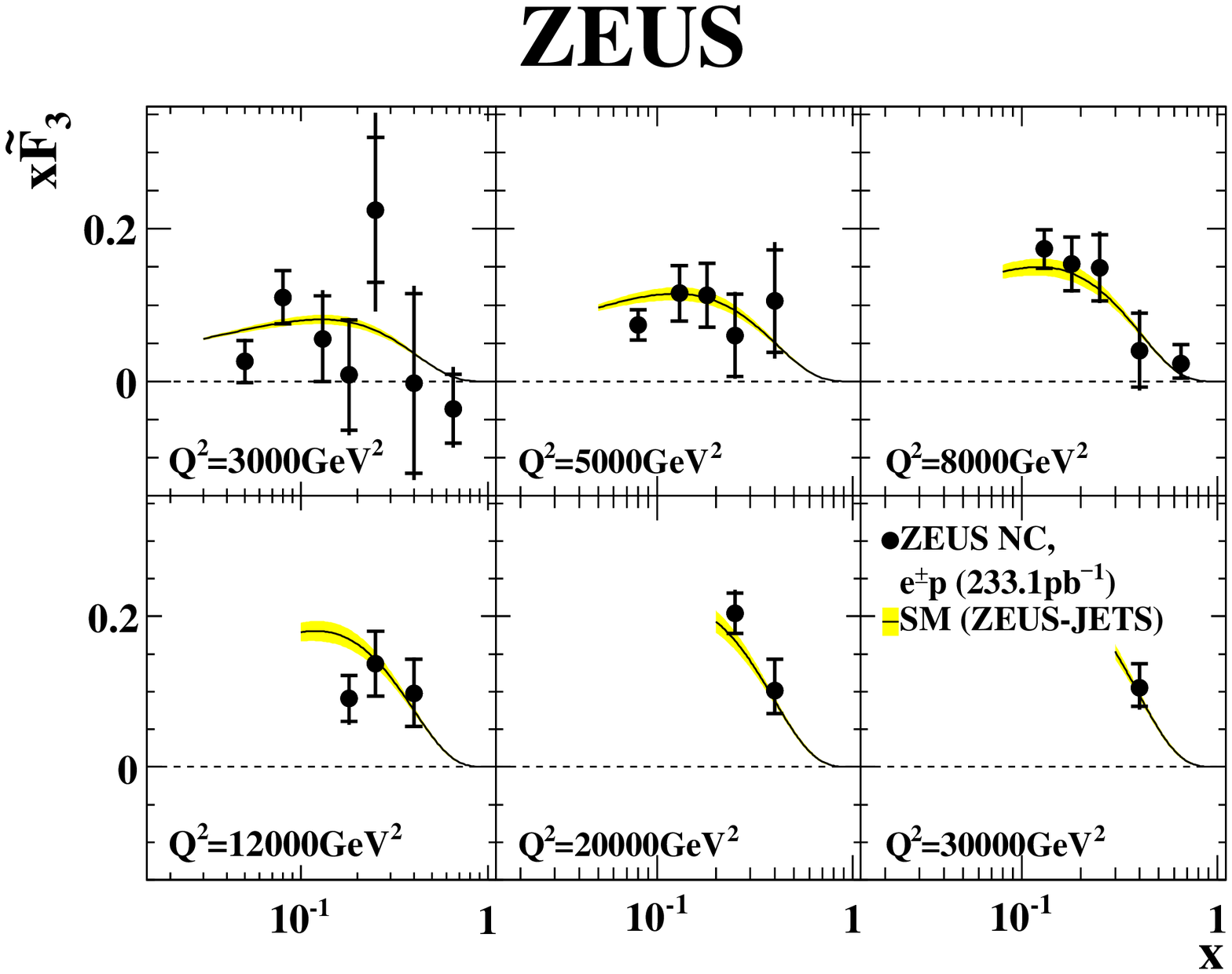}
\end{center}
\caption{
  The structure function $x\tilde{F_3}$ plotted
  as a function of $x$ at fixed-$Q^{2}$.
  The closed circles represent the ZEUS data.
  The inner error bars show the statistical uncertainty
  while the outer ones show the statistical and systematic uncertainties
  added in quadrature.
  The curves show the predictions of the SM
  evaluated using the ZEUS-JETS PDFs
  with the shaded band indicating the uncertainties.
}
\label{fig-xf3}
\vfill
\end{figure}

\Fig{xf3}
shows the structure function $x\tilde{F_3}$ obtained from the
unpolarised $e^- p$ and $e^+ p$ reduced cross sections in the high-$Q^2$ region.
The results are also given in Table~\ref{tab:xF3}.
The reduced cross sections visibly differ
and are well described by the SM predictions.

\begin{figure}[p]
\vfill
\begin{center}
\includegraphics[width=6in]{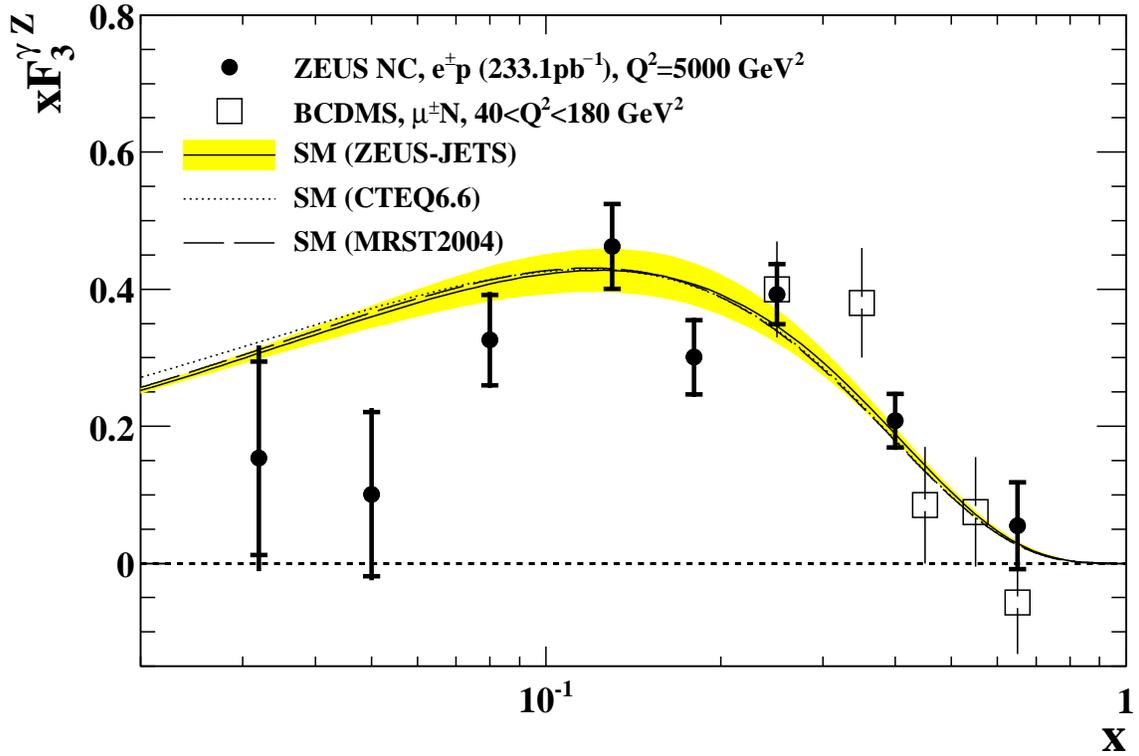} 
\end{center}
\caption{
  The structure function $xF_{3}^{\gamma Z}$ extrapolated
  to a single $Q^2$ value of $5\,000\gev^2$ and plotted as a function of $x$. 
  The closed circles represent data points in which
  the inner error bars show the statistical uncertainty
  while the outer bars show the statistical and systematic uncertainties
  added in quadrature.
  The curves show the predictions of the SM
  evaluated using several PDFs : ZEUS-JETS (shaded band shows the uncertainties),
  CTEQ6.6 and MRST2004.
  The measurements by the BCDMS collaboration are shown
  as open squares.
}
\label{fig-xf3_gz}
\vfill

\end{figure}

The structure function $xF_{3}^{\gamma Z}$ has little dependence on $Q^{2}$
and so the measurements from $1\,500 < Q^2 < 30\,000 \gev^{2}$ 
have been extrapolated to $5\,000 \gev^{2}$, and then averaged,
to obtain a higher statistical significance.
The structure function $xF_{3}^{\gamma Z}$ measured at $Q^2 = 5\,000 \gev^{2}$,
tabulated in Table~\ref{tab:xF3gz},
is shown in \fig{xf3_gz}.
It is well described by the SM predictions.
The integral of $xF^{\gamma Z}_{3}$
in the region of $0.032 < x < 0.65 $ is

\be
\int^{0.65}_{0.032} \frac{dx}{x}xF^{\gamma Z}_{3}
= 0.80 \pm 0.08(\text{stat.}) \pm 0.03(\text{syst.}).
\ee
This value is consistent with the SM prediction
of $0.94\pm 0.02$.
\Fig{xf3_gz} also shows the measurement of $xF^{\gamma Z}_{3}$
obtained by the BCDMS collaboration
from NC muon-carbon scattering at lower energies,
which was extracted over the kinematic range
$40 < Q^2 < 180 \gev^2$ and $0.2 < x < 0.7$~\cite{pl:b140:142}.
The measurements presented in this paper extend the $x$ range
for $xF_3^{\gamma Z}$ data down to $x \sim 0.03$,
well below the range of the BCDMS measurements.
Furthermore, they are extracted at higher $Q^2$ values
where perturbative QCD calculations are more reliable.
Moreover, this measurement is also free
from heavy-target corrections and isospin-symmetry assumptions
which are inherent in previous fixed-target measurements.
Therefore, at Bjorken $x$ values from $\sim 10^{-2}$ to $0.65$,
this measurement adds valuable information
to the global fits~\cite{epj:c12:375,epj:c23:73}
for parton distribution functions.

\subsection{Polarised cross sections}
\begin{figure}[p]
\vfill
\begin{center}
\includegraphics[width=6in]{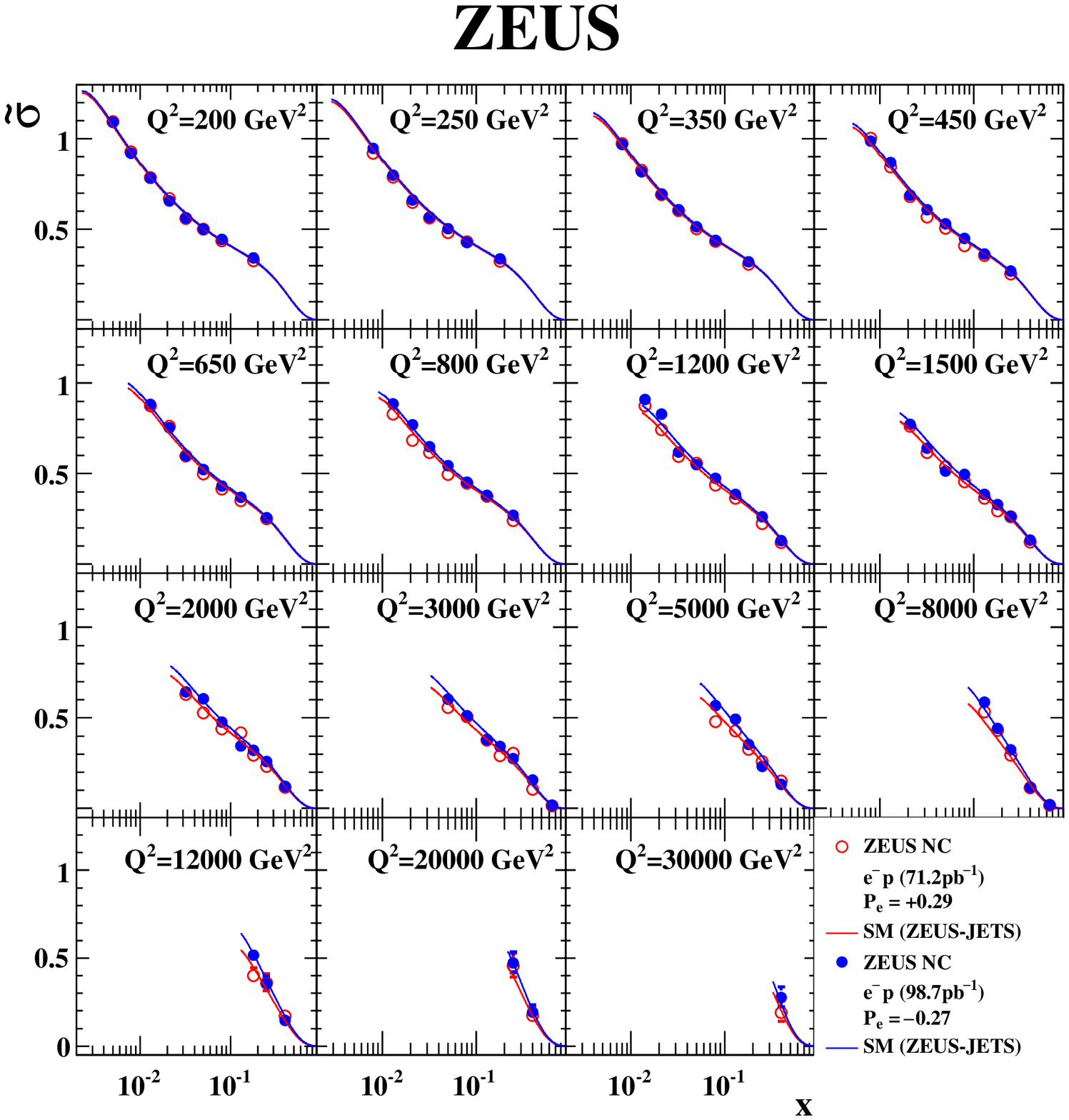}
\end{center}
\caption{
  The $e^- p$ NC DIS reduced cross section $\sitil$
  for positively and negatively polarised beams
  plotted as a function of $x$ at fixed $Q^2$.
  The closed (open) circles represent the ZEUS data
  for negative (positive) polarisation.
  The inner error bars show the statistical uncertainty
  while the outer bars show the statistical and systematic uncertainties
  added in quadrature.
  The curves show the predictions of the SM
  evaluated using the ZEUS-JETS PDFs. 
}
\label{fig-red_pol}
\vfill
\end{figure}

At HERA during 2005 and 2006
longitudinal polarisation effects in $e p$ DIS
become significant at the electroweak scale,
where the contributions of both $\gamma$ and $Z$ exchange
to the cross section are comparable.
The reduced cross sections
for positive and negative longitudinal polarisations,
tabulated in Tables~\ref{tab:ds2dxdq2Rh_1}, \ref{tab:ds2dxdq2RhSys_1},
\ref{tab:ds2dxdq2Lh_1} and \ref{tab:ds2dxdq2LhSys_1},
are shown separately in \fig{red_pol}
and are well described by the SM evaluated using the ZEUS-JETS PDFs.

\begin{figure}[p]
\vfill
\begin{center}
\includegraphics[width=6in]{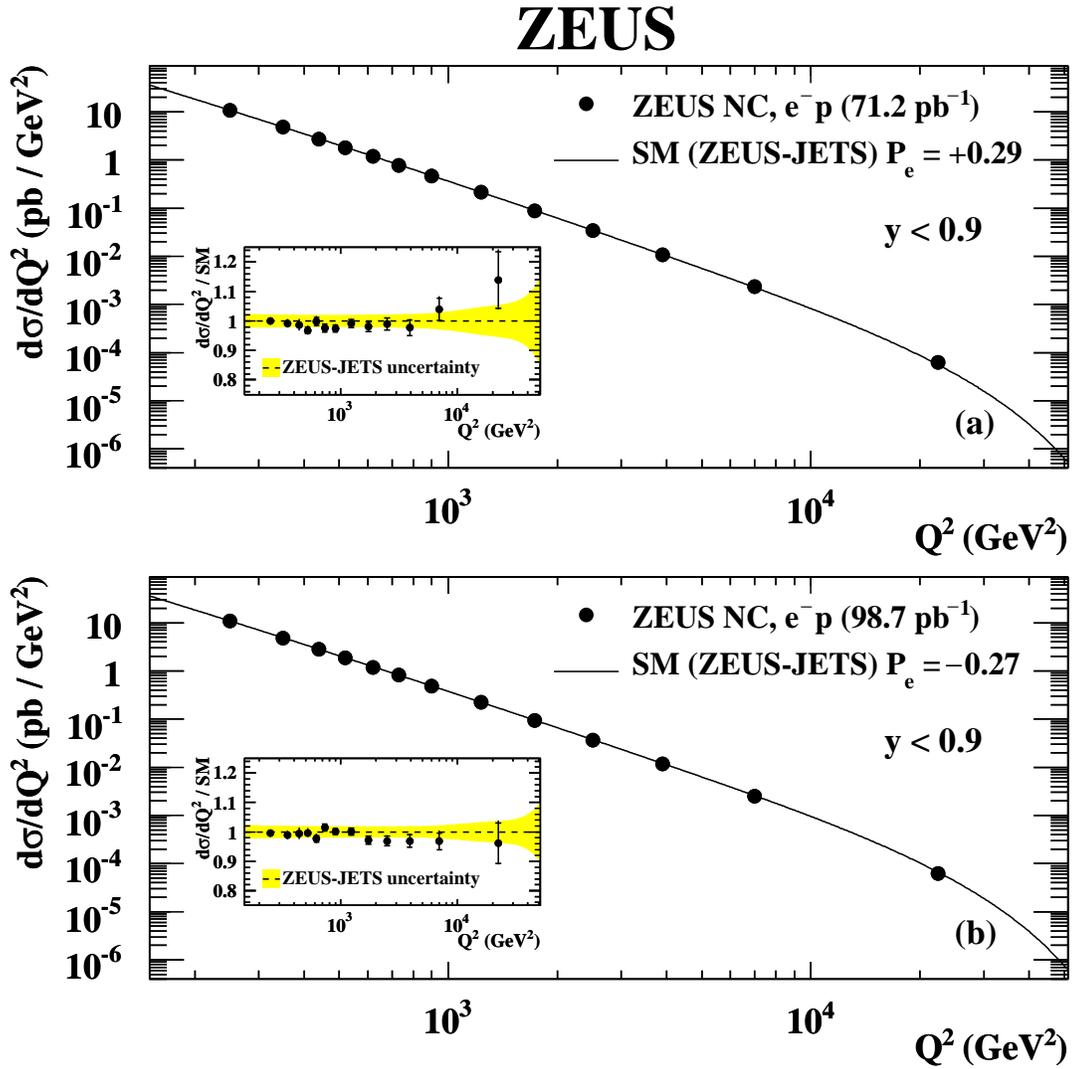}
\end{center}
\caption{
  The $e^- p$ NC DIS cross section $d\sigma/dQ^2$
  for (a) positive and (b) negative polarisation.
  The inset shows the ratio to the SM prediction.
  The closed circles represent the ZEUS data.
  The inner error bars show the statistical uncertainty
  while the outer bars show the statistical and systematic uncertainties
  added in quadrature.
  The curves show the predictions of the SM
  evaluated using the ZEUS-JETS PDFs.
}
\label{fig-dsdq2}
\vfill
\end{figure}

At high $Q^2$, a difference between the positively and negatively
polarised cross sections is predicted.
To demonstrate this effect,
the single-differential cross-section $d\sigma/dQ^{2}$ for $y < 0.9$,
tabulated in Tables~\ref{tab:dsdq2Rh} and \ref{tab:dsdq2RhSys},
was measured for positive and negative beam polarisations separately
and is shown in \fig{dsdq2}.
Both measurements are well described by the SM prediction.

\begin{figure}[p]
\vfill
\begin{center}
\includegraphics[width=6in]{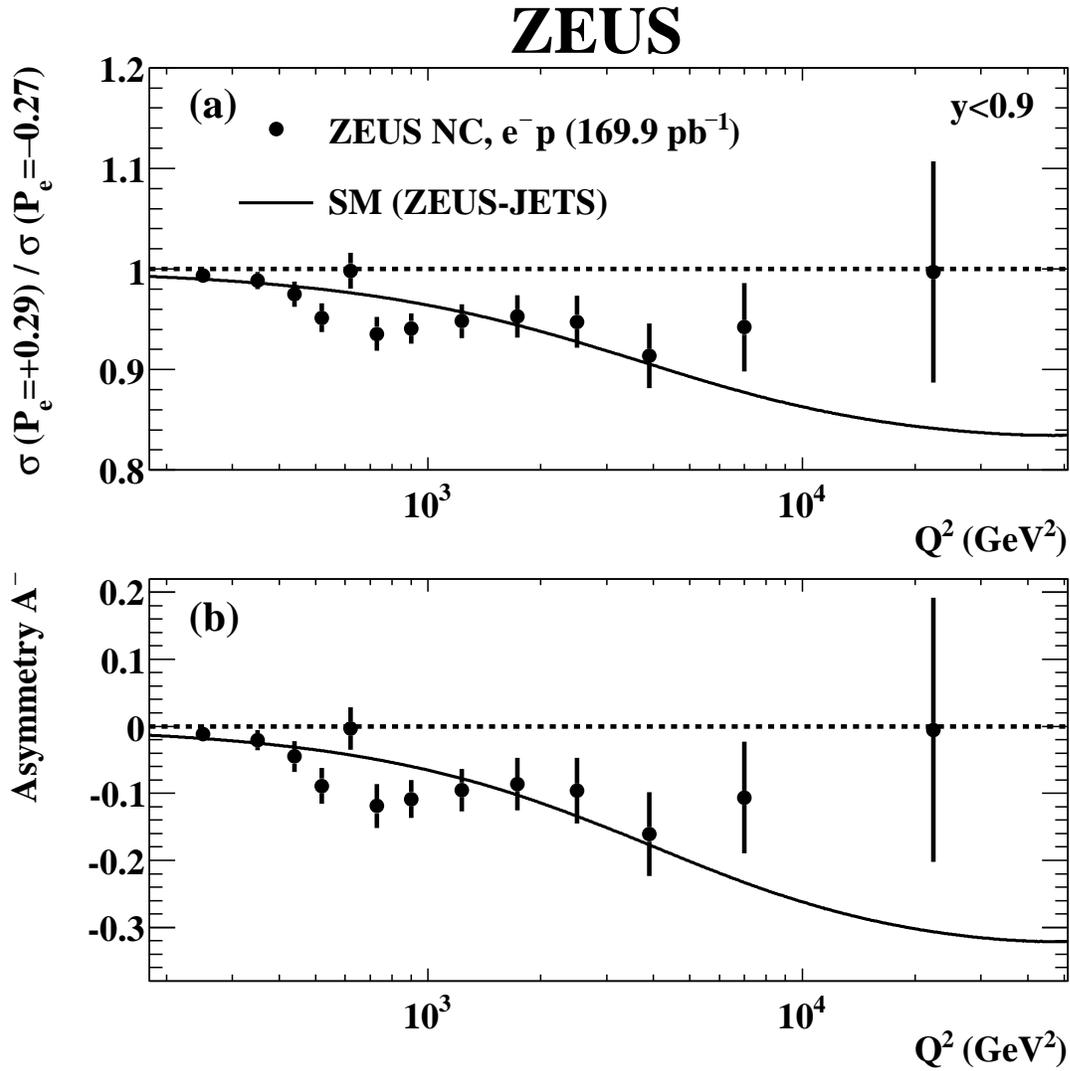}
\end{center}
\caption{
  The ratio of $d\sigma/dQ^{2}$ using positive and negative
  polarisation in (a),
  and the polarisation asymmetry $A^{-}$ as a function of $Q^2$ in (b).
  The closed circles represent ZEUS data.
  Only statistical uncertainties are considered
  as the systematic uncertainties are assumed to cancel.
  The curves show the predictions of the SM
  evaluated using the ZEUS-JETS PDFs.
}
\label{fig-asym}
\vfill
\end{figure}

The ratio of measured cross sections
for the two different polarisation states
are shown in \fig{asym} (a).
The difference between the two polarisation states
is clearly visible at higher $Q^{2}$.
The asymmetry $A^{-}$ (see Eq. \ref{eqn:asymMeas})
extracted from these measurements
is tabulated in Table~\ref{tab:asym}
and is shown in \fig{asym} (b),
where only statistical uncertainties are considered
as the systematic uncertainties are assumed to cancel.
The results compare well to the SM prediction.
The deviation of $A^{-}$ from zero shows the difference
in the behaviour of the two polarisation states
and is clear evidence of parity violation.

The effect of $\gamma$/$Z$ interference is quantified
by calculating the $\chi^2$ per degree of freedom of $A^{-}$
with respect both to zero and the SM prediction
using the ZEUS-JETS PDFs.
The $\chi^2/\rm{d.o.f.}$ with respect to zero is determined to be 5.5,
whereas the $\chi^2/\rm{d.o.f.}$ with respect to the SM prediction is 1.5.
Thus parity violation in $e p$ NC DIS at very small distances is
demonstrated at scales down to $10^{-18}$ m. 
At large $Q^{2}$ where the $u$ quark dominates the PDF
it is expected that $A^{-} \simeq 2 a_e v_u e_u/e_u^2 \simeq 0.3$.
The cross sections obtained from NC DIS measurements
can be used to constrain the NC quark couplings
within PDF fits~\cite{proc:dis:2006:shima},
and the polarised electron beam data provide
sensitivity to quark vector couplings.
Therefore, this measurement is
a stringent test of the electroweak sector of the Standard Model.

\section{Summary}
\label{sec-sum}

The cross sections for neutral current deep inelastic scattering in $e^-p$
collisions with a longitudinally polarised electron beam have been measured.
The measurements are based on a data sample with an integrated luminosity of
$169.9 \pb^{-1}$ collected with the ZEUS detector at HERA from 2005 to 2006
at a centre-of-mass energy of $318 \gev$.
The accessible range in $Q^2$ extended to $Q^2 = 50\,000 \gev^2$
and has allowed a stringent test of electroweak effects in the Standard Model.

The single-differential cross-sections with respect to $Q^2$, $x$ and $y$
are presented for $Q^2 > 185 \gev^2$ and $y < 0.9 $,
where the data obtained with negatively and positively polarised beams
are combined.
The cross sections $d\sigma/dx$ and $d\sigma/dy$
are also measured for $Q^2 > 3\,000 \gev^2$ and $y < 0.9 $.  
The reduced cross sections are measured in the kinematic range
$200 < Q^2 < 30\,000 \gev^2$ and $0.005 < x < 0.65$ at zero polarisation
by correcting the residual polarisation of the combined data sample.
These measurements are combined with previously measured
$e^+p$ neutral current cross sections to extract $x\tilde{F_3}$.
In addition, the interference structure function $xF_{3}^{\gamma Z}$
is extracted at an average value of $Q^{2} = 5\,000 \gev^2$.

The reduced cross-sections and
the single-differential cross-section $d\sigma/dQ^2$
have also been measured separately for positive and negative values
of the longitudinal polarisation of the electron beam.
Parity violation is observed through the polarisation asymmetry $A^-$.
The measured cross sections confirm the predictions of the Standard Model
and provide strong constraints at the electroweak scale.

\section*{Acknowledgements}

We appreciate the contributions to the construction and maintenance
of the ZEUS detector of many people who are not listed as authors.
The HERA machine group and the DESY computing staff
are especially acknowledged for their success
in providing excellent operation of the collider
and the data-analysis environment.
We thank the DESY directorate for their strong support and encouragement.

\vfill\eject

{
\def\bibname{\Large\bf References}
\def\refname{\Large\bf References}
\pagestyle{plain}
\ifzeusbst
  \bibliographystyle{./BiBTeX/bst/l4z_default}
\fi
\ifzdrftbst
  \bibliographystyle{./BiBTeX/bst/l4z_draft}
\fi
\ifzbstepj
  \bibliographystyle{./BiBTeX/bst/l4z_epj}
\fi
\ifzbstnp
  \bibliographystyle{./BiBTeX/bst/l4z_np}
\fi
\ifzbstpl
  \bibliographystyle{./BiBTeX/bst/l4z_pl}
\fi
{\raggedright
\bibliography{./BiBTeX/user/syn.bib,%
              ./BiBTeX/bib/l4z_articles.bib,%
              ./BiBTeX/bib/l4z_books.bib,%
              ./BiBTeX/bib/l4z_conferences.bib,%
              ./BiBTeX/bib/l4z_h1.bib,%
              ./BiBTeX/bib/l4z_misc.bib,%
              ./BiBTeX/bib/l4z_old.bib,%
              ./BiBTeX/bib/l4z_preprints.bib,%
              ./BiBTeX/bib/l4z_replaced.bib,%
              ./BiBTeX/bib/l4z_temporary.bib,%
              ./BiBTeX/bib/l4z_zeus.bib}}
}
\vfill\eject

\newpage
\begin{table}
\begin{scriptsize}
\begin{center} \begin{tabular}[t]{|rcr|r|rll|r|r|} \hline
\multicolumn{3}{|c|}{$Q^2$ range} & \multicolumn{1}{c|}{$Q^2_c$} & \multicolumn{3}{c|}{$d\sigma / dQ^{2}$} & \multicolumn{1}{c|}{$N_{\text{data}}$} & \multicolumn{1}{c|}{$N^{\text{MC}}_{\text{bg}}$}\\
\multicolumn{3}{|c|}{($\gev^{2}$)} & \multicolumn{1}{c|}{($\gev^{2}$)} & \multicolumn{3}{c|}{($\pb / \gev^{2}$)} & & \\ \hline \hline
185 & -- & 210 & 195 & (1.96 & $\pm$ 0.01 & $^{+0.02} _{-0.02}$ )$\cdot 10^{1}$ & 58678 & 65.4 \\
210 & -- & 240 & 220 & (1.47 & $\pm$ 0.01 & $^{+0.02} _{-0.02}$ )$\cdot 10^{1}$ & 51660 & 83.5 \\
240 & -- & 270 & 255 & (1.01 & $\pm$ 0.01 & $^{+0.01} _{-0.01}$ )$\cdot 10^{1}$ & 37659 & 52.4 \\
270 & -- & 300 & 285 & 7.89 & $\pm$ 0.05 & $^{+0.11} _{-0.09}$ & 29355 & 51.1 \\
300 & -- & 340 & 320 & 5.93 & $\pm$ 0.04 & $^{+0.08} _{-0.07}$ & 29377 & 53.3 \\
340 & -- & 380 & 360 & 4.51 & $\pm$ 0.03 & $^{+0.06} _{-0.05}$ & 22298 & 55.0 \\
380 & -- & 430 & 400 & 3.52 & $\pm$ 0.03 & $^{+0.05} _{-0.04}$ & 21008 & 59.4 \\
430 & -- & 480 & 450 & 2.63 & $\pm$ 0.02 & $^{+0.06} _{-0.05}$ & 15502 & 54.9 \\
480 & -- & 540 & 510 & 1.92 & $\pm$ 0.02 & $^{+0.03} _{-0.02}$ & 13332 & 45.5 \\
540 & -- & 600 & 570 & 1.45 & $\pm$ 0.02 & $^{+0.02} _{-0.01}$ & 9290 & 49.4 \\
600 & -- & 670 & 630 & 1.16 & $\pm$ 0.01 & $^{+0.02} _{-0.01}$ & 8405 & 48.6 \\
670 & -- & 740 & 700 & (8.98 & $\pm$ 0.11 & $^{+0.07} _{-0.07}$ )$\cdot 10^{-1}$ & 7483 & 20.6 \\
740 & -- & 820 & 780 & (6.88 & $\pm$ 0.08 & $^{+0.06} _{-0.06}$ )$\cdot 10^{-1}$ & 7608 & 26.3 \\
820 & -- & 900 & 860 & (5.33 & $\pm$ 0.07 & $^{+0.04} _{-0.04}$ )$\cdot 10^{-1}$ & 6287 & 33.5 \\
900 & -- & 990 & 940 & (4.27 & $\pm$ 0.06 & $^{+0.05} _{-0.04}$ )$\cdot 10^{-1}$ & 5690 & 32.1 \\
990 & -- & 1080 & 1030 & (3.41 & $\pm$ 0.05 & $^{+0.05} _{-0.05}$ )$\cdot 10^{-1}$ & 4644 & 29.4 \\
1080 & -- & 1200 & 1130 & (2.75 & $\pm$ 0.04 & $^{+0.03} _{-0.03}$ )$\cdot 10^{-1}$ & 4907 & 30.5 \\
1200 & -- & 1350 & 1270 & (2.02 & $\pm$ 0.03 & $^{+0.01} _{-0.01}$ )$\cdot 10^{-1}$ & 4645 & 24.4 \\
1350 & -- & 1500 & 1420 & (1.52 & $\pm$ 0.03 & $^{+0.02} _{-0.01}$ )$\cdot 10^{-1}$ & 3499 & 31.8 \\
1500 & -- & 1700 & 1590 & (1.12 & $\pm$ 0.02 & $^{+0.01} _{-0.01}$ )$\cdot 10^{-1}$ & 3452 & 27.6 \\
1700 & -- & 1900 & 1790 & (8.40 & $\pm$ 0.17 & $^{+0.09} _{-0.09}$ )$\cdot 10^{-2}$ & 2611 & 18.8 \\
1900 & -- & 2100 & 1990 & (6.25 & $\pm$ 0.14 & $^{+0.09} _{-0.10}$ )$\cdot 10^{-2}$ & 1957 & 7.2 \\
2100 & -- & 2600 & 2300 & (4.39 & $\pm$ 0.08 & $^{+0.03} _{-0.03}$ )$\cdot 10^{-2}$ & 3315 & 13.3 \\
2600 & -- & 3200 & 2800 & (2.65 & $\pm$ 0.06 & $^{+0.04} _{-0.02}$ )$\cdot 10^{-2}$ & 2345 & 21.9 \\
3200 & -- & 3900 & 3500 & (1.48 & $\pm$ 0.04 & $^{+0.01} _{-0.01}$ )$\cdot 10^{-2}$ & 1597 & 4.5 \\
3900 & -- & 4700 & 4200 & (9.32 & $\pm$ 0.28 & $^{+0.18} _{-0.08}$ )$\cdot 10^{-3}$ & 1111 & 2.9 \\
4700 & -- & 5600 & 5100 & (5.08 & $\pm$ 0.19 & $^{+0.11} _{-0.06}$ )$\cdot 10^{-3}$ & 708 & 2.8 \\
5600 & -- & 6600 & 6050 & (3.81 & $\pm$ 0.16 & $^{+0.10} _{-0.09}$ )$\cdot 10^{-3}$ & 586 & 4.4 \\
6600 & -- & 7800 & 7100 & (2.23 & $\pm$ 0.11 & $^{+0.07} _{-0.04}$ )$\cdot 10^{-3}$ & 401 & 1.5 \\
7800 & -- & 9200 & 8400 & (1.59 & $\pm$ 0.09 & $^{+0.06} _{-0.02}$ )$\cdot 10^{-3}$ & 331 & 0.0 \\
9200 & -- & 12800 & 10800 & (6.90 & $\pm$ 0.36 & $^{+0.19} _{-0.05}$ )$\cdot 10^{-4}$ & 369 & 1.4 \\
12800 & -- & 18100 & 15200 & (2.45 & $\pm$ 0.18 & $^{+0.07} _{-0.09}$ )$\cdot 10^{-4}$ & 193 & 1.4 \\
18100 & -- & 25600 & 21500 & (7.99 & $^{+0.93} _{-0.85}$ & $^{+0.65} _{-0.24}$ )$\cdot 10^{-5}$ & 97 & 3.0 \\
25600 & -- & 51200 & 36200 & (6.62 & $^{+1.28} _{-1.08}$ & $^{+0.45} _{-0.33}$ )$\cdot 10^{-6}$ & 37 & 0.0 \\

\hline
\end{tabular}
\end{center}
\caption[]
{The single differential cross section $d\sigma / dQ^{2}$ ($y < 0.9$)
for the reaction $e^{-}p \rightarrow e^{-}X$ ($\mathcal{L} = 169.9 \pbi, P_{e} = -0.03$).
The bin range, bin centre ($Q^2_c$) and measured cross section
corrected to the electroweak Born level are shown.
The first (second) error on the cross section
corresponds to the statistical (systematic) uncertainties.
The number of observed data events ($N_{\text{data}}$)
and simulated background events ($N^{\text{MC}}_{\text{bg}}$) are also shown.}
\label{tab:dsdq2Total}
\end{scriptsize}
\end{table}

\newpage
\begin{table}\begin{scriptsize}
\begin{center} \begin{tabular}[t]{|r|rl|c|c||c|c|c|c|c|c|c|c|} \hline
\multicolumn{1}{|c|}{$Q^2_c$} & \multicolumn{2}{c|}{$d\sigma / dQ^{2}$} & \multicolumn{1}{c|}{stat.} & \multicolumn{1}{c||}{sys.} & \multicolumn{1}{c|}{$\delta_{1}$} & \multicolumn{1}{c|}{$\delta_{2}$} & \multicolumn{1}{c|}{$\delta_{3}$} & \multicolumn{1}{c|}{$\delta_{4}$} & \multicolumn{1}{c|}{$\delta_{5}$} & \multicolumn{1}{c|}{$\delta_{6}$} & \multicolumn{1}{c|}{$\delta_{7}$} & \multicolumn{1}{c|}{$\delta_{8} - \delta_{13}$} \\
\multicolumn{1}{|c|}{($\gev^{2}$)} & \multicolumn{2}{c|}{($\pb / \gev^{2}$)} & \multicolumn{1}{c|}{(\%)} & \multicolumn{1}{c||}{(\%)} & \multicolumn{1}{c|}{(\%)} & \multicolumn{1}{c|}{(\%)} & \multicolumn{1}{c|}{(\%)} & \multicolumn{1}{c|}{(\%)} & \multicolumn{1}{c|}{(\%)} & \multicolumn{1}{c|}{(\%)} & \multicolumn{1}{c|}{(\%)} & \multicolumn{1}{c|}{(\%)} \\ \hline \hline
195 & 1.96 & $\cdot 10^{1}$ & $\pm$ 0.4 & $^{+1.2} _{-1.1}$ & $^{+0.0} _{-0.1}$ & $^{-0.5} _{+0.6}$ & $^{+0.4} _{-0.4}$ & $^{+0.1} _{-0.1}$ & $^{-0.1} _{+0.1}$ & $^{+0.1} _{+0.2}$ & $^{+0.0} _{-0.0}$ & $^{+0.9} _{-0.9}$ \\
220 & 1.47 & $\cdot 10^{1}$ & $\pm$ 0.5 & $^{+1.3} _{-1.2}$ & $^{+0.4} _{-0.0}$ & $^{-0.5} _{+0.6}$ & $^{+0.4} _{-0.4}$ & $^{+0.6} _{-0.6}$ & $^{-0.0} _{+0.1}$ & $^{-0.0} _{+0.2}$ & $^{+0.1} _{-0.1}$ & $^{+0.6} _{-0.8}$ \\
255 & 1.01 & $\cdot 10^{1}$ & $\pm$ 0.6 & $^{+1.5} _{-1.3}$ & $^{+0.6} _{-0.0}$ & $^{-0.5} _{+0.6}$ & $^{+0.8} _{-0.8}$ & $^{+0.6} _{-0.6}$ & $^{-0.1} _{+0.2}$ & $^{-0.1} _{+0.2}$ & $^{+0.1} _{-0.1}$ & $^{+0.7} _{-0.7}$ \\
285 & 7.89 & & $\pm$ 0.6 & $^{+1.4} _{-1.1}$ & $^{+0.7} _{-0.0}$ & $^{-0.5} _{+0.7}$ & $^{+0.2} _{-0.2}$ & $^{+0.7} _{-0.7}$ & $^{-0.1} _{+0.2}$ & $^{-0.1} _{+0.1}$ & $^{+0.1} _{-0.1}$ & $^{+0.6} _{-0.6}$ \\
320 & 5.93 & & $\pm$ 0.6 & $^{+1.4} _{-1.2}$ & $^{+0.5} _{+0.0}$ & $^{-0.5} _{+0.7}$ & $^{+0.6} _{-0.6}$ & $^{+0.5} _{-0.5}$ & $^{-0.1} _{+0.3}$ & $^{-0.2} _{+0.2}$ & $^{+0.1} _{-0.1}$ & $^{+0.7} _{-0.7}$ \\
360 & 4.51 & & $\pm$ 0.7 & $^{+1.4} _{-1.2}$ & $^{+0.8} _{-0.1}$ & $^{-0.5} _{+0.6}$ & $^{+0.4} _{-0.4}$ & $^{+0.7} _{-0.7}$ & $^{+0.1} _{+0.1}$ & $^{-0.2} _{+0.1}$ & $^{+0.1} _{-0.1}$ & $^{+0.6} _{-0.7}$ \\
400 & 3.52 & & $\pm$ 0.7 & $^{+1.5} _{-1.2}$ & $^{+0.5} _{-0.0}$ & $^{-0.5} _{+0.6}$ & $^{-0.1} _{+0.1}$ & $^{+0.9} _{-0.9}$ & $^{-0.0} _{+0.0}$ & $^{-0.4} _{+0.5}$ & $^{+0.1} _{-0.1}$ & $^{+0.6} _{-0.5}$ \\
450 & 2.63 & & $\pm$ 0.8 & $^{+2.1} _{-1.9}$ & $^{+0.5} _{-0.0}$ & $^{-0.5} _{+0.6}$ & $^{+0.9} _{-0.9}$ & $^{+1.4} _{-1.4}$ & $^{-0.2} _{+0.5}$ & $^{-0.3} _{+0.7}$ & $^{+0.1} _{-0.1}$ & $^{+0.6} _{-0.7}$ \\
510 & 1.92 & & $\pm$ 0.9 & $^{+1.5} _{-1.0}$ & $^{+1.0} _{-0.0}$ & $^{-0.6} _{+0.7}$ & $^{+0.3} _{-0.3}$ & $^{+0.2} _{-0.2}$ & $^{-0.1} _{+0.1}$ & $^{-0.3} _{+0.4}$ & $^{+0.1} _{-0.1}$ & $^{+0.7} _{-0.6}$ \\
570 & 1.45 & & $\pm$ 1.1 & $^{+1.4} _{-0.9}$ & $^{+0.8} _{-0.1}$ & $^{-0.5} _{+0.7}$ & $^{+0.3} _{-0.3}$ & $^{+0.1} _{-0.1}$ & $^{-0.2} _{+0.4}$ & $^{+0.1} _{+0.2}$ & $^{+0.2} _{-0.2}$ & $^{+0.7} _{-0.6}$ \\
630 & 1.16 & & $\pm$ 1.1 & $^{+1.5} _{-1.1}$ & $^{+1.0} _{-0.2}$ & $^{-0.7} _{+0.7}$ & $^{+0.0} _{-0.0}$ & $^{-0.7} _{+0.7}$ & $^{+0.1} _{-0.1}$ & $^{+0.1} _{+0.2}$ & $^{+0.2} _{-0.2}$ & $^{+0.2} _{-0.3}$ \\
700 & 8.98 & $\cdot 10^{-1}$ & $\pm$ 1.2 & $^{+0.8} _{-0.7}$ & $^{+0.1} _{-0.2}$ & $^{-0.6} _{+0.7}$ & $^{+0.1} _{-0.1}$ & $^{+0.1} _{-0.1}$ & $^{+0.1} _{+0.2}$ & $^{+0.3} _{-0.1}$ & $^{+0.1} _{-0.1}$ & $^{+0.3} _{-0.4}$ \\
780 & 6.88 & $\cdot 10^{-1}$ & $\pm$ 1.2 & $^{+0.9} _{-0.9}$ & $^{+0.0} _{-0.4}$ & $^{-0.5} _{+0.6}$ & $^{+0.5} _{-0.5}$ & $^{+0.5} _{-0.5}$ & $^{+0.1} _{+0.0}$ & $^{+0.1} _{+0.0}$ & $^{+0.1} _{-0.1}$ & $^{+0.2} _{-0.3}$ \\
860 & 5.33 & $\cdot 10^{-1}$ & $\pm$ 1.3 & $^{+0.8} _{-0.8}$ & $^{+0.4} _{+0.0}$ & $^{-0.4} _{+0.5}$ & $^{-0.4} _{+0.4}$ & $^{-0.1} _{+0.1}$ & $^{-0.3} _{-0.3}$ & $^{-0.1} _{-0.1}$ & $^{+0.2} _{-0.2}$ & $^{+0.3} _{-0.4}$ \\
940 & 4.27 & $\cdot 10^{-1}$ & $\pm$ 1.4 & $^{+1.1} _{-0.9}$ & $^{+0.0} _{-0.2}$ & $^{-0.4} _{+0.5}$ & $^{-0.4} _{+0.4}$ & $^{+0.6} _{-0.6}$ & $^{+0.4} _{-0.1}$ & $^{-0.0} _{-0.1}$ & $^{+0.2} _{-0.2}$ & $^{+0.4} _{-0.2}$ \\
1030 & 3.41 & $\cdot 10^{-1}$ & $\pm$ 1.5 & $^{+1.5} _{-1.4}$ & $^{+0.4} _{-0.1}$ & $^{-0.3} _{+0.5}$ & $^{+0.7} _{-0.7}$ & $^{+1.1} _{-1.1}$ & $^{+0.1} _{+0.2}$ & $^{-0.1} _{-0.4}$ & $^{+0.3} _{-0.3}$ & $^{+0.4} _{-0.4}$ \\
1130 & 2.75 & $\cdot 10^{-1}$ & $\pm$ 1.5 & $^{+1.2} _{-1.2}$ & $^{+0.2} _{-0.3}$ & $^{-0.4} _{+0.4}$ & $^{+0.9} _{-0.9}$ & $^{-0.1} _{+0.1}$ & $^{-0.1} _{+0.4}$ & $^{+0.1} _{-0.4}$ & $^{+0.2} _{-0.3}$ & $^{+0.3} _{-0.4}$ \\
1270 & 2.02 & $\cdot 10^{-1}$ & $\pm$ 1.5 & $^{+0.7} _{-0.7}$ & $^{+0.0} _{-0.1}$ & $^{-0.3} _{+0.4}$ & $^{-0.4} _{+0.4}$ & $^{-0.1} _{+0.1}$ & $^{+0.1} _{-0.1}$ & $^{-0.1} _{-0.2}$ & $^{+0.3} _{-0.2}$ & $^{+0.3} _{-0.4}$ \\
1420 & 1.52 & $\cdot 10^{-1}$ & $\pm$ 1.7 & $^{+1.0} _{-0.9}$ & $^{+0.5} _{+0.0}$ & $^{-0.3} _{+0.4}$ & $^{+0.1} _{-0.1}$ & $^{-0.6} _{+0.6}$ & $^{+0.1} _{-0.2}$ & $^{-0.1} _{-0.0}$ & $^{+0.4} _{-0.4}$ & $^{+0.4} _{-0.4}$ \\
1590 & 1.12 & $\cdot 10^{-1}$ & $\pm$ 1.7 & $^{+1.0} _{-0.6}$ & $^{+0.6} _{-0.1}$ & $^{-0.3} _{+0.3}$ & $^{+0.1} _{-0.1}$ & $^{-0.2} _{+0.2}$ & $^{+0.3} _{+0.0}$ & $^{+0.0} _{+0.3}$ & $^{+0.4} _{-0.3}$ & $^{+0.3} _{-0.3}$ \\
1790 & 8.40 & $\cdot 10^{-2}$ & $\pm$ 2.0 & $^{+1.0} _{-1.1}$ & $^{+0.5} _{-0.1}$ & $^{-0.3} _{+0.3}$ & $^{+0.6} _{-0.6}$ & $^{-0.5} _{+0.5}$ & $^{+0.2} _{-0.1}$ & $^{-0.5} _{-0.4}$ & $^{+0.3} _{-0.3}$ & $^{+0.3} _{-0.4}$ \\
1990 & 6.25 & $\cdot 10^{-2}$ & $\pm$ 2.3 & $^{+1.5} _{-1.6}$ & $^{+0.1} _{-0.7}$ & $^{-0.3} _{+0.4}$ & $^{-0.5} _{+0.5}$ & $^{+1.3} _{-1.3}$ & $^{+0.1} _{+0.0}$ & $^{+0.1} _{-0.2}$ & $^{+0.1} _{-0.2}$ & $^{+0.3} _{-0.3}$ \\
2300 & 4.39 & $\cdot 10^{-2}$ & $\pm$ 1.8 & $^{+0.7} _{-0.6}$ & $^{+0.2} _{+0.0}$ & $^{-0.2} _{+0.3}$ & $^{+0.2} _{-0.2}$ & $^{+0.1} _{-0.1}$ & $^{-0.2} _{-0.0}$ & $^{+0.3} _{+0.1}$ & $^{+0.3} _{-0.2}$ & $^{+0.3} _{-0.4}$ \\
2800 & 2.65 & $\cdot 10^{-2}$ & $\pm$ 2.1 & $^{+1.6} _{-0.7}$ & $^{+1.4} _{-0.3}$ & $^{-0.2} _{+0.3}$ & $^{-0.2} _{+0.2}$ & $^{+0.4} _{-0.4}$ & $^{+0.1} _{+0.0}$ & $^{+0.3} _{+0.1}$ & $^{+0.4} _{-0.4}$ & $^{+0.4} _{-0.2}$ \\
3500 & 1.48 & $\cdot 10^{-2}$ & $\pm$ 2.5 & $^{+0.8} _{-0.8}$ & $^{+0.2} _{+0.0}$ & $^{-0.2} _{+0.3}$ & $^{-0.5} _{+0.5}$ & $^{-0.2} _{+0.2}$ & $^{-0.1} _{-0.0}$ & $^{+0.0} _{-0.4}$ & $^{+0.3} _{-0.1}$ & $^{+0.4} _{-0.4}$ \\
4200 & 9.32 & $\cdot 10^{-3}$ & $\pm$ 3.0 & $^{+1.9} _{-0.8}$ & $^{+1.8} _{-0.3}$ & $^{-0.2} _{+0.3}$ & $^{+0.4} _{-0.4}$ & $^{-0.0} _{+0.0}$ & $^{+0.1} _{-0.0}$ & $^{-0.4} _{+0.0}$ & $^{+0.1} _{-0.2}$ & $^{+0.2} _{-0.4}$ \\
5100 & 5.08 & $\cdot 10^{-3}$ & $\pm$ 3.8 & $^{+2.2} _{-1.1}$ & $^{+1.9} _{-0.3}$ & $^{-0.2} _{+0.2}$ & $^{+0.9} _{-0.9}$ & $^{-0.0} _{+0.0}$ & $^{+0.0} _{+0.0}$ & $^{+0.1} _{+0.2}$ & $^{+0.2} _{-0.2}$ & $^{+0.5} _{-0.5}$ \\
6050 & 3.81 & $\cdot 10^{-3}$ & $\pm$ 4.2 & $^{+2.5} _{-2.4}$ & $^{+1.0} _{+0.0}$ & $^{-0.2} _{+0.3}$ & $^{-2.1} _{+2.1}$ & $^{-0.4} _{+0.4}$ & $^{+0.0} _{+0.0}$ & $^{-0.8} _{-0.1}$ & $^{+0.8} _{-0.3}$ & $^{+0.4} _{-0.4}$ \\
7100 & 2.23 & $\cdot 10^{-3}$ & $\pm$ 5.0 & $^{+3.0} _{-1.9}$ & $^{+2.4} _{-0.1}$ & $^{-0.3} _{+0.3}$ & $^{+1.4} _{-1.4}$ & $^{+0.9} _{-0.9}$ & $^{+0.0} _{+0.0}$ & $^{-0.8} _{+0.5}$ & $^{+0.2} _{-0.1}$ & $^{+0.4} _{-0.5}$ \\
8400 & 1.59 & $\cdot 10^{-3}$ & $\pm$ 5.5 & $^{+3.7} _{-1.2}$ & $^{+3.7} _{-0.9}$ & $^{-0.3} _{+0.4}$ & $^{-0.2} _{+0.2}$ & $^{+0.1} _{-0.1}$ & $^{+0.0} _{+0.0}$ & $^{-0.0} _{-0.0}$ & $^{+0.2} _{+0.0}$ & $^{+0.3} _{-0.8}$ \\
10800 & 6.90 & $\cdot 10^{-4}$ & $\pm$ 5.2 & $^{+2.8} _{-0.7}$ & $^{+2.7} _{+0.0}$ & $^{-0.3} _{+0.4}$ & $^{-0.2} _{+0.2}$ & $^{+0.1} _{-0.1}$ & $^{+0.0} _{+0.0}$ & $^{-0.2} _{-0.1}$ & $^{+0.3} _{-0.2}$ & $^{+0.4} _{-0.5}$ \\
15200 & 2.45 & $\cdot 10^{-4}$ & $\pm$ 7.2 & $^{+2.8} _{-3.7}$ & $^{+1.4} _{-0.7}$ & $^{-1.1} _{+0.3}$ & $^{-2.1} _{+2.1}$ & $^{+0.7} _{-0.7}$ & $^{+0.0} _{+0.0}$ & $^{-0.9} _{-2.6}$ & $^{+0.6} _{-0.3}$ & $^{+0.5} _{-0.6}$ \\
21500 & 7.99 & $\cdot 10^{-5}$ & $^{+11.7} _{-10.6}$ & $^{+8.1} _{-2.9}$ & $^{+7.8} _{-1.4}$ & $^{-0.2} _{+0.2}$ & $^{+0.1} _{-0.1}$ & $^{+1.1} _{-1.1}$ & $^{+0.0} _{+0.0}$ & $^{-1.7} _{+0.4}$ & $^{+1.5} _{-1.3}$ & $^{+1.0} _{-1.0}$ \\
36200 & 6.62 & $\cdot 10^{-6}$ & $^{+19.3} _{-16.4}$ & $^{+6.7} _{-4.9}$ & $^{+6.5} _{+0.0}$ & $^{-0.3} _{+0.3}$ & $^{-0.9} _{+0.9}$ & $^{+0.8} _{-0.8}$ & $^{+0.0} _{+0.0}$ & $^{+0.1} _{-3.0}$ & $^{+0.0} _{-3.5}$ & $^{+1.4} _{-1.4}$ \\

\hline
\end{tabular}
\end{center}
\caption[]
{Systematic uncertainties with bin-to-bin correlations
for $d\sigma / dQ^{2}$ ($y < 0.9$) for the reaction
$e^{-}p \rightarrow e^{-}X$ ($\mathcal{L} = 169.9 \pbi, P_{e} = -0.03$).
The left four columns of the table contain
the bin centre ($Q^2_c$), the measured cross section,
the statistical uncertainty and the total systematic uncertainty.
The right eight columns of the table list
the bin-to-bin correlated systematic uncertainties
for $\delta_{1} - \delta_{7}$,
and the systematic uncertainties
summed in quadrature for $\delta_{8} - \delta_{13}$,
as defined in the section \ref{sec-sys}.
The upper and lower correlated uncertainties correspond to
a positive or negative variation of a cut value for example.
However, if this is not possible for a particular systematic,
the uncertainty is symmetrised.}
\label{tab:dsdq2TotalSys}
\end{scriptsize}\end{table}

\newpage
\begin{table}
\begin{scriptsize}
\begin{center} \begin{tabular}[t]{|r|rcl|rl|rll|r|r|} \hline
\multicolumn{1}{|c|}{$Q^2~>$} & \multicolumn{3}{c|}{$x$ range} & \multicolumn{2}{c|}{$x_c$} & \multicolumn{3}{c|}{$d\sigma / dx$} & \multicolumn{1}{c|}{$N_{\text{data}}$} & \multicolumn{1}{c|}{$N^{\text{MC}}_{\text{bg}}$}\\
\multicolumn{1}{|c|}{($\gev^2$)} & \multicolumn{3}{c|}{} & \multicolumn{2}{c|}{} & \multicolumn{3}{c|}{($\pb$)} & & \\
\hline \hline
185 & (0.63 & -- & 1.00 )$\cdot 10^{-2}$ & 0.794 & $\cdot 10^{-2}$ & (8.66 & $\pm$ 0.05 & $^{+0.15} _{-0.13}$ )$\cdot 10^{4}$ & 39875 & 243.8 \\
& (0.10 & -- & 0.16 )$\cdot 10^{-1}$ & 0.126 & $\cdot 10^{-1}$ & (5.80 & $\pm$ 0.03 & $^{+0.08} _{-0.07}$ )$\cdot 10^{4}$ & 48561 & 202.6 \\
& (0.16 & -- & 0.25 )$\cdot 10^{-1}$ & 0.200 & $\cdot 10^{-1}$ & (3.60 & $\pm$ 0.02 & $^{+0.04} _{-0.03}$ )$\cdot 10^{4}$ & 49042 & 122.0 \\
& (0.25 & -- & 0.40 )$\cdot 10^{-1}$ & 0.316 & $\cdot 10^{-1}$ & (2.11 & $\pm$ 0.01 & $^{+0.05} _{-0.04}$ )$\cdot 10^{4}$ & 49989 & 58.2 \\
& (0.40 & -- & 0.63 )$\cdot 10^{-1}$ & 0.501 & $\cdot 10^{-1}$ & (1.24 & $\pm$ 0.01 & $^{+0.03} _{-0.03}$ )$\cdot 10^{4}$ & 41427 & 14.5 \\
& (0.63 & -- & 1.00 )$\cdot 10^{-1}$ & 0.794 & $\cdot 10^{-1}$ & (7.05 & $\pm$ 0.04 & $^{+0.18} _{-0.18}$ )$\cdot 10^{3}$ & 37564 & 8.6 \\
& 0.10 & -- & 0.16 & 0.126 & & (3.96 & $\pm$ 0.02 & $^{+0.14} _{-0.14}$ )$\cdot 10^{3}$ & 34201 & 4.3 \\
& 0.16 & -- & 0.25 & 0.200 & & (2.03 & $\pm$ 0.02 & $^{+0.09} _{-0.09}$ )$\cdot 10^{3}$ & 19029 & 2.9 \\
\hline
3000 & (0.40 & -- & 0.63 )$\cdot 10^{-1}$ & 0.501 & $\cdot 10^{-1}$ & (1.89 & $\pm$ 0.08 & $^{+0.08} _{-0.04}$ )$\cdot 10^{2}$ & 640 & 7.4 \\
& (0.63 & -- & 1.00 )$\cdot 10^{-1}$ & 0.794 & $\cdot 10^{-1}$ & (2.11 & $\pm$ 0.06 & $^{+0.03} _{-0.02}$ )$\cdot 10^{2}$ & 1211 & 5.8 \\
& 0.10 & -- & 0.16 & 0.126 & & (1.62 & $\pm$ 0.04 & $^{+0.02} _{-0.01}$ )$\cdot 10^{2}$ & 1522 & 4.3 \\
& 0.16 & -- & 0.25 & 0.200 & & (9.37 & $\pm$ 0.26 & $^{+0.17} _{-0.15}$ )$\cdot 10^{1}$ & 1306 & 2.9 \\
& 0.25 & -- & 0.40 & 0.316 & & (4.21 & $\pm$ 0.14 & $^{+0.04} _{-0.04}$ )$\cdot 10^{1}$ & 941 & 1.5 \\
& 0.40 & -- & 0.75 & 0.687 & & 1.42 & $\pm$ 0.07 & $^{+0.06} _{-0.05}$ & 381 & 0.0 \\

\hline
\end{tabular}
\end{center}
\caption[]
{The single differential cross section $d\sigma / dx$ ($y<0.9$)
for $Q^2 > 185 \gev^2$ and $Q^2 > 3\,000 \gev^2$
for the reaction $e^{-}p \rightarrow e^{-}X$ ($\mathcal{L} = 169.9 \pbi, P_{e} = -0.03$).
The $Q^2$ and bin range, bin centre ($x_c$) and measured cross section
corrected to the electroweak Born level are shown.
The first (second) error on the cross section
corresponds to the statistical (systematic) uncertainties.
The number of observed data events ($N_{\text{data}}$)
and simulated background events ($N^{\text{MC}}_{\text{bg}}$) are also shown.}
\label{tab:dsdxTotal}
\end{scriptsize}
\end{table}

\begin{table}
\begin{scriptsize}
\begin{center} \begin{tabular}[t]{|r|rl|lr|c|c||c|c|c|c|c|c|c|c|} \hline
\multicolumn{1}{|c|}{$Q^{2}~>$} &\multicolumn{2}{c|}{$x_c$} & \multicolumn{2}{c|}{$d\sigma / dx$} & \multicolumn{1}{c|}{stat.} & \multicolumn{1}{c||}{sys.} & \multicolumn{1}{c|}{$\delta_{1}$} & \multicolumn{1}{c|}{$\delta_{2}$} & \multicolumn{1}{c|}{$\delta_{3}$} & \multicolumn{1}{c|}{$\delta_{4}$} & \multicolumn{1}{c|}{$\delta_{5}$} & \multicolumn{1}{c|}{$\delta_{6}$} & \multicolumn{1}{c|}{$\delta_{7}$} & \multicolumn{1}{c|}{$\delta_{8} - \delta_{13}$} \\
\multicolumn{1}{|c|}{$(\gev^{2})$} & \multicolumn{2}{c|}{} & \multicolumn{2}{c|}{($\pb$)} & \multicolumn{1}{c|}{(\%)} & \multicolumn{1}{c||}{(\%)} & \multicolumn{1}{c|}{(\%)} & \multicolumn{1}{c|}{(\%)} & \multicolumn{1}{c|}{(\%)} & \multicolumn{1}{c|}{(\%)} & \multicolumn{1}{c|}{(\%)} & \multicolumn{1}{c|}{(\%)} & \multicolumn{1}{c|}{(\%)} & \multicolumn{1}{c|}{(\%)} \\ \hline \hline
185 & 0.794 & $\cdot 10^{-2}$ & 8.66 & $\cdot 10^{4}$ & $\pm$ 0.5 & $^{+1.7} _{-1.5}$ & $^{+0.6} _{-0.1}$ & $^{-0.7} _{+0.8}$ & $^{+0.4} _{-0.4}$ & $^{-1.0} _{+1.0}$ & $^{+0.0} _{+0.0}$ & $^{-0.1} _{+0.4}$ & $^{+0.2} _{-0.2}$ & $^{+0.7} _{-0.7}$ \\
& 0.126 & $\cdot 10^{-1}$ & 5.80 & $\cdot 10^{4}$ & $\pm$ 0.5 & $^{+1.4} _{-1.3}$ & $^{+0.4} _{-0.0}$ & $^{-0.5} _{+0.6}$ & $^{-0.0} _{+0.0}$ & $^{-1.0} _{+1.0}$ & $^{+0.0} _{+0.0}$ & $^{-0.2} _{+0.3}$ & $^{+0.2} _{-0.2}$ & $^{+0.6} _{-0.6}$ \\
& 0.200 & $\cdot 10^{-1}$ & 3.60 & $\cdot 10^{4}$ & $\pm$ 0.5 & $^{+1.1} _{-0.9}$ & $^{+0.5} _{-0.0}$ & $^{-0.4} _{+0.5}$ & $^{-0.4} _{+0.4}$ & $^{+0.4} _{-0.4}$ & $^{+0.0} _{+0.0}$ & $^{-0.2} _{+0.2}$ & $^{+0.1} _{-0.1}$ & $^{+0.6} _{-0.6}$ \\
& 0.316 & $\cdot 10^{-1}$ & 2.11 & $\cdot 10^{4}$ & $\pm$ 0.5 & $^{+2.2} _{-2.1}$ & $^{+0.4} _{-0.0}$ & $^{-0.3} _{+0.4}$ & $^{-1.0} _{+1.0}$ & $^{+1.7} _{-1.7}$ & $^{+0.0} _{+0.0}$ & $^{-0.1} _{+0.2}$ & $^{+0.1} _{-0.0}$ & $^{+0.6} _{-0.7}$ \\
& 0.501 & $\cdot 10^{-1}$ & 1.24 & $\cdot 10^{4}$ & $\pm$ 0.5 & $^{+2.3} _{-2.2}$ & $^{+0.3} _{-0.0}$ & $^{-0.2} _{+0.3}$ & $^{-1.3} _{+1.3}$ & $^{+1.7} _{-1.7}$ & $^{+0.0} _{+0.0}$ & $^{-0.0} _{+0.2}$ & $^{+0.0} _{-0.0}$ & $^{+0.6} _{-0.6}$ \\
& 0.794 & $\cdot 10^{-1}$ & 7.05 & $\cdot 10^{3}$ & $\pm$ 0.5 & $^{+2.5} _{-2.5}$ & $^{+0.3} _{-0.0}$ & $^{-0.1} _{+0.3}$ & $^{+1.8} _{-1.8}$ & $^{+1.5} _{-1.5}$ & $^{+0.0} _{+0.0}$ & $^{+0.0} _{+0.0}$ & $^{+0.0} _{-0.0}$ & $^{+0.6} _{-0.7}$ \\
& 0.126 & & 3.96 & $\cdot 10^{3}$ & $\pm$ 0.6 & $^{+3.6} _{-3.6}$ & $^{+0.3} _{+0.0}$ & $^{-0.1} _{+0.2}$ & $^{+3.1} _{-3.1}$ & $^{+1.6} _{-1.6}$ & $^{-0.1} _{+0.0}$ & $^{-0.1} _{+0.1}$ & $^{+0.0} _{-0.0}$ & $^{+0.6} _{-0.6}$ \\
& 0.200 & & 2.03 & $\cdot 10^{3}$ & $\pm$ 0.8 & $^{+4.7} _{-4.5}$ & $^{+0.3} _{-0.0}$ & $^{-0.1} _{+0.2}$ & $^{+4.4} _{-4.4}$ & $^{+0.5} _{-0.5}$ & $^{-0.3} _{+1.3}$ & $^{-0.0} _{+0.1}$ & $^{+0.0} _{-0.0}$ & $^{+0.6} _{-0.6}$ \\
\hline
3000 & 0.501 & $\cdot 10^{-1}$ & 1.89 & $\cdot 10^{2}$ & $\pm$ 4.0 & $^{+4.2} _{-2.1}$ & $^{+3.8} _{+0.0}$ & $^{-0.2} _{+0.1}$ & $^{-1.7} _{+1.7}$ & $^{-0.1} _{+0.1}$ & $^{+0.0} _{+0.0}$ & $^{-0.4} _{-0.9}$ & $^{+0.6} _{-0.5}$ & $^{+0.7} _{-0.5}$ \\
& 0.794 & $\cdot 10^{-1}$ & 2.11 & $\cdot 10^{2}$ & $\pm$ 2.9 & $^{+1.4} _{-1.1}$ & $^{+1.0} _{-0.3}$ & $^{-0.2} _{+0.3}$ & $^{+0.5} _{-0.5}$ & $^{+0.6} _{-0.6}$ & $^{+0.0} _{+0.0}$ & $^{-0.5} _{-0.3}$ & $^{+0.5} _{-0.2}$ & $^{+0.3} _{-0.4}$ \\
& 0.126 & & 1.62 & $\cdot 10^{2}$ & $\pm$ 2.6 & $^{+1.1} _{-0.5}$ & $^{+1.0} _{-0.0}$ & $^{-0.2} _{+0.3}$ & $^{+0.0} _{-0.0}$ & $^{-0.1} _{+0.1}$ & $^{+0.0} _{+0.0}$ & $^{-0.4} _{-0.1}$ & $^{+0.1} _{-0.1}$ & $^{+0.2} _{-0.2}$ \\
& 0.200 & & 9.37 & $\cdot 10^{1}$ & $\pm$ 2.8 & $^{+1.8} _{-1.6}$ & $^{+1.0} _{-0.1}$ & $^{-0.3} _{+0.2}$ & $^{-1.5} _{+1.5}$ & $^{+0.1} _{-0.1}$ & $^{+0.0} _{+0.0}$ & $^{+0.1} _{-0.3}$ & $^{+0.2} _{-0.1}$ & $^{+0.3} _{-0.3}$ \\
& 0.316 & & 4.21 & $\cdot 10^{1}$ & $\pm$ 3.3 & $^{+0.9} _{-1.0}$ & $^{+0.6} _{-0.5}$ & $^{-0.2} _{+0.2}$ & $^{-0.2} _{+0.2}$ & $^{-0.3} _{+0.3}$ & $^{+0.0} _{+0.0}$ & $^{+0.1} _{-0.5}$ & $^{+0.1} _{-0.1}$ & $^{+0.4} _{-0.5}$ \\
& 0.687 & & 1.42 & & $\pm$ 5.1 & $^{+4.3} _{-3.9}$ & $^{+1.9} _{+0.0}$ & $^{-0.4} _{+0.6}$ & $^{+3.6} _{-3.6}$ & $^{-0.3} _{+0.3}$ & $^{+0.3} _{+0.1}$ & $^{-1.1} _{+1.1}$ & $^{+0.0} _{-0.2}$ & $^{+0.8} _{-0.9}$ \\

\hline
\end{tabular}
\end{center}
\caption[]
{Systematic uncertainties with bin-to-bin correlations
for $d\sigma / dx$ ($y<0.9$)
for $Q^2 > 185 \gev^2$ and $Q^2 > 3\,000 \gev^2$ for the reaction
$e^{-}p \rightarrow e^{-}X$ ($\mathcal{L} = 169.9 \pbi, P_{e} = -0.03$).
The left five columns of the table contain
the $Q^2$ range, bin centre ($x_c$), the measured cross section,
the statistical uncertainty and the total systematic uncertainty.
The right eight columns of the table list
the bin-to-bin correlated systematic uncertainties
for $\delta_{1} - \delta_{7}$,
and the systematic uncertainties
summed in quadrature for $\delta_{8} - \delta_{13}$,
as defined in the section \ref{sec-sys}.
The upper and lower correlated uncertainties correspond to
a positive or negative variation of a cut value for example.
However, if this is not possible for a particular systematic,
the uncertainty is symmetrised.}
\label{tab:dsdxTotalSys}
\end{scriptsize}
\end{table}

\newpage
\begin{table}
\begin{scriptsize}
\begin{center} \begin{tabular}[t]{|r|rcl|c|rll|r|r|} \hline
\multicolumn{1}{|c|}{$Q^{2}~>$} & \multicolumn{3}{c|}{$y$ range} & \multicolumn{1}{c|}{$y_c$} & \multicolumn{3}{c|}{$d\sigma / dy$} & \multicolumn{1}{c|}{$N_{\text{data}}$} & \multicolumn{1}{c|}{$N^{\text{MC}}_{\text{bg}}$}\\
\multicolumn{1}{|c|}{($\gev^{2}$)} & \multicolumn{3}{c|}{} & \multicolumn{1}{c|}{} & \multicolumn{3}{c|}{($\pb$)} & & \\
\hline \hline
185  & 0.00 & -- & 0.05 & 0.025 & (1.62 & $\pm$ 0.01 & $^{+0.07} _{-0.07}$ )$\cdot 10^{4}$ & 77160 & 0.0 \\
 & 0.05 & -- & 0.10 & 0.075 & (8.04 & $\pm$ 0.03 & $^{+0.20} _{-0.20}$ )$\cdot 10^{3}$ & 63212 & 0.0 \\
 & 0.10 & -- & 0.15 & 0.125 & (5.65 & $\pm$ 0.03 & $^{+0.07} _{-0.06}$ )$\cdot 10^{3}$ & 44182 & 0.0 \\
 & 0.15 & -- & 0.20 & 0.175 & (4.33 & $\pm$ 0.03 & $^{+0.05} _{-0.04}$ )$\cdot 10^{3}$ & 32450 & 2.9 \\
 & 0.20 & -- & 0.25 & 0.225 & (3.54 & $\pm$ 0.02 & $^{+0.03} _{-0.03}$ )$\cdot 10^{3}$ & 25546 & 18.8 \\
 & 0.25 & -- & 0.30 & 0.275 & (2.97 & $\pm$ 0.02 & $^{+0.04} _{-0.04}$ )$\cdot 10^{3}$ & 20660 & 11.4 \\
 & 0.30 & -- & 0.35 & 0.325 & (2.59 & $\pm$ 0.02 & $^{+0.04} _{-0.03}$ )$\cdot 10^{3}$ & 17536 & 34.0 \\
 & 0.35 & -- & 0.40 & 0.375 & (2.23 & $\pm$ 0.02 & $^{+0.03} _{-0.02}$ )$\cdot 10^{3}$ & 14668 & 53.5 \\
 & 0.40 & -- & 0.45 & 0.425 & (1.98 & $\pm$ 0.02 & $^{+0.03} _{-0.02}$ )$\cdot 10^{3}$ & 12363 & 69.7 \\
 & 0.45 & -- & 0.50 & 0.475 & (1.76 & $\pm$ 0.02 & $^{+0.03} _{-0.02}$ )$\cdot 10^{3}$ & 10889 & 101.1 \\
 & 0.50 & -- & 0.55 & 0.525 & (1.60 & $\pm$ 0.02 & $^{+0.03} _{-0.02}$ )$\cdot 10^{3}$ & 9651 & 108.2 \\
 & 0.55 & -- & 0.60 & 0.575 & (1.47 & $\pm$ 0.02 & $^{+0.03} _{-0.03}$ )$\cdot 10^{3}$ & 8568 & 81.8 \\
 & 0.60 & -- & 0.65 & 0.625 & (1.33 & $\pm$ 0.02 & $^{+0.04} _{-0.04}$ )$\cdot 10^{3}$ & 7248 & 99.8 \\
 & 0.65 & -- & 0.70 & 0.675 & (1.21 & $\pm$ 0.02 & $^{+0.04} _{-0.04}$ )$\cdot 10^{3}$ & 5936 & 95.1 \\
 & 0.70 & -- & 0.75 & 0.725 & (1.12 & $\pm$ 0.02 & $^{+0.06} _{-0.05}$ )$\cdot 10^{3}$ & 4376 & 83.4 \\
\hline
3000  & 0.05 & -- & 0.10 & 0.075 & (3.46 & $\pm$ 0.22 & $^{+0.07} _{-0.06}$ )$\cdot 10^{1}$ & 242 & 0.0 \\
 & 0.10 & -- & 0.15 & 0.125 & (6.09 & $\pm$ 0.29 & $^{+0.21} _{-0.18}$ )$\cdot 10^{1}$ & 452 & 0.0 \\
 & 0.15 & -- & 0.20 & 0.175 & (6.49 & $\pm$ 0.29 & $^{+0.10} _{-0.10}$ )$\cdot 10^{1}$ & 493 & 0.0 \\
 & 0.20 & -- & 0.25 & 0.225 & (6.25 & $\pm$ 0.28 & $^{+0.05} _{-0.07}$ )$\cdot 10^{1}$ & 505 & 0.0 \\
 & 0.25 & -- & 0.30 & 0.275 & (6.39 & $\pm$ 0.28 & $^{+0.11} _{-0.07}$ )$\cdot 10^{1}$ & 525 & 0.0 \\
 & 0.30 & -- & 0.35 & 0.325 & (6.15 & $\pm$ 0.28 & $^{+0.12} _{-0.07}$ )$\cdot 10^{1}$ & 501 & 1.4 \\
 & 0.35 & -- & 0.40 & 0.375 & (5.22 & $\pm$ 0.25 & $^{+0.13} _{-0.07}$ )$\cdot 10^{1}$ & 425 & 0.0 \\
 & 0.40 & -- & 0.45 & 0.425 & (5.34 & $\pm$ 0.26 & $^{+0.18} _{-0.17}$ )$\cdot 10^{1}$ & 439 & 0.0 \\
 & 0.45 & -- & 0.50 & 0.475 & (5.19 & $\pm$ 0.25 & $^{+0.14} _{-0.11}$ )$\cdot 10^{1}$ & 421 & 0.0 \\
 & 0.50 & -- & 0.55 & 0.525 & (4.14 & $\pm$ 0.23 & $^{+0.12} _{-0.13}$ )$\cdot 10^{1}$ & 332 & 2.9 \\
 & 0.55 & -- & 0.60 & 0.575 & (4.28 & $\pm$ 0.23 & $^{+0.05} _{-0.07}$ )$\cdot 10^{1}$ & 335 & 0.0 \\
 & 0.60 & -- & 0.65 & 0.625 & (3.72 & $\pm$ 0.22 & $^{+0.08} _{-0.18}$ )$\cdot 10^{1}$ & 286 & 0.0 \\
 & 0.65 & -- & 0.70 & 0.675 & (3.93 & $\pm$ 0.23 & $^{+0.09} _{-0.20}$ )$\cdot 10^{1}$ & 301 & 3.0 \\
 & 0.70 & -- & 0.75 & 0.725 & (2.92 & $\pm$ 0.20 & $^{+0.08} _{-0.08}$ )$\cdot 10^{1}$ & 218 & 1.5 \\
 & 0.75 & -- & 0.80 & 0.775 & (2.98 & $\pm$ 0.20 & $^{+0.10} _{-0.10}$ )$\cdot 10^{1}$ & 219 & 1.5 \\
 & 0.80 & -- & 0.85 & 0.825 & (2.85 & $\pm$ 0.20 & $^{+0.39} _{-0.12}$ )$\cdot 10^{1}$ & 201 & 4.4 \\
 & 0.85 & -- & 0.90 & 0.875 & (2.57 & $\pm$ 0.21 & $^{+0.14} _{-0.43}$ )$\cdot 10^{1}$ & 173 & 8.7 \\

\hline
\end{tabular}
\end{center}
\caption[]
{The single differential cross section $d\sigma / dy$
for $Q^2 > 185 \gev^2$ and $Q^2 > 3\,000 \gev^2$
for the reaction $e^{-}p \rightarrow e^{-}X$ ($\mathcal{L} = 169.9 \pbi, P_{e} = -0.03$).
The $Q^2$ and bin range, bin centre ($y_c$) and measured cross section
corrected to the electroweak Born level are shown.
The first (second) error on the cross section
corresponds to the statistical (systematic) uncertainties.
The number of observed data events ($N_{\text{data}}$)
and simulated background events ($N^{\text{MC}}_{\text{bg}}$) are also shown.}
\label{tab:dsdyTotal}
\end{scriptsize}
\end{table}

\begin{table}
\begin{scriptsize}
\begin{center} \begin{tabular}[t]{|r|c|lr|c|c||c|c|c|c|c|c|c|c|} \hline
\multicolumn{1}{|c|}{$Q^{2}~>$} &\multicolumn{1}{c|}{$y_c$} & \multicolumn{2}{c|}{$d\sigma / dy$} & \multicolumn{1}{c|}{stat.} & \multicolumn{1}{c||}{sys.} & \multicolumn{1}{c|}{$\delta_{1}$} & \multicolumn{1}{c|}{$\delta_{2}$} & \multicolumn{1}{c|}{$\delta_{3}$} & \multicolumn{1}{c|}{$\delta_{4}$} & \multicolumn{1}{c|}{$\delta_{5}$} & \multicolumn{1}{c|}{$\delta_{6}$} & \multicolumn{1}{c|}{$\delta_{7}$} & \multicolumn{1}{c|}{$\delta_{8} - \delta_{13}$} \\
\multicolumn{1}{|c|}{$(\gev^{2})$} & \multicolumn{1}{c|}{} & \multicolumn{2}{c|}{($\pb$)} & \multicolumn{1}{c|}{(\%)} & \multicolumn{1}{c||}{(\%)} & \multicolumn{1}{c|}{(\%)} & \multicolumn{1}{c|}{(\%)} & \multicolumn{1}{c|}{(\%)} & \multicolumn{1}{c|}{(\%)} & \multicolumn{1}{c|}{(\%)} & \multicolumn{1}{c|}{(\%)} & \multicolumn{1}{c|}{(\%)} & \multicolumn{1}{c|}{(\%)} \\ \hline \hline
185 & 0.025 & 1.62 & $\cdot 10^{4}$ & $\pm$ 0.4 & $^{+4.1} _{-4.1}$ & $^{+0.1} _{-0.0}$ & $^{-0.1} _{+0.2}$ & $^{+3.6} _{-3.6}$ & $^{+1.7} _{-1.7}$ & $^{-0.1} _{+0.5}$ & $^{-0.0} _{+0.2}$ & $^{+0.0} _{+0.0}$ & $^{+0.7} _{-0.8}$ \\
& 0.075 & 8.04 & $\cdot 10^{3}$ & $\pm$ 0.4 & $^{+2.5} _{-2.5}$ & $^{+0.4} _{-0.0}$ & $^{-0.2} _{+0.3}$ & $^{-1.5} _{+1.5}$ & $^{+1.9} _{-1.9}$ & $^{+0.0} _{-0.0}$ & $^{+0.0} _{+0.2}$ & $^{+0.0} _{+0.0}$ & $^{+0.7} _{-0.7}$ \\
& 0.125 & 5.65 & $\cdot 10^{3}$ & $\pm$ 0.5 & $^{+1.2} _{-1.1}$ & $^{+0.5} _{-0.0}$ & $^{-0.3} _{+0.4}$ & $^{-0.5} _{+0.5}$ & $^{+0.7} _{-0.7}$ & $^{+0.0} _{+0.0}$ & $^{-0.1} _{+0.2}$ & $^{+0.0} _{+0.0}$ & $^{+0.6} _{-0.6}$ \\
& 0.175 & 4.33 & $\cdot 10^{3}$ & $\pm$ 0.6 & $^{+1.1} _{-1.0}$ & $^{+0.4} _{-0.0}$ & $^{-0.4} _{+0.5}$ & $^{-0.2} _{+0.2}$ & $^{-0.6} _{+0.6}$ & $^{+0.0} _{+0.0}$ & $^{+0.0} _{+0.2}$ & $^{+0.0} _{-0.0}$ & $^{+0.5} _{-0.7}$ \\
& 0.225 & 3.54 & $\cdot 10^{3}$ & $\pm$ 0.7 & $^{+1.0} _{-0.8}$ & $^{+0.4} _{+0.0}$ & $^{-0.4} _{+0.5}$ & $^{+0.1} _{-0.1}$ & $^{-0.4} _{+0.4}$ & $^{+0.0} _{+0.0}$ & $^{-0.1} _{+0.2}$ & $^{+0.0} _{-0.0}$ & $^{+0.6} _{-0.5}$ \\
& 0.275 & 2.97 & $\cdot 10^{3}$ & $\pm$ 0.7 & $^{+1.2} _{-1.2}$ & $^{+0.4} _{-0.0}$ & $^{-0.4} _{+0.5}$ & $^{-0.2} _{+0.2}$ & $^{-0.8} _{+0.8}$ & $^{+0.0} _{+0.0}$ & $^{-0.2} _{+0.2}$ & $^{+0.0} _{-0.0}$ & $^{+0.6} _{-0.7}$ \\
& 0.325 & 2.59 & $\cdot 10^{3}$ & $\pm$ 0.8 & $^{+1.4} _{-1.2}$ & $^{+0.7} _{-0.1}$ & $^{-0.4} _{+0.4}$ & $^{+0.2} _{-0.2}$ & $^{-1.0} _{+1.0}$ & $^{+0.0} _{+0.0}$ & $^{-0.1} _{+0.4}$ & $^{+0.1} _{-0.1}$ & $^{+0.5} _{-0.6}$ \\
& 0.375 & 2.23 & $\cdot 10^{3}$ & $\pm$ 0.9 & $^{+1.1} _{-0.8}$ & $^{+0.7} _{-0.0}$ & $^{-0.3} _{+0.5}$ & $^{-0.0} _{+0.0}$ & $^{+0.1} _{-0.1}$ & $^{+0.0} _{+0.0}$ & $^{-0.2} _{+0.3}$ & $^{+0.1} _{-0.1}$ & $^{+0.6} _{-0.7}$ \\
& 0.425 & 1.98 & $\cdot 10^{3}$ & $\pm$ 1.0 & $^{+1.4} _{-1.0}$ & $^{+0.9} _{-0.0}$ & $^{-0.3} _{+0.5}$ & $^{+0.8} _{-0.8}$ & $^{-0.1} _{+0.1}$ & $^{+0.0} _{+0.0}$ & $^{-0.1} _{-0.0}$ & $^{+0.2} _{-0.2}$ & $^{+0.5} _{-0.5}$ \\
& 0.475 & 1.76 & $\cdot 10^{3}$ & $\pm$ 1.0 & $^{+1.9} _{-1.0}$ & $^{+1.7} _{-0.2}$ & $^{-0.3} _{+0.5}$ & $^{-0.3} _{+0.3}$ & $^{+0.1} _{-0.1}$ & $^{+0.0} _{+0.0}$ & $^{+0.1} _{+0.2}$ & $^{+0.4} _{-0.4}$ & $^{+0.6} _{-0.7}$ \\
& 0.525 & 1.60 & $\cdot 10^{3}$ & $\pm$ 1.1 & $^{+1.6} _{-1.3}$ & $^{+0.9} _{-0.0}$ & $^{-0.5} _{+0.6}$ & $^{+0.6} _{-0.6}$ & $^{-0.8} _{+0.8}$ & $^{+0.0} _{+0.0}$ & $^{-0.3} _{+0.2}$ & $^{+0.5} _{-0.5}$ & $^{+0.5} _{-0.4}$ \\
& 0.575 & 1.47 & $\cdot 10^{3}$ & $\pm$ 1.2 & $^{+2.3} _{-2.3}$ & $^{+0.1} _{-0.3}$ & $^{-0.7} _{+0.9}$ & $^{+0.8} _{-0.8}$ & $^{-1.8} _{+1.8}$ & $^{+0.0} _{+0.0}$ & $^{-0.1} _{+0.1}$ & $^{+0.4} _{-0.4}$ & $^{+0.4} _{-0.5}$ \\
& 0.625 & 1.33 & $\cdot 10^{3}$ & $\pm$ 1.3 & $^{+3.3} _{-3.2}$ & $^{+0.5} _{-0.1}$ & $^{-1.5} _{+1.6}$ & $^{-1.8} _{+1.8}$ & $^{-2.1} _{+2.1}$ & $^{+0.0} _{+0.0}$ & $^{-0.3} _{+0.1}$ & $^{+0.6} _{-0.6}$ & $^{+0.4} _{-0.4}$ \\
& 0.675 & 1.21 & $\cdot 10^{3}$ & $\pm$ 1.4 & $^{+3.5} _{-3.2}$ & $^{+0.4} _{-0.9}$ & $^{-2.7} _{+3.2}$ & $^{+0.6} _{-0.6}$ & $^{-0.8} _{+0.8}$ & $^{+0.0} _{+0.0}$ & $^{-0.1} _{+0.4}$ & $^{+0.7} _{-0.7}$ & $^{+0.4} _{-0.5}$ \\
& 0.725 & 1.12 & $\cdot 10^{3}$ & $\pm$ 1.6 & $^{+5.7} _{-4.7}$ & $^{+0.2} _{-0.1}$ & $^{-4.3} _{+5.3}$ & $^{+0.1} _{-0.1}$ & $^{-1.6} _{+1.6}$ & $^{+0.0} _{+0.0}$ & $^{-0.1} _{+0.2}$ & $^{+0.8} _{-0.8}$ & $^{+0.8} _{-0.8}$ \\
\hline
3000 & 0.075 & 3.46 & $\cdot 10^{1}$ & $\pm$ 6.5 & $^{+1.9} _{-1.7}$ & $^{+0.1} _{-0.0}$ & $^{-0.4} _{+0.5}$ & $^{-0.2} _{+0.2}$ & $^{-0.3} _{+0.3}$ & $^{+1.3} _{-0.1}$ & $^{-1.0} _{+0.5}$ & $^{+0.0} _{+0.0}$ & $^{+1.2} _{-1.3}$ \\
& 0.125 & 6.09 & $\cdot 10^{1}$ & $\pm$ 4.7 & $^{+3.4} _{-2.9}$ & $^{+1.4} _{-0.0}$ & $^{-0.2} _{+0.3}$ & $^{+2.5} _{-2.5}$ & $^{-1.3} _{+1.3}$ & $^{+0.0} _{+0.0}$ & $^{+1.0} _{-0.1}$ & $^{+0.0} _{+0.0}$ & $^{+0.4} _{-0.6}$ \\
& 0.175 & 6.49 & $\cdot 10^{1}$ & $\pm$ 4.5 & $^{+1.5} _{-1.5}$ & $^{+0.1} _{-0.4}$ & $^{-0.3} _{+0.3}$ & $^{-1.3} _{+1.3}$ & $^{-0.5} _{+0.5}$ & $^{+0.0} _{+0.0}$ & $^{+0.1} _{-0.2}$ & $^{+0.0} _{+0.0}$ & $^{+0.5} _{-0.3}$ \\
& 0.225 & 6.25 & $\cdot 10^{1}$ & $\pm$ 4.5 & $^{+0.8} _{-1.2}$ & $^{+0.2} _{-0.5}$ & $^{-0.2} _{+0.3}$ & $^{+0.5} _{-0.5}$ & $^{+0.2} _{-0.2}$ & $^{+0.0} _{+0.0}$ & $^{-0.9} _{-0.1}$ & $^{+0.0} _{+0.0}$ & $^{+0.5} _{-0.3}$ \\
& 0.275 & 6.39 & $\cdot 10^{1}$ & $\pm$ 4.4 & $^{+1.8} _{-1.1}$ & $^{+1.5} _{-0.1}$ & $^{-0.2} _{+0.2}$ & $^{-0.9} _{+0.9}$ & $^{-0.1} _{+0.1}$ & $^{+0.0} _{+0.0}$ & $^{-0.6} _{-0.4}$ & $^{+0.0} _{+0.0}$ & $^{+0.4} _{-0.3}$ \\
& 0.325 & 6.15 & $\cdot 10^{1}$ & $\pm$ 4.5 & $^{+2.0} _{-1.2}$ & $^{+1.6} _{-0.1}$ & $^{-0.2} _{+0.3}$ & $^{+0.5} _{-0.5}$ & $^{+0.8} _{-0.8}$ & $^{+0.0} _{+0.0}$ & $^{+0.5} _{+0.2}$ & $^{+0.1} _{-0.1}$ & $^{+0.4} _{-0.6}$ \\
& 0.375 & 5.22 & $\cdot 10^{1}$ & $\pm$ 4.9 & $^{+2.5} _{-1.3}$ & $^{+2.3} _{-0.2}$ & $^{-0.2} _{+0.3}$ & $^{+0.6} _{-0.6}$ & $^{+0.8} _{-0.8}$ & $^{+0.0} _{+0.0}$ & $^{+0.1} _{+0.0}$ & $^{+0.0} _{+0.0}$ & $^{+0.4} _{-0.7}$ \\
& 0.425 & 5.34 & $\cdot 10^{1}$ & $\pm$ 4.8 & $^{+3.3} _{-3.1}$ & $^{+1.0} _{-0.3}$ & $^{-0.2} _{+0.3}$ & $^{-3.0} _{+3.0}$ & $^{+0.9} _{-0.9}$ & $^{+0.0} _{+0.0}$ & $^{-0.3} _{-0.3}$ & $^{+0.0} _{+0.0}$ & $^{+0.4} _{-0.4}$ \\
& 0.475 & 5.19 & $\cdot 10^{1}$ & $\pm$ 4.9 & $^{+2.7} _{-2.1}$ & $^{+2.1} _{+0.0}$ & $^{-0.2} _{+0.3}$ & $^{+1.5} _{-1.5}$ & $^{+0.9} _{-0.9}$ & $^{+0.0} _{+0.0}$ & $^{-0.0} _{-0.9}$ & $^{+0.0} _{+0.0}$ & $^{+0.3} _{-0.7}$ \\
& 0.525 & 4.14 & $\cdot 10^{1}$ & $\pm$ 5.5 & $^{+3.0} _{-3.2}$ & $^{+0.2} _{-0.8}$ & $^{-0.3} _{+0.3}$ & $^{-2.8} _{+2.8}$ & $^{+0.0} _{-0.0}$ & $^{+0.0} _{+0.0}$ & $^{-0.5} _{+0.4}$ & $^{+0.4} _{-0.4}$ & $^{+0.9} _{-1.0}$ \\
& 0.575 & 4.28 & $\cdot 10^{1}$ & $\pm$ 5.5 & $^{+1.2} _{-1.5}$ & $^{+0.4} _{-0.6}$ & $^{-0.2} _{+0.2}$ & $^{+1.1} _{-1.1}$ & $^{+0.0} _{-0.0}$ & $^{+0.0} _{+0.0}$ & $^{-0.8} _{-0.2}$ & $^{+0.0} _{+0.0}$ & $^{+0.3} _{-0.4}$ \\
& 0.625 & 3.72 & $\cdot 10^{1}$ & $\pm$ 5.9 & $^{+2.1} _{-4.9}$ & $^{+0.9} _{-4.5}$ & $^{-0.2} _{+0.3}$ & $^{-1.8} _{+1.8}$ & $^{-0.6} _{+0.6}$ & $^{+0.0} _{+0.0}$ & $^{-0.6} _{+0.0}$ & $^{+0.0} _{+0.0}$ & $^{+0.5} _{-0.3}$ \\
& 0.675 & 3.93 & $\cdot 10^{1}$ & $\pm$ 5.9 & $^{+2.2} _{-5.0}$ & $^{+1.8} _{-4.9}$ & $^{-0.2} _{+0.2}$ & $^{+0.9} _{-0.9}$ & $^{+0.5} _{-0.5}$ & $^{+0.0} _{+0.0}$ & $^{-0.6} _{+0.2}$ & $^{+0.4} _{-0.4}$ & $^{+0.7} _{-0.6}$ \\
& 0.725 & 2.92 & $\cdot 10^{1}$ & $\pm$ 6.8 & $^{+2.8} _{-2.8}$ & $^{+0.1} _{-2.0}$ & $^{-0.2} _{+0.2}$ & $^{-1.5} _{+1.5}$ & $^{-1.0} _{+1.0}$ & $^{+0.0} _{+0.0}$ & $^{-0.7} _{+1.8}$ & $^{+0.3} _{-0.3}$ & $^{+1.2} _{-0.7}$ \\
& 0.775 & 2.98 & $\cdot 10^{1}$ & $\pm$ 6.8 & $^{+3.4} _{-3.4}$ & $^{+3.1} _{+0.0}$ & $^{-0.2} _{+0.2}$ & $^{+0.9} _{-0.9}$ & $^{+0.5} _{-0.5}$ & $^{+0.0} _{+0.0}$ & $^{-0.6} _{-3.1}$ & $^{+0.3} _{-0.3}$ & $^{+0.7} _{-0.7}$ \\
& 0.825 & 2.85 & $\cdot 10^{1}$ & $\pm$ 7.2 & $^{+13.8} _{-4.1}$ & $^{+13.7} _{-0.5}$ & $^{-0.1} _{+0.2}$ & $^{+0.2} _{-0.2}$ & $^{-0.0} _{+0.0}$ & $^{+0.0} _{+0.0}$ & $^{+1.6} _{-3.9}$ & $^{+1.1} _{-0.9}$ & $^{+0.6} _{-0.5}$ \\
& 0.875 & 2.57 & $\cdot 10^{1}$ & $\pm$ 8.2 & $^{+5.3} _{-16.8}$ & $^{+2.1} _{-16.1}$ & $^{-1.5} _{+0.5}$ & $^{+0.6} _{-0.6}$ & $^{+0.2} _{-0.2}$ & $^{+0.0} _{+0.0}$ & $^{-3.9} _{-0.1}$ & $^{+4.6} _{-2.1}$ & $^{+1.4} _{-2.0}$ \\

\hline
\end{tabular}
\end{center}
\caption[]
{Systematic uncertainties with bin-to-bin correlations for $d\sigma / dy$
for $Q^2 > 185 \gev^2$ and $Q^2 > 3\,000 \gev^2$ for the reaction
$e^{-}p \rightarrow e^{-}X$ ($\mathcal{L} = 169.9 \pbi, P_{e} = -0.03$).
The left five columns of the table contain
the $Q^2$ range, bin centre ($y_c$), the measured cross section,
the statistical uncertainty and the total systematic uncertainty.
The right eight columns of the table list
the bin-to-bin correlated systematic uncertainties
for $\delta_{1} - \delta_{7}$,
and the systematic uncertainties
summed in quadrature for $\delta_{8} - \delta_{13}$,
as defined in the section \ref{sec-sys}.
The upper and lower correlated uncertainties correspond to
a positive or negative variation of a cut value for example.
However, if this is not possible for a particular systematic,
the uncertainty is symmetrised.}
\label{tab:dsdyTotalSys}
\end{scriptsize}
\end{table}

\newpage
\begin{table} \begin{scriptsize} \begin{center} \begin{tabular}[t]{|rcr|r|lcl|l|rll|r|r|} \hline
\multicolumn{3}{|c|}{$Q^2$ range} & \multicolumn{1}{c|}{$Q^2_c$} & \multicolumn{3}{|c|}{$x$ range} & \multicolumn{1}{c|}{$x_c$} & \multicolumn{3}{c|}{$\tilde{\sigma}$} & \multicolumn{1}{c|}{$N_{\text{data}}$} & \multicolumn{1}{c|}{$N^{\text{MC}}_{\text{bg}}$} \\
\multicolumn{3}{|c|}{($\gev^{2}$)} & \multicolumn{1}{c|}{($\gev^{2}$)} & \multicolumn{3}{|c|}{} & \multicolumn{1}{c|}{} & \multicolumn{3}{c|}{} & & \\ \hline \hline
185 & -- & 240 & 200 & 0.0037 & -- & 0.006 & 0.005 & 1.09 & $\pm$ 0.01 & $^{+0.01} _{-0.01}$ & 14079 & 88.0 \\
& & & & 0.006 & -- & 0.01 & 0.008 & 0.92 & $\pm$ 0.01 & $^{+0.02} _{-0.02}$ & 16389 & 23.3 \\
& & & & 0.01 & -- & 0.017 & 0.013 & 0.78 & $\pm$ 0.01 & $^{+0.01} _{-0.01}$ & 17628 & 2.9 \\
& & & & 0.017 & -- & 0.025 & 0.021 & 0.66 & $\pm$ 0.01 & $^{+0.01} _{-0.01}$ & 12598 & 0.0 \\
& & & & 0.025 & -- & 0.037 & 0.032 & 0.56 & $\pm$ 0.01 & $^{+0.02} _{-0.02}$ & 11904 & 0.0 \\
& & & & 0.037 & -- & 0.06 & 0.05 & 0.50 & $\pm$ 0.01 & $^{+0.01} _{-0.01}$ & 11002 & 0.0 \\
& & & & 0.06 & -- & 0.12 & 0.08 & 0.44 & $\pm$ 0.00 & $^{+0.02} _{-0.02}$ & 14235 & 0.0 \\
& & & & 0.12 & -- & 0.25 & 0.18 & 0.33 & $\pm$ 0.00 & $^{+0.02} _{-0.02}$ & 7372 & 0.0 \\
\hline
240 & -- & 310 & 250 & 0.006 & -- & 0.01 & 0.008 & 0.94 & $\pm$ 0.01 & $^{+0.01} _{-0.01}$ & 11341 & 26.1 \\
& & & & 0.01 & -- & 0.017 & 0.013 & 0.79 & $\pm$ 0.01 & $^{+0.01} _{-0.01}$ & 12602 & 6.0 \\
& & & & 0.017 & -- & 0.025 & 0.021 & 0.66 & $\pm$ 0.01 & $^{+0.01} _{-0.01}$ & 8935 & 1.5 \\
& & & & 0.025 & -- & 0.037 & 0.032 & 0.56 & $\pm$ 0.01 & $^{+0.02} _{-0.02}$ & 8745 & 0.0 \\
& & & & 0.037 & -- & 0.06 & 0.05 & 0.49 & $\pm$ 0.01 & $^{+0.02} _{-0.02}$ & 8494 & 0.0 \\
& & & & 0.06 & -- & 0.12 & 0.08 & 0.43 & $\pm$ 0.00 & $^{+0.02} _{-0.02}$ & 10350 & 0.0 \\
& & & & 0.12 & -- & 0.25 & 0.18 & 0.33 & $\pm$ 0.00 & $^{+0.02} _{-0.02}$ & 6985 & 0.0 \\
\hline
310 & -- & 410 & 350 & 0.006 & -- & 0.01 & 0.008 & 0.97 & $\pm$ 0.01 & $^{+0.02} _{-0.01}$ & 6903 & 71.5 \\
& & & & 0.01 & -- & 0.017 & 0.013 & 0.82 & $\pm$ 0.01 & $^{+0.02} _{-0.01}$ & 9505 & 8.8 \\
& & & & 0.017 & -- & 0.025 & 0.021 & 0.69 & $\pm$ 0.01 & $^{+0.01} _{-0.00}$ & 7274 & 0.0 \\
& & & & 0.025 & -- & 0.037 & 0.032 & 0.61 & $\pm$ 0.01 & $^{+0.01} _{-0.01}$ & 7071 & 0.0 \\
& & & & 0.037 & -- & 0.06 & 0.05 & 0.51 & $\pm$ 0.01 & $^{+0.01} _{-0.01}$ & 7509 & 0.0 \\
& & & & 0.06 & -- & 0.12 & 0.08 & 0.44 & $\pm$ 0.01 & $^{+0.01} _{-0.01}$ & 8417 & 0.0 \\
& & & & 0.12 & -- & 0.25 & 0.18 & 0.31 & $\pm$ 0.00 & $^{+0.02} _{-0.02}$ & 6571 & 0.0 \\
\hline
410 & -- & 530 & 450 & 0.006 & -- & 0.01 & 0.008 & 0.99 & $\pm$ 0.01 & $^{+0.02} _{-0.02}$ & 5334 & 84.2 \\
& & & & 0.01 & -- & 0.017 & 0.013 & 0.86 & $\pm$ 0.01 & $^{+0.01} _{-0.01}$ & 4719 & 16.1 \\
& & & & 0.017 & -- & 0.025 & 0.021 & 0.69 & $\pm$ 0.01 & $^{+0.01} _{-0.01}$ & 3668 & 1.3 \\
& & & & 0.025 & -- & 0.037 & 0.032 & 0.59 & $\pm$ 0.01 & $^{+0.01} _{-0.01}$ & 4173 & 0.0 \\
& & & & 0.037 & -- & 0.06 & 0.05 & 0.52 & $\pm$ 0.01 & $^{+0.01} _{-0.01}$ & 5225 & 0.0 \\
& & & & 0.06 & -- & 0.1 & 0.08 & 0.43 & $\pm$ 0.01 & $^{+0.01} _{-0.01}$ & 4249 & 0.0 \\
& & & & 0.1 & -- & 0.17 & 0.13 & 0.36 & $\pm$ 0.01 & $^{+0.02} _{-0.02}$ & 3727 & 0.0 \\
& & & & 0.17 & -- & 0.3 & 0.25 & 0.26 & $\pm$ 0.01 & $^{+0.01} _{-0.01}$ & 2819 & 0.0 \\

\hline
\end{tabular}
\end{center}
\end{scriptsize}
\caption[]
{The reduced cross section $\tilde{\sigma}$ for the reaction
$e^{-}p \rightarrow e^{-}X$ ($\mathcal{L} = 169.9 \pbi, P_{e} = 0$).
The bin range, bin centre ($Q^2_c$ and $x_c$)
and measured cross section corrected to the electroweak Born level are shown.
The first (second) error on the cross section
corresponds to the statistical (systematic) uncertainties.
The number of observed data events ($N_{\text{data}}$)
and simulated background events ($N^{\text{MC}}_{\text{bg}}$) are also shown.
This table has two continuations.}
\label{tab:ds2dxdq2Total_1}
\end{table}

\begin{table} \begin{scriptsize} \begin{center} \begin{tabular}[t]{|rcr|r|lcl|l|rll|r|r|} \hline
\multicolumn{3}{|c|}{$Q^2$ range} & \multicolumn{1}{c|}{$Q^2_c$} & \multicolumn{3}{|c|}{$x$ range} & \multicolumn{1}{c|}{$x_c$} & \multicolumn{3}{c|}{$\tilde{\sigma}$} & \multicolumn{1}{c|}{$N_{\text{data}}$} & \multicolumn{1}{c|}{$N^{\text{MC}}_{\text{bg}}$} \\
\multicolumn{3}{|c|}{($\gev^{2}$)} & \multicolumn{1}{c|}{($\gev^{2}$)} & \multicolumn{3}{|c|}{} & \multicolumn{1}{c|}{} & \multicolumn{3}{c|}{} & & \\ \hline \hline
530 & -- & 710 & 650 & 0.01 & -- & 0.017 & 0.013 & 0.88 & $\pm$ 0.01 & $^{+0.01} _{-0.01}$ & 5347 & 59.5 \\
& & & & 0.017 & -- & 0.025 & 0.021 & 0.76 & $\pm$ 0.01 & $^{+0.01} _{-0.01}$ & 3568 & 5.7 \\
& & & & 0.025 & -- & 0.037 & 0.032 & 0.60 & $\pm$ 0.01 & $^{+0.01} _{-0.01}$ & 2740 & 1.3 \\
& & & & 0.037 & -- & 0.06 & 0.05 & 0.51 & $\pm$ 0.01 & $^{+0.01} _{-0.00}$ & 2908 & 0.0 \\
& & & & 0.06 & -- & 0.1 & 0.08 & 0.42 & $\pm$ 0.01 & $^{+0.01} _{-0.01}$ & 2418 & 0.0 \\
& & & & 0.1 & -- & 0.17 & 0.13 & 0.36 & $\pm$ 0.01 & $^{+0.01} _{-0.01}$ & 2201 & 0.0 \\
& & & & 0.17 & -- & 0.3 & 0.25 & 0.25 & $\pm$ 0.01 & $^{+0.02} _{-0.02}$ & 1887 & 0.0 \\
\hline
710 & -- & 900 & 800 & 0.009 & -- & 0.017 & 0.013 & 0.86 & $\pm$ 0.02 & $^{+0.01} _{-0.02}$ & 3339 & 49.4 \\
& & & & 0.017 & -- & 0.025 & 0.021 & 0.73 & $\pm$ 0.02 & $^{+0.01} _{-0.01}$ & 2360 & 3.1 \\
& & & & 0.025 & -- & 0.037 & 0.032 & 0.63 & $\pm$ 0.01 & $^{+0.01} _{-0.01}$ & 2453 & 7.4 \\
& & & & 0.037 & -- & 0.06 & 0.05 & 0.52 & $\pm$ 0.01 & $^{+0.01} _{-0.01}$ & 2719 & 0.0 \\
& & & & 0.06 & -- & 0.1 & 0.08 & 0.45 & $\pm$ 0.01 & $^{+0.01} _{-0.01}$ & 2332 & 0.0 \\
& & & & 0.1 & -- & 0.17 & 0.13 & 0.38 & $\pm$ 0.01 & $^{+0.00} _{-0.00}$ & 1903 & 0.0 \\
& & & & 0.17 & -- & 0.3 & 0.25 & 0.26 & $\pm$ 0.01 & $^{+0.01} _{-0.01}$ & 1405 & 0.0 \\
\hline
900 & -- & 1300 & 1200 & 0.01 & -- & 0.017 & 0.014 & 0.89 & $\pm$ 0.02 & $^{+0.02} _{-0.03}$ & 2217 & 75.7 \\
& & & & 0.017 & -- & 0.025 & 0.021 & 0.79 & $\pm$ 0.02 & $^{+0.01} _{-0.01}$ & 2553 & 23.2 \\
& & & & 0.025 & -- & 0.037 & 0.032 & 0.61 & $\pm$ 0.01 & $^{+0.01} _{-0.01}$ & 2478 & 11.7 \\
& & & & 0.037 & -- & 0.06 & 0.05 & 0.55 & $\pm$ 0.01 & $^{+0.01} _{-0.01}$ & 3197 & 1.3 \\
& & & & 0.06 & -- & 0.1 & 0.08 & 0.46 & $\pm$ 0.01 & $^{+0.01} _{-0.01}$ & 3039 & 0.0 \\
& & & & 0.1 & -- & 0.17 & 0.13 & 0.38 & $\pm$ 0.01 & $^{+0.00} _{-0.00}$ & 2451 & 0.0 \\
& & & & 0.17 & -- & 0.3 & 0.25 & 0.25 & $\pm$ 0.01 & $^{+0.01} _{-0.01}$ & 1872 & 0.0 \\
& & & & 0.3 & -- & 0.53 & 0.4 & 0.12 & $\pm$ 0.01 & $^{+0.01} _{-0.01}$ & 641 & 0.0 \\
\hline
1300 & -- & 1800 & 1500 & 0.017 & -- & 0.025 & 0.021 & 0.76 & $\pm$ 0.02 & $^{+0.02} _{-0.02}$ & 1273 & 58.3 \\
& & & & 0.025 & -- & 0.037 & 0.032 & 0.63 & $\pm$ 0.02 & $^{+0.00} _{-0.00}$ & 1402 & 7.1 \\
& & & & 0.037 & -- & 0.06 & 0.05 & 0.52 & $\pm$ 0.01 & $^{+0.00} _{-0.00}$ & 1756 & 1.4 \\
& & & & 0.06 & -- & 0.1 & 0.08 & 0.48 & $\pm$ 0.01 & $^{+0.00} _{-0.01}$ & 1887 & 0.0 \\
& & & & 0.1 & -- & 0.15 & 0.13 & 0.38 & $\pm$ 0.01 & $^{+0.01} _{-0.01}$ & 1201 & 0.0 \\
& & & & 0.15 & -- & 0.23 & 0.18 & 0.31 & $\pm$ 0.01 & $^{+0.01} _{-0.01}$ & 1028 & 0.0 \\
& & & & 0.23 & -- & 0.35 & 0.25 & 0.26 & $\pm$ 0.01 & $^{+0.00} _{-0.01}$ & 729 & 0.0 \\
& & & & 0.35 & -- & 0.53 & 0.4 & 0.13 & $\pm$ 0.01 & $^{+0.01} _{-0.01}$ & 318 & 0.0 \\
\hline
1800 & -- & 2500 & 2000 & 0.023 & -- & 0.037 & 0.032 & 0.64 & $\pm$ 0.02 & $^{+0.02} _{-0.01}$ & 977 & 19.1 \\
& & & & 0.037 & -- & 0.06 & 0.05 & 0.57 & $\pm$ 0.02 & $^{+0.01} _{-0.00}$ & 1229 & 0.0 \\
& & & & 0.06 & -- & 0.1 & 0.08 & 0.46 & $\pm$ 0.01 & $^{+0.01} _{-0.01}$ & 1246 & 1.3 \\
& & & & 0.1 & -- & 0.15 & 0.13 & 0.37 & $\pm$ 0.01 & $^{+0.00} _{-0.00}$ & 883 & 0.0 \\
& & & & 0.15 & -- & 0.23 & 0.18 & 0.31 & $\pm$ 0.01 & $^{+0.00} _{-0.01}$ & 723 & 0.0 \\
& & & & 0.23 & -- & 0.35 & 0.25 & 0.25 & $\pm$ 0.01 & $^{+0.00} _{-0.00}$ & 502 & 0.0 \\
& & & & 0.35 & -- & 0.53 & 0.4 & 0.12 & $\pm$ 0.01 & $^{+0.00} _{-0.00}$ & 239 & 0.0 \\

\hline
\end{tabular}
\end{center}
\end{scriptsize}
\vspace{0.5cm} \hspace{5.4cm} {\bf Table 7:}~~~{\it Continuation 1.}
\label{tab:ds2dxdq2Total_2}
\end{table}

\begin{table} \begin{scriptsize} \begin{center} \begin{tabular}[t]{|rcr|r|lcl|l|rll|r|r|} \hline
\multicolumn{3}{|c|}{$Q^2$ range} & \multicolumn{1}{c|}{$Q^2_c$} & \multicolumn{3}{|c|}{$x$ range} & \multicolumn{1}{c|}{$x_c$} & \multicolumn{3}{c|}{$\tilde{\sigma}$} & \multicolumn{1}{c|}{$N_{\text{data}}$} & \multicolumn{1}{c|}{$N^{\text{MC}}_{\text{bg}}$} \\
\multicolumn{3}{|c|}{($\gev^{2}$)} & \multicolumn{1}{c|}{($\gev^{2}$)} & \multicolumn{3}{|c|}{} & \multicolumn{1}{c|}{} & \multicolumn{3}{c|}{} & & \\ \hline \hline
2500 & -- & 3500 & 3000 & 0.037 & -- & 0.06 & 0.05 & 0.58 & $\pm$ 0.02 & $^{+0.01} _{-0.01}$ & 777 & 9.0 \\
& & & & 0.06 & -- & 0.1 & 0.08 & 0.51 & $\pm$ 0.02 & $^{+0.01} _{-0.01}$ & 902 & 1.5 \\
& & & & 0.1 & -- & 0.15 & 0.13 & 0.38 & $\pm$ 0.02 & $^{+0.00} _{-0.00}$ & 623 & 0.0 \\
& & & & 0.15 & -- & 0.23 & 0.18 & 0.32 & $\pm$ 0.01 & $^{+0.00} _{-0.00}$ & 528 & 0.0 \\
& & & & 0.23 & -- & 0.35 & 0.25 & 0.29 & $\pm$ 0.01 & $^{+0.02} _{-0.02}$ & 427 & 0.0 \\
& & & & 0.35 & -- & 0.53 & 0.4 & 0.13 & $\pm$ 0.01 & $^{+0.01} _{-0.00}$ & 185 & 0.0 \\
& & & & 0.53 & -- & 0.75 & 0.65 & 0.02 & $^{+0.00} _{-0.00}$ & $^{+0.00} _{-0.00}$ & 62 & 0.0 \\
\hline
3500 & -- & 5600 & 5000 & 0.04 & -- & 0.1 & 0.08 & 0.53 & $\pm$ 0.02 & $^{+0.01} _{-0.00}$ & 1001 & 7.3 \\
& & & & 0.1 & -- & 0.15 & 0.13 & 0.46 & $\pm$ 0.02 & $^{+0.00} _{-0.00}$ & 610 & 1.4 \\
& & & & 0.15 & -- & 0.23 & 0.18 & 0.34 & $\pm$ 0.02 & $^{+0.00} _{-0.00}$ & 499 & 0.0 \\
& & & & 0.23 & -- & 0.35 & 0.25 & 0.24 & $\pm$ 0.01 & $^{+0.00} _{-0.00}$ & 318 & 0.0 \\
& & & & 0.35 & -- & 0.53 & 0.4 & 0.14 & $\pm$ 0.01 & $^{+0.01} _{-0.01}$ & 176 & 0.0 \\
\hline
5600 & -- & 9000 & 8000 & 0.07 & -- & 0.15 & 0.13 & 0.56 & $\pm$ 0.02 & $^{+0.01} _{-0.01}$ & 582 & 4.4 \\
& & & & 0.15 & -- & 0.23 & 0.18 & 0.43 & $\pm$ 0.02 & $^{+0.01} _{-0.01}$ & 346 & 0.0 \\
& & & & 0.23 & -- & 0.35 & 0.25 & 0.31 & $\pm$ 0.02 & $^{+0.01} _{-0.01}$ & 232 & 0.0 \\
& & & & 0.35 & -- & 0.53 & 0.4 & 0.11 & $^{+0.01} _{-0.01}$ & $^{+0.01} _{-0.01}$ & 87 & 0.0 \\
& & & & 0.53 & -- & 0.75 & 0.65 & 0.02 & $^{+0.00} _{-0.00}$ & $^{+0.00} _{-0.00}$ & 35 & 0.0 \\
\hline
9000 & -- & 15000 & 12000 & 0.09 & -- & 0.23 & 0.18 & 0.46 & $\pm$ 0.03 & $^{+0.01} _{-0.02}$ & 275 & 1.4 \\
& & & & 0.23 & -- & 0.35 & 0.25 & 0.35 & $\pm$ 0.03 & $^{+0.01} _{-0.01}$ & 149 & 0.0 \\
& & & & 0.35 & -- & 0.53 & 0.4 & 0.16 & $^{+0.02} _{-0.02}$ & $^{+0.01} _{-0.01}$ & 72 & 0.0 \\
\hline
15000 & -- & 25000 & 20000 & 0.15 & -- & 0.35 & 0.25 & 0.46 & $\pm$ 0.04 & $^{+0.03} _{-0.02}$ & 129 & 4.4 \\
& & & & 0.35 & -- & 0.75 & 0.4 & 0.18 & $^{+0.03} _{-0.03}$ & $^{+0.01} _{-0.01}$ & 48 & 0.0 \\
\hline
25000 & -- & 50000 & 30000 & 0.25 & -- & 0.75 & 0.4 & 0.24 & $^{+0.04} _{-0.04}$ & $^{+0.02} _{-0.01}$ & 42 & 0.0 \\

\hline
\end{tabular}
\end{center}
\end{scriptsize}
\vspace{0.5cm} \hspace{5.4cm} {\bf Table 7:}~~~{\it Continuation 2.}
\label{tab:ds2dxdq2Total_3}
\end{table}

\begin{table} \begin{scriptsize} \begin{center} \begin{tabular}[t]{|r|l|c|c|c||c|c|c|c|c|c|c|c|} \hline
\multicolumn{1}{|c|}{$Q^2_c$} & \multicolumn{1}{c|}{$x_c$} & \multicolumn{1}{c|}{$\tilde{\sigma}$} & \multicolumn{1}{c|}{stat.} & \multicolumn{1}{c||}{sys.} & \multicolumn{1}{c|}{$\delta_{1}$} & \multicolumn{1}{c|}{$\delta_{2}$} & \multicolumn{1}{c|}{$\delta_{3}$} & \multicolumn{1}{c|}{$\delta_{4}$} & \multicolumn{1}{c|}{$\delta_{5}$} & \multicolumn{1}{c|}{$\delta_{6}$} & \multicolumn{1}{c|}{$\delta_{7}$} & \multicolumn{1}{c|}{$\delta_{8} - \delta_{13}$}\\
\multicolumn{1}{|c|}{($\gev^{2}$)} & & & \multicolumn{1}{c|}{(\%)} & \multicolumn{1}{c||}{(\%)} & \multicolumn{1}{c|}{(\%)} & \multicolumn{1}{c|}{(\%)} & \multicolumn{1}{c|}{(\%)} & \multicolumn{1}{c|}{(\%)} & \multicolumn{1}{c|}{(\%)} & \multicolumn{1}{c|}{(\%)} & \multicolumn{1}{c|}{(\%)} & \multicolumn{1}{c|}{(\%)} \\ \hline \hline
200 & 0.005 & 1.09 & $\pm$ 0.9 & $^{+1.3} _{-1.0}$ & $^{+0.7} _{-0.0}$ & $^{-0.4} _{+0.6}$ & $^{+0.4} _{-0.4}$ & $^{-0.4} _{+0.4}$ & $^{+0.0} _{+0.0}$ & $^{+0.1} _{-0.1}$ & $^{+0.3} _{-0.3}$ & $^{+0.7} _{-0.7}$ \\
& 0.008 & 0.92 & $\pm$ 0.8 & $^{+1.9} _{-1.9}$ & $^{+0.3} _{-0.0}$ & $^{-0.4} _{+0.5}$ & $^{+0.2} _{-0.2}$ & $^{-1.6} _{+1.6}$ & $^{+0.0} _{+0.0}$ & $^{+0.1} _{+0.4}$ & $^{+0.1} _{-0.1}$ & $^{+0.8} _{-1.0}$ \\
& 0.013 & 0.78 & $\pm$ 0.8 & $^{+1.7} _{-1.6}$ & $^{+0.2} _{+0.0}$ & $^{-0.5} _{+0.5}$ & $^{-0.2} _{+0.2}$ & $^{-1.3} _{+1.3}$ & $^{+0.0} _{+0.0}$ & $^{+0.0} _{+0.4}$ & $^{+0.0} _{-0.0}$ & $^{+0.8} _{-0.9}$ \\
& 0.021 & 0.66 & $\pm$ 1.0 & $^{+2.1} _{-2.1}$ & $^{+0.4} _{+0.0}$ & $^{-0.3} _{+0.5}$ & $^{-1.6} _{+1.6}$ & $^{+0.9} _{-0.9}$ & $^{+0.0} _{+0.0}$ & $^{-0.1} _{+0.1}$ & $^{+0.0} _{+0.0}$ & $^{+0.8} _{-0.9}$ \\
& 0.032 & 0.56 & $\pm$ 1.0 & $^{+4.0} _{-4.0}$ & $^{+0.1} _{-0.0}$ & $^{-0.3} _{+0.4}$ & $^{-1.5} _{+1.5}$ & $^{+3.6} _{-3.6}$ & $^{+0.0} _{+0.0}$ & $^{+0.2} _{+0.4}$ & $^{+0.0} _{+0.0}$ & $^{+0.8} _{-0.8}$ \\
& 0.05 & 0.50 & $\pm$ 1.0 & $^{+2.5} _{-2.4}$ & $^{+0.0} _{-0.3}$ & $^{-0.2} _{+0.3}$ & $^{-0.5} _{+0.5}$ & $^{+2.2} _{-2.2}$ & $^{+0.0} _{+0.0}$ & $^{+0.0} _{+0.3}$ & $^{+0.0} _{+0.0}$ & $^{+1.0} _{-0.9}$ \\
& 0.08 & 0.44 & $\pm$ 0.9 & $^{+5.5} _{-5.5}$ & $^{+0.0} _{-0.1}$ & $^{-0.2} _{+0.2}$ & $^{+5.0} _{-5.0}$ & $^{+2.0} _{-2.0}$ & $^{-0.0} _{+0.0}$ & $^{-0.0} _{+0.1}$ & $^{+0.0} _{+0.0}$ & $^{+0.8} _{-1.0}$ \\
& 0.18 & 0.33 & $\pm$ 1.3 & $^{+5.4} _{-5.4}$ & $^{+0.1} _{+0.0}$ & $^{-0.1} _{+0.2}$ & $^{+5.3} _{-5.3}$ & $^{+0.5} _{-0.5}$ & $^{+0.7} _{+0.2}$ & $^{-0.1} _{+0.1}$ & $^{+0.0} _{+0.0}$ & $^{+0.7} _{-0.9}$ \\
\hline
250 & 0.008 & 0.94 & $\pm$ 1.0 & $^{+1.5} _{-1.3}$ & $^{+0.6} _{-0.0}$ & $^{-0.3} _{+0.4}$ & $^{+0.5} _{-0.5}$ & $^{-0.9} _{+0.9}$ & $^{+0.0} _{+0.0}$ & $^{-0.2} _{+0.3}$ & $^{+0.1} _{-0.1}$ & $^{+0.7} _{-0.7}$ \\
& 0.013 & 0.79 & $\pm$ 1.0 & $^{+1.7} _{-1.5}$ & $^{+0.7} _{+0.0}$ & $^{-0.5} _{+0.6}$ & $^{+0.0} _{-0.0}$ & $^{-1.3} _{+1.3}$ & $^{+0.0} _{+0.0}$ & $^{-0.1} _{+0.2}$ & $^{+0.0} _{-0.0}$ & $^{+0.5} _{-0.7}$ \\
& 0.021 & 0.66 & $\pm$ 1.1 & $^{+1.6} _{-1.4}$ & $^{+0.7} _{-0.0}$ & $^{-0.3} _{+0.5}$ & $^{-0.1} _{+0.1}$ & $^{+1.1} _{-1.1}$ & $^{+0.0} _{+0.0}$ & $^{-0.2} _{+0.2}$ & $^{+0.0} _{-0.0}$ & $^{+0.8} _{-0.7}$ \\
& 0.032 & 0.56 & $\pm$ 1.2 & $^{+3.3} _{-3.3}$ & $^{+0.6} _{-0.0}$ & $^{-0.3} _{+0.4}$ & $^{-0.8} _{+0.8}$ & $^{+3.0} _{-3.0}$ & $^{+0.0} _{+0.0}$ & $^{+0.0} _{+0.1}$ & $^{+0.0} _{+0.0}$ & $^{+0.7} _{-0.9}$ \\
& 0.05 & 0.49 & $\pm$ 1.2 & $^{+3.4} _{-3.3}$ & $^{+0.4} _{+0.0}$ & $^{-0.1} _{+0.3}$ & $^{-2.4} _{+2.4}$ & $^{+2.2} _{-2.2}$ & $^{+0.0} _{+0.0}$ & $^{-0.0} _{-0.1}$ & $^{+0.0} _{+0.0}$ & $^{+0.7} _{-0.6}$ \\
& 0.08 & 0.43 & $\pm$ 1.1 & $^{+5.2} _{-5.1}$ & $^{+0.7} _{-0.0}$ & $^{-0.1} _{+0.3}$ & $^{+4.6} _{-4.6}$ & $^{+2.2} _{-2.2}$ & $^{+0.0} _{+0.0}$ & $^{-0.0} _{+0.1}$ & $^{+0.0} _{+0.0}$ & $^{+0.7} _{-0.7}$ \\
& 0.18 & 0.33 & $\pm$ 1.3 & $^{+4.8} _{-4.7}$ & $^{+0.2} _{+0.0}$ & $^{+0.0} _{+0.2}$ & $^{+4.1} _{-4.1}$ & $^{+2.0} _{-2.0}$ & $^{+0.3} _{+1.0}$ & $^{+0.0} _{+0.2}$ & $^{+0.0} _{+0.0}$ & $^{+0.8} _{-0.8}$ \\
\hline
350 & 0.008 & 0.97 & $\pm$ 1.3 & $^{+2.1} _{-1.5}$ & $^{+1.6} _{-0.1}$ & $^{-0.5} _{+0.6}$ & $^{+0.9} _{-0.9}$ & $^{-0.4} _{+0.4}$ & $^{+0.0} _{+0.0}$ & $^{-0.6} _{+0.4}$ & $^{+0.4} _{-0.4}$ & $^{+0.6} _{-0.8}$ \\
& 0.013 & 0.82 & $\pm$ 1.1 & $^{+1.9} _{-1.2}$ & $^{+1.4} _{-0.0}$ & $^{-0.2} _{+0.5}$ & $^{-0.2} _{+0.2}$ & $^{-1.0} _{+1.0}$ & $^{+0.0} _{+0.0}$ & $^{-0.2} _{+0.5}$ & $^{+0.0} _{-0.0}$ & $^{+0.7} _{-0.6}$ \\
& 0.021 & 0.69 & $\pm$ 1.3 & $^{+0.9} _{-0.6}$ & $^{+0.4} _{-0.1}$ & $^{-0.3} _{+0.4}$ & $^{+0.3} _{-0.3}$ & $^{+0.0} _{-0.0}$ & $^{+0.0} _{+0.0}$ & $^{-0.2} _{-0.0}$ & $^{+0.0} _{+0.0}$ & $^{+0.6} _{-0.5}$ \\
& 0.032 & 0.61 & $\pm$ 1.3 & $^{+1.9} _{-1.8}$ & $^{+0.8} _{-0.0}$ & $^{-0.2} _{+0.4}$ & $^{-1.6} _{+1.6}$ & $^{+0.2} _{-0.2}$ & $^{+0.0} _{+0.0}$ & $^{-0.3} _{+0.1}$ & $^{+0.0} _{+0.0}$ & $^{+0.6} _{-0.7}$ \\
& 0.05 & 0.51 & $\pm$ 1.2 & $^{+2.7} _{-2.6}$ & $^{+0.6} _{-0.0}$ & $^{-0.1} _{+0.2}$ & $^{-1.9} _{+1.9}$ & $^{+1.6} _{-1.6}$ & $^{+0.0} _{+0.0}$ & $^{-0.1} _{+0.2}$ & $^{+0.0} _{+0.0}$ & $^{+0.5} _{-0.7}$ \\
& 0.08 & 0.44 & $\pm$ 1.2 & $^{+2.0} _{-2.0}$ & $^{+0.3} _{+0.0}$ & $^{-0.1} _{+0.2}$ & $^{+1.1} _{-1.1}$ & $^{+1.6} _{-1.6}$ & $^{+0.0} _{+0.0}$ & $^{+0.0} _{+0.1}$ & $^{+0.0} _{+0.0}$ & $^{+0.7} _{-0.7}$ \\
& 0.18 & 0.31 & $\pm$ 1.3 & $^{+6.4} _{-6.3}$ & $^{+0.4} _{+0.0}$ & $^{-0.0} _{+0.2}$ & $^{+5.7} _{-5.7}$ & $^{+2.6} _{-2.6}$ & $^{-0.1} _{+1.2}$ & $^{+0.1} _{+0.2}$ & $^{+0.0} _{+0.0}$ & $^{+0.7} _{-0.6}$ \\
\hline
450 & 0.008 & 0.99 & $\pm$ 1.5 & $^{+2.4} _{-2.1}$ & $^{+0.7} _{-0.3}$ & $^{-1.4} _{+1.6}$ & $^{+1.3} _{-1.3}$ & $^{-0.5} _{+0.5}$ & $^{+0.0} _{+0.0}$ & $^{-0.2} _{+0.3}$ & $^{+0.6} _{-0.6}$ & $^{+0.5} _{-0.4}$ \\
& 0.013 & 0.86 & $\pm$ 1.5 & $^{+1.1} _{-1.1}$ & $^{+0.1} _{-0.1}$ & $^{-0.5} _{+0.6}$ & $^{-0.4} _{+0.4}$ & $^{+0.2} _{-0.2}$ & $^{+0.0} _{+0.0}$ & $^{-0.6} _{+0.6}$ & $^{+0.1} _{-0.1}$ & $^{+0.5} _{-0.6}$ \\
& 0.021 & 0.69 & $\pm$ 1.7 & $^{+1.8} _{-1.7}$ & $^{+0.2} _{-0.1}$ & $^{-0.4} _{+0.5}$ & $^{-0.1} _{+0.1}$ & $^{+1.3} _{-1.3}$ & $^{+0.0} _{+0.0}$ & $^{-0.7} _{+0.9}$ & $^{+0.0} _{-0.0}$ & $^{+0.8} _{-0.8}$ \\
& 0.032 & 0.59 & $\pm$ 1.6 & $^{+1.9} _{-1.7}$ & $^{+0.8} _{-0.1}$ & $^{-0.2} _{+0.4}$ & $^{+0.0} _{-0.0}$ & $^{+1.4} _{-1.4}$ & $^{+0.0} _{+0.0}$ & $^{-0.5} _{+0.4}$ & $^{+0.0} _{+0.0}$ & $^{+1.0} _{-0.9}$ \\
& 0.05 & 0.52 & $\pm$ 1.4 & $^{+2.5} _{-2.4}$ & $^{+0.7} _{-0.0}$ & $^{+0.0} _{+0.2}$ & $^{-1.6} _{+1.6}$ & $^{+1.6} _{-1.6}$ & $^{+0.0} _{+0.0}$ & $^{-0.4} _{+0.6}$ & $^{+0.0} _{+0.0}$ & $^{+0.7} _{-0.7}$ \\
& 0.08 & 0.43 & $\pm$ 1.6 & $^{+2.1} _{-1.9}$ & $^{+0.3} _{+0.0}$ & $^{+0.1} _{+0.2}$ & $^{-0.7} _{+0.7}$ & $^{+1.7} _{-1.7}$ & $^{+0.0} _{+0.0}$ & $^{+0.1} _{+0.6}$ & $^{+0.0} _{+0.0}$ & $^{+0.7} _{-0.5}$ \\
& 0.13 & 0.36 & $\pm$ 1.7 & $^{+4.9} _{-4.9}$ & $^{+0.6} _{+0.0}$ & $^{+0.1} _{+0.1}$ & $^{+4.4} _{-4.4}$ & $^{+1.9} _{-1.9}$ & $^{+0.0} _{+0.0}$ & $^{-0.4} _{+0.4}$ & $^{+0.0} _{+0.0}$ & $^{+0.5} _{-0.6}$ \\
& 0.25 & 0.26 & $\pm$ 2.0 & $^{+5.6} _{-5.5}$ & $^{+1.1} _{+0.0}$ & $^{+0.2} _{+0.1}$ & $^{+5.2} _{-5.2}$ & $^{+1.1} _{-1.1}$ & $^{-0.9} _{+0.8}$ & $^{-0.1} _{+0.6}$ & $^{+0.0} _{+0.0}$ & $^{+0.8} _{-0.8}$ \\

\hline
\end{tabular}
\end{center}
\end{scriptsize}
\caption[]
{Systematic uncertainties with bin-to-bin correlations
for the reduced cross section $\tilde{\sigma}$
for the reaction $e^{-}p \rightarrow e^{-}X$ ($\mathcal{L} = 169.9 \pbi, P_{e} = 0$).
The left five columns of the table contain
the bin centres, $Q^{2}_{c}$ and $x_{c}$, the measured cross section,
the statistical uncertainty and the total systematic uncertainty.
The right eight columns of the table list
the bin-to-bin correlated systematic uncertainties
for $\delta_{1} - \delta_{7}$,
and the systematic uncertainties
summed in quadrature for $\delta_{8} - \delta_{13}$,
as defined in the section \ref{sec-sys}.
The upper and lower correlated uncertainties correspond to
a positive or negative variation of a cut value for example.
However, if this is not possible for a particular systematic,
the uncertainty is symmetrised.
This table has two continuations.}
\label{tab:ds2dxdq2TotalSys_1}
\end{table}

\begin{table} \begin{scriptsize} \begin{center} \begin{tabular}[t]{|r|l|c|c|c||c|c|c|c|c|c|c|c|} \hline
\multicolumn{1}{|c|}{$Q^2_c$} & \multicolumn{1}{c|}{$x_c$} & \multicolumn{1}{c|}{$\tilde{\sigma}$} & \multicolumn{1}{c|}{stat.} & \multicolumn{1}{c||}{sys.} & \multicolumn{1}{c|}{$\delta_{1}$} & \multicolumn{1}{c|}{$\delta_{2}$} & \multicolumn{1}{c|}{$\delta_{3}$} & \multicolumn{1}{c|}{$\delta_{4}$} & \multicolumn{1}{c|}{$\delta_{5}$} & \multicolumn{1}{c|}{$\delta_{6}$} & \multicolumn{1}{c|}{$\delta_{7}$} & \multicolumn{1}{c|}{$\delta_{8} - \delta_{13}$}\\
\multicolumn{1}{|c|}{($\gev^{2}$)} & & & \multicolumn{1}{c|}{(\%)} & \multicolumn{1}{c||}{(\%)} & \multicolumn{1}{c|}{(\%)} & \multicolumn{1}{c|}{(\%)} & \multicolumn{1}{c|}{(\%)} & \multicolumn{1}{c|}{(\%)} & \multicolumn{1}{c|}{(\%)} & \multicolumn{1}{c|}{(\%)} & \multicolumn{1}{c|}{(\%)} & \multicolumn{1}{c|}{(\%)} \\ \hline \hline
650 & 0.013 & 0.88 & $\pm$ 1.4 & $^{+1.5} _{-0.8}$ & $^{+1.2} _{-0.2}$ & $^{-0.5} _{+0.6}$ & $^{-0.0} _{+0.0}$ & $^{-0.1} _{+0.1}$ & $^{+0.0} _{+0.0}$ & $^{-0.2} _{+0.3}$ & $^{+0.5} _{-0.5}$ & $^{+0.4} _{-0.3}$ \\
& 0.021 & 0.76 & $\pm$ 1.7 & $^{+1.1} _{-1.1}$ & $^{+0.0} _{-0.1}$ & $^{-0.4} _{+0.5}$ & $^{-0.3} _{+0.3}$ & $^{-0.9} _{+0.9}$ & $^{+0.0} _{+0.0}$ & $^{-0.2} _{+0.0}$ & $^{+0.1} _{-0.1}$ & $^{+0.2} _{-0.2}$ \\
& 0.032 & 0.60 & $\pm$ 2.0 & $^{+1.2} _{-1.0}$ & $^{+0.5} _{-0.2}$ & $^{-0.4} _{+0.5}$ & $^{-0.5} _{+0.5}$ & $^{+0.7} _{-0.7}$ & $^{+0.0} _{+0.0}$ & $^{+0.5} _{+0.5}$ & $^{+0.0} _{-0.0}$ & $^{+0.2} _{-0.2}$ \\
& 0.05 & 0.51 & $\pm$ 1.9 & $^{+1.0} _{-0.9}$ & $^{+0.1} _{-0.0}$ & $^{-0.3} _{+0.4}$ & $^{-0.4} _{+0.4}$ & $^{+0.7} _{-0.7}$ & $^{+0.0} _{+0.0}$ & $^{+0.1} _{+0.0}$ & $^{+0.0} _{+0.0}$ & $^{+0.5} _{-0.4}$ \\
& 0.08 & 0.42 & $\pm$ 2.1 & $^{+2.4} _{-2.3}$ & $^{+0.7} _{-0.0}$ & $^{-0.1} _{+0.3}$ & $^{-1.8} _{+1.8}$ & $^{+1.0} _{-1.0}$ & $^{+0.0} _{+0.0}$ & $^{+0.1} _{-0.0}$ & $^{+0.0} _{+0.0}$ & $^{+0.9} _{-1.1}$ \\
& 0.13 & 0.36 & $\pm$ 2.2 & $^{+2.7} _{-2.4}$ & $^{+0.9} _{-0.0}$ & $^{-0.1} _{+0.2}$ & $^{+2.1} _{-2.1}$ & $^{-0.8} _{+0.8}$ & $^{+0.0} _{+0.0}$ & $^{+0.1} _{+0.4}$ & $^{+0.0} _{+0.0}$ & $^{+1.0} _{-0.9}$ \\
& 0.25 & 0.25 & $\pm$ 2.4 & $^{+6.6} _{-6.5}$ & $^{+1.0} _{-0.0}$ & $^{-0.1} _{+0.3}$ & $^{+6.3} _{-6.3}$ & $^{+1.0} _{-1.0}$ & $^{-0.8} _{+0.2}$ & $^{-0.1} _{+0.5}$ & $^{+0.0} _{+0.0}$ & $^{+1.0} _{-1.1}$ \\
\hline
800 & 0.013 & 0.86 & $\pm$ 1.8 & $^{+1.7} _{-1.9}$ & $^{+0.4} _{-1.0}$ & $^{-0.8} _{+1.0}$ & $^{+0.4} _{-0.4}$ & $^{-1.0} _{+1.0}$ & $^{+0.0} _{+0.0}$ & $^{-0.1} _{+0.0}$ & $^{+0.6} _{-0.6}$ & $^{+0.3} _{-0.4}$ \\
& 0.021 & 0.73 & $\pm$ 2.1 & $^{+1.1} _{-0.9}$ & $^{+0.5} _{-0.1}$ & $^{-0.4} _{+0.5}$ & $^{+0.6} _{-0.6}$ & $^{-0.2} _{+0.2}$ & $^{+0.0} _{+0.0}$ & $^{+0.0} _{+0.5}$ & $^{+0.1} _{-0.1}$ & $^{+0.4} _{-0.4}$ \\
& 0.032 & 0.63 & $\pm$ 2.1 & $^{+1.4} _{-1.4}$ & $^{+0.0} _{-0.4}$ & $^{-0.4} _{+0.5}$ & $^{-0.8} _{+0.8}$ & $^{-0.9} _{+0.9}$ & $^{+0.0} _{+0.0}$ & $^{+0.3} _{+0.0}$ & $^{+0.1} _{-0.1}$ & $^{+0.2} _{-0.4}$ \\
& 0.05 & 0.52 & $\pm$ 2.0 & $^{+2.2} _{-2.2}$ & $^{+0.3} _{+0.0}$ & $^{-0.3} _{+0.4}$ & $^{+0.7} _{-0.7}$ & $^{+2.0} _{-2.0}$ & $^{+0.0} _{+0.0}$ & $^{-0.4} _{+0.1}$ & $^{+0.0} _{+0.0}$ & $^{+0.3} _{-0.4}$ \\
& 0.08 & 0.45 & $\pm$ 2.1 & $^{+2.6} _{-2.6}$ & $^{+0.5} _{-0.0}$ & $^{-0.2} _{+0.3}$ & $^{-2.5} _{+2.5}$ & $^{+0.1} _{-0.1}$ & $^{+0.0} _{+0.0}$ & $^{+0.5} _{-0.7}$ & $^{+0.0} _{+0.0}$ & $^{+0.3} _{-0.4}$ \\
& 0.13 & 0.38 & $\pm$ 2.4 & $^{+1.1} _{-1.0}$ & $^{+0.5} _{-0.0}$ & $^{-0.2} _{+0.3}$ & $^{+0.7} _{-0.7}$ & $^{+0.6} _{-0.6}$ & $^{+0.0} _{+0.0}$ & $^{+0.1} _{-0.4}$ & $^{+0.0} _{+0.0}$ & $^{+0.2} _{-0.2}$ \\
& 0.25 & 0.26 & $\pm$ 2.8 & $^{+3.7} _{-3.4}$ & $^{+0.3} _{+0.0}$ & $^{-0.2} _{+0.2}$ & $^{+3.3} _{-3.3}$ & $^{+0.8} _{-0.8}$ & $^{+1.2} _{-0.1}$ & $^{+0.7} _{-0.5}$ & $^{+0.0} _{+0.0}$ & $^{+0.2} _{-0.3}$ \\
\hline
1200 & 0.014 & 0.89 & $\pm$ 2.3 & $^{+2.4} _{-3.2}$ & $^{+0.6} _{-2.2}$ & $^{-0.8} _{+1.0}$ & $^{-0.7} _{+0.7}$ & $^{-1.4} _{+1.4}$ & $^{+0.0} _{+0.0}$ & $^{-0.3} _{-0.8}$ & $^{+1.4} _{-1.4}$ & $^{+0.5} _{-0.2}$ \\
& 0.021 & 0.79 & $\pm$ 2.1 & $^{+1.9} _{-1.6}$ & $^{+0.9} _{-0.1}$ & $^{-0.5} _{+0.5}$ & $^{+1.5} _{-1.5}$ & $^{+0.0} _{-0.0}$ & $^{+0.0} _{+0.0}$ & $^{+0.1} _{+0.4}$ & $^{+0.4} _{-0.4}$ & $^{+0.3} _{-0.4}$ \\
& 0.032 & 0.61 & $\pm$ 2.1 & $^{+1.2} _{-1.1}$ & $^{+0.0} _{-0.1}$ & $^{-0.4} _{+0.5}$ & $^{-0.5} _{+0.5}$ & $^{-0.7} _{+0.7}$ & $^{+0.0} _{+0.0}$ & $^{+0.1} _{+0.1}$ & $^{+0.2} _{-0.2}$ & $^{+0.6} _{-0.6}$ \\
& 0.05 & 0.55 & $\pm$ 1.8 & $^{+1.3} _{-1.4}$ & $^{+0.1} _{-0.0}$ & $^{-0.3} _{+0.3}$ & $^{-0.1} _{+0.1}$ & $^{+1.2} _{-1.2}$ & $^{+0.0} _{+0.0}$ & $^{-0.3} _{-0.4}$ & $^{+0.0} _{-0.0}$ & $^{+0.4} _{-0.4}$ \\
& 0.08 & 0.46 & $\pm$ 1.9 & $^{+1.4} _{-1.3}$ & $^{+0.3} _{-0.0}$ & $^{-0.2} _{+0.4}$ & $^{-0.3} _{+0.3}$ & $^{+1.2} _{-1.2}$ & $^{+0.0} _{+0.0}$ & $^{+0.3} _{-0.3}$ & $^{+0.0} _{+0.0}$ & $^{+0.4} _{-0.3}$ \\
& 0.13 & 0.38 & $\pm$ 2.1 & $^{+1.1} _{-1.2}$ & $^{+0.1} _{-0.0}$ & $^{-0.1} _{+0.2}$ & $^{-0.9} _{+0.9}$ & $^{+0.4} _{-0.4}$ & $^{+0.0} _{+0.0}$ & $^{+0.1} _{-0.5}$ & $^{+0.0} _{+0.0}$ & $^{+0.3} _{-0.4}$ \\
& 0.25 & 0.25 & $\pm$ 2.4 & $^{+2.4} _{-2.3}$ & $^{+0.3} _{+0.0}$ & $^{-0.2} _{+0.3}$ & $^{+2.0} _{-2.0}$ & $^{+1.0} _{-1.0}$ & $^{+0.0} _{+0.0}$ & $^{-0.3} _{-0.1}$ & $^{+0.0} _{+0.0}$ & $^{+0.5} _{-0.4}$ \\
& 0.4 & 0.12 & $\pm$ 4.1 & $^{+8.1} _{-5.7}$ & $^{+0.0} _{-0.1}$ & $^{-0.4} _{+0.4}$ & $^{+5.6} _{-5.6}$ & $^{+0.6} _{-0.6}$ & $^{+5.8} _{+3.4}$ & $^{+0.5} _{-1.0}$ & $^{+0.0} _{+0.0}$ & $^{+0.5} _{-0.5}$ \\
\hline
1500 & 0.021 & 0.76 & $\pm$ 3.0 & $^{+2.6} _{-2.3}$ & $^{+0.9} _{-0.2}$ & $^{-0.6} _{+0.4}$ & $^{-0.4} _{+0.4}$ & $^{-0.1} _{+0.1}$ & $^{+0.0} _{+0.0}$ & $^{-1.0} _{+0.1}$ & $^{+2.3} _{-1.9}$ & $^{+0.4} _{-0.4}$ \\
& 0.032 & 0.63 & $\pm$ 2.7 & $^{+0.7} _{-0.6}$ & $^{+0.3} _{-0.1}$ & $^{-0.3} _{+0.3}$ & $^{-0.1} _{+0.1}$ & $^{-0.3} _{+0.3}$ & $^{+0.0} _{+0.0}$ & $^{-0.1} _{-0.1}$ & $^{+0.2} _{-0.2}$ & $^{+0.4} _{-0.3}$ \\
& 0.05 & 0.52 & $\pm$ 2.4 & $^{+0.9} _{-0.8}$ & $^{+0.2} _{-0.0}$ & $^{-0.3} _{+0.4}$ & $^{+0.2} _{-0.2}$ & $^{+0.4} _{-0.4}$ & $^{+0.0} _{+0.0}$ & $^{-0.4} _{+0.5}$ & $^{+0.0} _{-0.0}$ & $^{+0.3} _{-0.4}$ \\
& 0.08 & 0.48 & $\pm$ 2.3 & $^{+1.0} _{-1.1}$ & $^{+0.2} _{-0.0}$ & $^{-0.3} _{+0.4}$ & $^{-0.9} _{+0.9}$ & $^{-0.2} _{+0.2}$ & $^{+0.0} _{+0.0}$ & $^{-0.4} _{+0.1}$ & $^{+0.0} _{+0.0}$ & $^{+0.3} _{-0.5}$ \\
& 0.13 & 0.38 & $\pm$ 2.9 & $^{+2.1} _{-1.8}$ & $^{+0.8} _{-0.0}$ & $^{-0.1} _{+0.3}$ & $^{-1.3} _{+1.3}$ & $^{-1.1} _{+1.1}$ & $^{+0.0} _{+0.0}$ & $^{+0.8} _{-0.2}$ & $^{+0.0} _{+0.0}$ & $^{+0.3} _{-0.4}$ \\
& 0.18 & 0.31 & $\pm$ 3.2 & $^{+1.8} _{-1.7}$ & $^{+0.6} _{-0.1}$ & $^{-0.3} _{+0.2}$ & $^{+0.0} _{-0.0}$ & $^{-1.6} _{+1.6}$ & $^{+0.0} _{+0.0}$ & $^{-0.3} _{-0.2}$ & $^{+0.0} _{+0.0}$ & $^{+0.4} _{-0.4}$ \\
& 0.25 & 0.26 & $\pm$ 3.8 & $^{+1.9} _{-1.9}$ & $^{+0.0} _{-0.3}$ & $^{-0.3} _{+0.3}$ & $^{+1.7} _{-1.7}$ & $^{+0.6} _{-0.6}$ & $^{+0.1} _{+0.0}$ & $^{-0.1} _{-0.3}$ & $^{+0.0} _{+0.0}$ & $^{+0.3} _{-0.3}$ \\
& 0.4 & 0.13 & $\pm$ 5.7 & $^{+8.8} _{-4.3}$ & $^{+0.6} _{+0.0}$ & $^{-0.5} _{+0.3}$ & $^{+2.0} _{-2.0}$ & $^{-0.9} _{+0.9}$ & $^{+8.4} _{-3.6}$ & $^{+0.3} _{-0.4}$ & $^{+0.0} _{+0.0}$ & $^{+0.6} _{-0.5}$ \\
\hline
2000 & 0.032 & 0.64 & $\pm$ 3.3 & $^{+2.5} _{-1.4}$ & $^{+1.8} _{+0.0}$ & $^{-0.2} _{+0.2}$ & $^{+0.3} _{-0.3}$ & $^{+1.0} _{-1.0}$ & $^{+0.0} _{+0.0}$ & $^{+0.4} _{+0.7}$ & $^{+1.1} _{-0.8}$ & $^{+0.3} _{-0.4}$ \\
& 0.05 & 0.57 & $\pm$ 2.9 & $^{+1.0} _{-0.8}$ & $^{+0.5} _{-0.0}$ & $^{-0.3} _{+0.4}$ & $^{-0.3} _{+0.3}$ & $^{-0.6} _{+0.6}$ & $^{+0.0} _{+0.0}$ & $^{+0.1} _{+0.1}$ & $^{+0.0} _{+0.0}$ & $^{+0.3} _{-0.3}$ \\
& 0.08 & 0.46 & $\pm$ 2.9 & $^{+2.0} _{-2.0}$ & $^{+0.1} _{-0.1}$ & $^{-0.3} _{+0.4}$ & $^{-0.1} _{+0.1}$ & $^{+1.9} _{-1.9}$ & $^{+0.0} _{+0.0}$ & $^{+0.2} _{+0.0}$ & $^{+0.0} _{-0.0}$ & $^{+0.3} _{-0.3}$ \\
& 0.13 & 0.37 & $\pm$ 3.4 & $^{+0.8} _{-1.2}$ & $^{+0.1} _{-0.6}$ & $^{-0.2} _{+0.3}$ & $^{-0.5} _{+0.5}$ & $^{+0.5} _{-0.5}$ & $^{+0.0} _{+0.0}$ & $^{+0.1} _{-0.6}$ & $^{+0.0} _{+0.0}$ & $^{+0.3} _{-0.4}$ \\
& 0.18 & 0.31 & $\pm$ 3.8 & $^{+0.6} _{-1.9}$ & $^{+0.0} _{-0.7}$ & $^{-0.2} _{+0.2}$ & $^{-0.3} _{+0.3}$ & $^{+0.1} _{-0.1}$ & $^{+0.0} _{+0.0}$ & $^{-0.4} _{-1.6}$ & $^{+0.0} _{+0.0}$ & $^{+0.4} _{-0.6}$ \\
& 0.25 & 0.25 & $\pm$ 4.5 & $^{+1.7} _{-1.6}$ & $^{+0.3} _{-0.0}$ & $^{-0.2} _{+0.3}$ & $^{+1.5} _{-1.5}$ & $^{+0.5} _{-0.5}$ & $^{+0.0} _{+0.0}$ & $^{+0.4} _{+0.5}$ & $^{+0.0} _{+0.0}$ & $^{+0.5} _{-0.5}$ \\
& 0.4 & 0.12 & $\pm$ 6.5 & $^{+3.7} _{-3.9}$ & $^{+0.0} _{-0.2}$ & $^{-0.3} _{+0.4}$ & $^{+3.1} _{-3.1}$ & $^{+1.1} _{-1.1}$ & $^{-0.5} _{+1.0}$ & $^{+0.1} _{-1.3}$ & $^{+0.0} _{+0.0}$ & $^{+1.2} _{-1.4}$ \\

\hline
\end{tabular}
\end{center}
\end{scriptsize}
\vspace{0.5cm} \hspace{5.4cm} {\bf Table 8:}~~~{\it Continuation 1.}
\label{tab:ds2dxdq2TotalSys_2}
\end{table}

\begin{table} \begin{scriptsize} \begin{center} \begin{tabular}[t]{|r|l|c|c|c||c|c|c|c|c|c|c|c|} \hline
\multicolumn{1}{|c|}{$Q^2_c$} & \multicolumn{1}{c|}{$x_c$} & \multicolumn{1}{c|}{$\tilde{\sigma}$} & \multicolumn{1}{c|}{stat.} & \multicolumn{1}{c||}{sys.} & \multicolumn{1}{c|}{$\delta_{1}$} & \multicolumn{1}{c|}{$\delta_{2}$} & \multicolumn{1}{c|}{$\delta_{3}$} & \multicolumn{1}{c|}{$\delta_{4}$} & \multicolumn{1}{c|}{$\delta_{5}$} & \multicolumn{1}{c|}{$\delta_{6}$} & \multicolumn{1}{c|}{$\delta_{7}$} & \multicolumn{1}{c|}{$\delta_{8} - \delta_{13}$}\\
\multicolumn{1}{|c|}{($\gev^{2}$)} & & & \multicolumn{1}{c|}{(\%)} & \multicolumn{1}{c||}{(\%)} & \multicolumn{1}{c|}{(\%)} & \multicolumn{1}{c|}{(\%)} & \multicolumn{1}{c|}{(\%)} & \multicolumn{1}{c|}{(\%)} & \multicolumn{1}{c|}{(\%)} & \multicolumn{1}{c|}{(\%)} & \multicolumn{1}{c|}{(\%)} & \multicolumn{1}{c|}{(\%)} \\ \hline \hline
3000 & 0.05 & 0.58 & $\pm$ 3.7 & $^{+2.0} _{-1.1}$ & $^{+1.6} _{-0.1}$ & $^{-0.3} _{+0.4}$ & $^{+0.9} _{-0.9}$ & $^{-0.0} _{+0.0}$ & $^{+0.0} _{+0.0}$ & $^{+0.4} _{+0.2}$ & $^{+0.5} _{-0.5}$ & $^{+0.4} _{-0.3}$ \\
& 0.08 & 0.51 & $\pm$ 3.4 & $^{+1.1} _{-1.1}$ & $^{+0.2} _{-0.4}$ & $^{-0.2} _{+0.3}$ & $^{-0.4} _{+0.4}$ & $^{+0.8} _{-0.8}$ & $^{+0.0} _{+0.0}$ & $^{+0.5} _{-0.1}$ & $^{+0.1} _{-0.1}$ & $^{+0.3} _{-0.5}$ \\
& 0.13 & 0.38 & $\pm$ 4.0 & $^{+0.9} _{-0.7}$ & $^{+0.1} _{-0.1}$ & $^{-0.2} _{+0.3}$ & $^{+0.6} _{-0.6}$ & $^{-0.1} _{+0.1}$ & $^{+0.0} _{+0.0}$ & $^{-0.3} _{-0.3}$ & $^{+0.0} _{+0.0}$ & $^{+0.7} _{-0.1}$ \\
& 0.18 & 0.32 & $\pm$ 4.4 & $^{+1.2} _{-1.3}$ & $^{+0.0} _{-0.3}$ & $^{-0.2} _{+0.2}$ & $^{-0.9} _{+0.9}$ & $^{-0.6} _{+0.6}$ & $^{+0.0} _{+0.0}$ & $^{-0.5} _{+0.2}$ & $^{+0.0} _{+0.0}$ & $^{+0.4} _{-0.3}$ \\
& 0.25 & 0.29 & $\pm$ 4.9 & $^{+7.4} _{-7.3}$ & $^{+0.6} _{-0.0}$ & $^{-0.2} _{+0.2}$ & $^{+7.3} _{-7.3}$ & $^{+0.2} _{-0.2}$ & $^{+0.0} _{+0.0}$ & $^{+1.1} _{+0.3}$ & $^{+0.0} _{+0.0}$ & $^{+0.8} _{-0.6}$ \\
& 0.4 & 0.13 & $\pm$ 7.4 & $^{+3.8} _{-3.5}$ & $^{+1.3} _{-0.0}$ & $^{-0.2} _{+0.4}$ & $^{-3.4} _{+3.4}$ & $^{+0.5} _{-0.5}$ & $^{+0.4} _{-0.0}$ & $^{+0.8} _{-0.3}$ & $^{+0.0} _{+0.0}$ & $^{+0.6} _{-0.6}$ \\
& 0.65 & 0.02 & $^{+14.4} _{-12.7}$ & $^{+10.2} _{-10.4}$ & $^{+0.0} _{-0.6}$ & $^{-0.8} _{+1.0}$ & $^{-8.7} _{+8.7}$ & $^{+4.6} _{-4.6}$ & $^{+1.8} _{+0.7}$ & $^{-2.7} _{+0.9}$ & $^{+0.0} _{+0.0}$ & $^{+1.7} _{-2.0}$ \\
\hline
5000 & 0.08 & 0.53 & $\pm$ 3.2 & $^{+1.8} _{-0.9}$ & $^{+1.6} _{-0.3}$ & $^{-0.2} _{+0.2}$ & $^{-0.3} _{+0.3}$ & $^{+0.2} _{-0.2}$ & $^{+0.0} _{+0.0}$ & $^{-0.6} _{-0.4}$ & $^{+0.3} _{-0.3}$ & $^{+0.5} _{-0.3}$ \\
& 0.13 & 0.46 & $\pm$ 4.1 & $^{+1.0} _{-0.8}$ & $^{+0.8} _{+0.0}$ & $^{-0.2} _{+0.2}$ & $^{-0.4} _{+0.4}$ & $^{+0.3} _{-0.3}$ & $^{+0.0} _{+0.0}$ & $^{+0.1} _{-0.3}$ & $^{+0.1} _{-0.1}$ & $^{+0.2} _{-0.5}$ \\
& 0.18 & 0.34 & $\pm$ 4.5 & $^{+1.0} _{-0.5}$ & $^{+0.2} _{-0.1}$ & $^{-0.3} _{+0.2}$ & $^{+0.2} _{-0.2}$ & $^{+0.3} _{-0.3}$ & $^{+0.0} _{+0.0}$ & $^{-0.1} _{+0.7}$ & $^{+0.0} _{+0.0}$ & $^{+0.5} _{-0.3}$ \\
& 0.25 & 0.24 & $\pm$ 5.6 & $^{+1.8} _{-1.2}$ & $^{+0.7} _{-0.0}$ & $^{-0.2} _{+0.3}$ & $^{+0.7} _{-0.7}$ & $^{-0.7} _{+0.7}$ & $^{+0.0} _{+0.0}$ & $^{+1.2} _{-0.6}$ & $^{+0.0} _{+0.0}$ & $^{+0.4} _{-0.3}$ \\
& 0.4 & 0.14 & $\pm$ 7.6 & $^{+6.5} _{-6.4}$ & $^{+0.0} _{-0.3}$ & $^{-0.2} _{+0.3}$ & $^{+6.0} _{-6.0}$ & $^{-2.1} _{+2.1}$ & $^{+0.0} _{+0.0}$ & $^{+0.4} _{+1.3}$ & $^{+0.0} _{+0.0}$ & $^{+1.1} _{-1.1}$ \\
\hline
8000 & 0.13 & 0.56 & $\pm$ 4.2 & $^{+1.0} _{-1.9}$ & $^{+0.0} _{-1.0}$ & $^{-0.3} _{+0.3}$ & $^{-0.5} _{+0.5}$ & $^{-0.1} _{+0.1}$ & $^{+0.0} _{+0.0}$ & $^{-1.3} _{+0.3}$ & $^{+0.5} _{-0.3}$ & $^{+0.5} _{-0.5}$ \\
& 0.18 & 0.43 & $\pm$ 5.4 & $^{+2.6} _{-2.9}$ & $^{+0.1} _{-0.3}$ & $^{-0.2} _{+0.2}$ & $^{-2.4} _{+2.4}$ & $^{+0.6} _{-0.6}$ & $^{+0.0} _{+0.0}$ & $^{+0.5} _{-0.3}$ & $^{+0.0} _{+0.0}$ & $^{+0.6} _{-1.3}$ \\
& 0.25 & 0.31 & $\pm$ 6.6 & $^{+3.8} _{-3.9}$ & $^{+0.7} _{-0.1}$ & $^{-0.3} _{+0.2}$ & $^{-3.6} _{+3.6}$ & $^{+0.4} _{-0.4}$ & $^{+0.0} _{+0.0}$ & $^{-1.1} _{-0.6}$ & $^{+0.0} _{+0.0}$ & $^{+0.4} _{-0.4}$ \\
& 0.4 & 0.11 & $^{+11.9} _{-10.7}$ & $^{+7.1} _{-7.3}$ & $^{+0.1} _{-0.0}$ & $^{-0.3} _{+0.4}$ & $^{+7.0} _{-7.0}$ & $^{-0.6} _{+0.6}$ & $^{+0.0} _{+0.0}$ & $^{-0.3} _{+0.5}$ & $^{+0.0} _{+0.0}$ & $^{+0.6} _{-2.0}$ \\
& 0.65 & 0.02 & $^{+19.9} _{-16.9}$ & $^{+4.1} _{-4.2}$ & $^{+0.2} _{-1.4}$ & $^{-0.9} _{+0.9}$ & $^{+1.5} _{-1.5}$ & $^{+2.3} _{-2.3}$ & $^{+0.0} _{+0.0}$ & $^{-2.6} _{+2.8}$ & $^{+0.0} _{+0.0}$ & $^{+0.8} _{-0.6}$ \\
\hline
12000 & 0.18 & 0.46 & $\pm$ 6.1 & $^{+1.3} _{-3.8}$ & $^{+1.0} _{-3.6}$ & $^{-0.3} _{+0.4}$ & $^{-0.4} _{+0.4}$ & $^{-0.2} _{+0.2}$ & $^{+0.0} _{+0.0}$ & $^{-0.3} _{-0.8}$ & $^{+0.2} _{-0.3}$ & $^{+0.5} _{-0.6}$ \\
& 0.25 & 0.35 & $\pm$ 8.2 & $^{+2.8} _{-2.3}$ & $^{+2.4} _{-0.6}$ & $^{-0.2} _{+0.2}$ & $^{+0.3} _{-0.3}$ & $^{+0.8} _{-0.8}$ & $^{+0.0} _{+0.0}$ & $^{+0.3} _{-1.6}$ & $^{+0.0} _{+0.0}$ & $^{+1.0} _{-1.2}$ \\
& 0.4 & 0.16 & $^{+13.4} _{-11.8}$ & $^{+4.6} _{-3.8}$ & $^{+2.7} _{-0.2}$ & $^{-0.3} _{+0.4}$ & $^{+3.2} _{-3.2}$ & $^{-0.8} _{+0.8}$ & $^{+0.0} _{+0.0}$ & $^{-1.2} _{+1.5}$ & $^{+0.0} _{+0.0}$ & $^{+0.6} _{-1.3}$ \\
\hline
20000 & 0.25 & 0.46 & $\pm$ 9.2 & $^{+6.1} _{-3.6}$ & $^{+4.9} _{-1.8}$ & $^{-1.5} _{+0.2}$ & $^{-2.2} _{+2.2}$ & $^{+0.1} _{-0.1}$ & $^{+0.0} _{+0.0}$ & $^{-0.7} _{-0.2}$ & $^{+2.7} _{-1.4}$ & $^{+0.6} _{-0.5}$ \\
& 0.4 & 0.18 & $^{+16.6} _{-14.4}$ & $^{+4.9} _{-3.8}$ & $^{+3.6} _{+0.0}$ & $^{-0.4} _{+0.4}$ & $^{+0.9} _{-0.9}$ & $^{+3.0} _{-3.0}$ & $^{+0.0} _{+0.0}$ & $^{-1.8} _{-0.3}$ & $^{+0.0} _{+0.0}$ & $^{+1.1} _{-1.2}$ \\
\hline
30000 & 0.4 & 0.24 & $^{+17.9} _{-15.4}$ & $^{+6.9} _{-5.9}$ & $^{+6.7} _{+0.0}$ & $^{-0.2} _{+0.3}$ & $^{-0.8} _{+0.8}$ & $^{+0.7} _{-0.7}$ & $^{+0.0} _{+0.0}$ & $^{+0.1} _{-4.9}$ & $^{+0.0} _{-2.9}$ & $^{+1.3} _{-1.3}$ \\

\hline
\end{tabular}
\end{center}
\end{scriptsize}
\vspace{0.5cm} \hspace{5.4cm} {\bf Table 8:}~~~{\it Continuation 2.}
\label{tab:ds2dxdq2TotalSys_3}
\end{table}

\clearpage
\newpage
\begin{table} 
\begin{footnotesize}
\begin{center}
\begin{tabular}[t]{|rcr|r|lcl|l|rll|} \hline
\multicolumn{3}{|c|}{$Q^2$ range} & \multicolumn{1}{c|}{$Q^{2}_c$} & \multicolumn{3}{c|}{$x$ range} & \multicolumn{1}{c|}{$x_c$} & \multicolumn{3}{c|}{$x\tilde{F_3} \times 10$} \\
\multicolumn{3}{|c|}{($\gev^{2}$)} & \multicolumn{1}{c|}{($\gev^{2}$)} & \multicolumn{3}{c|}{} & \multicolumn{1}{c|}{} & \multicolumn{3}{c|}{} \\ \hline \hline
2500 & -- & 3500 & 3000 & 0.037 & -- & 0.06 & 0.05 & 0.26 & $\pm$ 0.28 & $^{+0.10} _{-0.08}$\\
& & & & 0.06 & -- & 0.1 & 0.08 & 1.10 & $\pm$ 0.35 & $^{+0.10} _{-0.11}$\\
& & & & 0.1 & -- & 0.15 & 0.13 & 0.56 & $\pm$ 0.56 & $^{+0.30} _{-0.10}$\\
& & & & 0.15 & -- & 0.23 & 0.18 & 0.08 & $\pm$ 0.72 & $^{+0.17} _{-0.32}$\\
& & & & 0.23 & -- & 0.35 & 0.25 & 2.25 & $\pm$ 0.95 & $^{+0.85} _{-0.93}$\\
& & & & 0.35 & -- & 0.53 & 0.4 & -0.03 & $\pm$ 1.18 & $^{+0.51} _{-0.47}$\\
& & & & 0.53 & -- & 0.75 & 0.65 & -0.36 & $^{+0.46} _{-0.44}$ & $^{+0.31} _{-0.23}$\\
\hline
3500 & -- & 5600 & 5000 & 0.04 & -- & 0.1 & 0.08 & 0.74 & $\pm$ 0.20 & $^{+0.09} _{-0.06}$\\
& & & & 0.1 & -- & 0.15 & 0.13 & 1.16 & $\pm$ 0.36 & $^{+0.07} _{-0.07}$\\
& & & & 0.15 & -- & 0.23 & 0.18 & 1.13 & $\pm$ 0.42 & $^{+0.09} _{-0.13}$\\
& & & & 0.23 & -- & 0.35 & 0.25 & 0.60 & $\pm$ 0.54 & $^{+0.17} _{-0.16}$\\
& & & & 0.35 & -- & 0.53 & 0.4 & 1.05 & $\pm$ 0.67 & $^{+0.38} _{-0.35}$\\
\hline
5600 & -- & 9000 & 8000 & 0.07 & -- & 0.15 & 0.13 & 1.74 & $\pm$ 0.25 & $^{+0.07} _{-0.09}$\\
& & & & 0.15 & -- & 0.23 & 0.18 & 1.54 & $\pm$ 0.35 & $^{+0.11} _{-0.14}$\\
& & & & 0.23 & -- & 0.35 & 0.25 & 1.49 & $\pm$ 0.43 & $^{+0.19} _{-0.18}$\\
& & & & 0.35 & -- & 0.53 & 0.4 & 0.40 & $^{+0.50} _{-0.48}$ & $^{+0.22} _{-0.21}$\\
& & & & 0.53 & -- & 0.75 & 0.65 & 0.23 & $^{+0.25} _{-0.19}$ & $^{+0.04} _{-0.05}$\\
\hline
9000 & -- & 15000 & 12000 & 0.09 & -- & 0.23 & 0.18 & 0.91 & $\pm$ 0.31 & $^{+0.07} _{-0.17}$\\
& & & & 0.23 & -- & 0.35 & 0.25 & 1.37 & $\pm$ 0.43 & $^{+0.09} _{-0.14}$\\
& & & & 0.35 & -- & 0.53 & 0.4 & 0.97 & $^{+0.46} _{-0.43}$ & $^{+0.14} _{-0.15}$\\
\hline
15000 & -- & 25000 & 20000 & 0.15 & -- & 0.35 & 0.25 & 2.04 & $\pm$ 0.27 & $^{+0.16} _{-0.09}$\\
& & & & 0.35 & -- & 0.75 & 0.4 & 1.02 & $^{+0.41} _{-0.31}$ & $^{+0.12} _{-0.11}$\\
\hline
25000 & -- & 50000 & 30000 & 0.25 & -- & 0.75 & 0.4 & 1.05 & $^{+0.32} _{-0.25}$ & $^{+0.10} _{-0.15}$\\

\hline\end{tabular}
\end{center}
\end{footnotesize}
\caption[]
{The structure function $x\tilde{F_3}$
extracted using the $e^{-}p$ data set
($\mathcal{L} = 169.9 \pbi, P_{e}=0$)
and previously published NC $e^{+}p$ DIS results
($\mathcal{L} = 63.2 \pbi, P_{e} = 0$).
The bin range and bin centre for $Q^2$ and $x$,
and measured $x\tilde{F_3}$ are shown.
The first (second) error on the measurement
refers to the statistical (systematic) uncertainties.}
\label{tab:xF3}
\end{table}

\begin{table} 
\begin{center}
\begin{tabular}[t]{|r|l|rll|}
\hline
\multicolumn{1}{|c|}{$Q^{2}$} & \multicolumn{1}{c|}{$x_c$} & \multicolumn{3}{c|}{$xF_{3}^{\gamma Z} \times 10$} \\
\multicolumn{1}{|c|}{($\gev^{2}$)} & & & &\\
\hline \hline
5000 & 0.032 & 1.53 & $\pm$ 1.41 & $\pm$ 0.80\\
 & 0.05 & 1.01 & $\pm$ 1.19 & $\pm$ 0.38\\
 & 0.08 & 3.26 & $\pm$ 0.66 & $\pm$ 0.24\\
 & 0.13 & 4.62 & $\pm$ 0.62 & $\pm$ 0.16\\
 & 0.18 & 3.01 & $\pm$ 0.54 & $\pm$ 0.19\\
 & 0.25 & 3.93 & $\pm$ 0.44 & $\pm$ 0.17\\
 & 0.4 & 2.08 & $\pm$ 0.39 & $\pm$ 0.15\\
 & 0.65 & 0.55 & $\pm$ 0.63 & $\pm$ 0.13\\

\hline
\end{tabular}
\end{center}
\caption[]
{The interference structure function $xF^{\gamma Z}_{3}$
evaluated at $Q^{2}=5\ 000 \gev^{2}$ for $x$ bins centred on $x_c$.
The first (second) error on the measurement
refers to the statistical (systematic) uncertainties.}
\label{tab:xF3gz}
\end{table}

\clearpage
\newpage
\begin{table} \begin{scriptsize} \begin{center} \begin{tabular}[t]{|rcr|r|lcl|l|rll|r|r|} \hline
\multicolumn{3}{|c|}{$Q^2$ range} & \multicolumn{1}{c|}{$Q^2_c$} & \multicolumn{3}{|c|}{$x$ range} & \multicolumn{1}{c|}{$x_c$} & \multicolumn{3}{c|}{$\tilde{\sigma}$} & \multicolumn{1}{c|}{$N_{\text{data}}$} & \multicolumn{1}{c|}{$N^{\text{MC}}_{\text{bg}}$} \\
\multicolumn{3}{|c|}{($\gev^{2}$)} & \multicolumn{1}{c|}{($\gev^{2}$)} & \multicolumn{3}{|c|}{} & \multicolumn{1}{c|}{} & \multicolumn{3}{c|}{} & & \\ \hline \hline
185 & -- & 240 & 200 & 0.0037 & -- & 0.006 & 0.005 & 1.09 & $\pm$ 0.01 & $^{+0.01} _{-0.01}$ & 5913 & 36.8 \\
& & & & 0.006 & -- & 0.01 & 0.008 & 0.93 & $\pm$ 0.01 & $^{+0.02} _{-0.02}$ & 6914 & 9.8 \\
& & & & 0.01 & -- & 0.017 & 0.013 & 0.78 & $\pm$ 0.01 & $^{+0.01} _{-0.01}$ & 7406 & 1.2 \\
& & & & 0.017 & -- & 0.025 & 0.021 & 0.67 & $\pm$ 0.01 & $^{+0.01} _{-0.01}$ & 5337 & 0.0 \\
& & & & 0.025 & -- & 0.037 & 0.032 & 0.56 & $\pm$ 0.01 & $^{+0.02} _{-0.02}$ & 4970 & 0.0 \\
& & & & 0.037 & -- & 0.06 & 0.05 & 0.50 & $\pm$ 0.01 & $^{+0.01} _{-0.01}$ & 4615 & 0.0 \\
& & & & 0.06 & -- & 0.12 & 0.08 & 0.43 & $\pm$ 0.01 & $^{+0.02} _{-0.02}$ & 5891 & 0.0 \\
& & & & 0.12 & -- & 0.25 & 0.18 & 0.33 & $\pm$ 0.01 & $^{+0.02} _{-0.02}$ & 3003 & 0.0 \\
\hline
240 & -- & 310 & 250 & 0.006 & -- & 0.01 & 0.008 & 0.92 & $\pm$ 0.01 & $^{+0.01} _{-0.01}$ & 4675 & 10.9 \\
& & & & 0.01 & -- & 0.017 & 0.013 & 0.79 & $\pm$ 0.01 & $^{+0.01} _{-0.01}$ & 5247 & 2.5 \\
& & & & 0.017 & -- & 0.025 & 0.021 & 0.65 & $\pm$ 0.01 & $^{+0.01} _{-0.01}$ & 3703 & 0.6 \\
& & & & 0.025 & -- & 0.037 & 0.032 & 0.56 & $\pm$ 0.01 & $^{+0.02} _{-0.02}$ & 3657 & 0.0 \\
& & & & 0.037 & -- & 0.06 & 0.05 & 0.48 & $\pm$ 0.01 & $^{+0.02} _{-0.02}$ & 3465 & 0.0 \\
& & & & 0.06 & -- & 0.12 & 0.08 & 0.43 & $\pm$ 0.01 & $^{+0.02} _{-0.02}$ & 4353 & 0.0 \\
& & & & 0.12 & -- & 0.25 & 0.18 & 0.32 & $\pm$ 0.01 & $^{+0.02} _{-0.02}$ & 2864 & 0.0 \\
\hline
310 & -- & 410 & 350 & 0.006 & -- & 0.01 & 0.008 & 0.97 & $\pm$ 0.02 & $^{+0.02} _{-0.01}$ & 2902 & 30.0 \\
& & & & 0.01 & -- & 0.017 & 0.013 & 0.83 & $\pm$ 0.01 & $^{+0.02} _{-0.01}$ & 4010 & 3.7 \\
& & & & 0.017 & -- & 0.025 & 0.021 & 0.69 & $\pm$ 0.01 & $^{+0.01} _{-0.00}$ & 3044 & 0.0 \\
& & & & 0.025 & -- & 0.037 & 0.032 & 0.60 & $\pm$ 0.01 & $^{+0.01} _{-0.01}$ & 2953 & 0.0 \\
& & & & 0.037 & -- & 0.06 & 0.05 & 0.50 & $\pm$ 0.01 & $^{+0.01} _{-0.01}$ & 3112 & 0.0 \\
& & & & 0.06 & -- & 0.12 & 0.08 & 0.43 & $\pm$ 0.01 & $^{+0.01} _{-0.01}$ & 3501 & 0.0 \\
& & & & 0.12 & -- & 0.25 & 0.18 & 0.31 & $\pm$ 0.01 & $^{+0.02} _{-0.02}$ & 2689 & 0.0 \\
\hline
410 & -- & 530 & 450 & 0.006 & -- & 0.01 & 0.008 & 1.00 & $\pm$ 0.02 & $^{+0.02} _{-0.02}$ & 2262 & 35.3 \\
& & & & 0.01 & -- & 0.017 & 0.013 & 0.84 & $\pm$ 0.02 & $^{+0.01} _{-0.01}$ & 1948 & 6.8 \\
& & & & 0.017 & -- & 0.025 & 0.021 & 0.68 & $\pm$ 0.02 & $^{+0.01} _{-0.01}$ & 1523 & 0.5 \\
& & & & 0.025 & -- & 0.037 & 0.032 & 0.57 & $\pm$ 0.01 & $^{+0.01} _{-0.01}$ & 1676 & 0.0 \\
& & & & 0.037 & -- & 0.06 & 0.05 & 0.50 & $\pm$ 0.01 & $^{+0.01} _{-0.01}$ & 2125 & 0.0 \\
& & & & 0.06 & -- & 0.1 & 0.08 & 0.41 & $\pm$ 0.01 & $^{+0.01} _{-0.01}$ & 1687 & 0.0 \\
& & & & 0.1 & -- & 0.17 & 0.13 & 0.35 & $\pm$ 0.01 & $^{+0.02} _{-0.02}$ & 1545 & 0.0 \\
& & & & 0.17 & -- & 0.3 & 0.25 & 0.25 & $\pm$ 0.01 & $^{+0.01} _{-0.01}$ & 1135 & 0.0 \\

\hline
\end{tabular}
\end{center}
\end{scriptsize}
\caption[]
{The reduced cross section $\tilde{\sigma}$ for the reaction
$e^{-}p \rightarrow e^{-}X$ ($\mathcal{L} = 71.2 \pbi, P_{e} = +0.29$).
The bin range, bin centre ($Q^2_c$ and $x_c$)
and measured cross section corrected to the electroweak Born level are shown.
The first (second) error on the cross section
corresponds to the statistical (systematic) uncertainties.
The number of observed data events ($N_{\text{data}}$)
and simulated background events ($N^{\text{MC}}_{\text{bg}}$) are also shown.
This table has two continuations.}
\label{tab:ds2dxdq2Rh_1}
\end{table}

\begin{table} \begin{scriptsize} \begin{center} \begin{tabular}[t]{|rcr|r|lcl|l|rll|r|r|} \hline
\multicolumn{3}{|c|}{$Q^2$ range} & \multicolumn{1}{c|}{$Q^2_c$} & \multicolumn{3}{|c|}{$x$ range} & \multicolumn{1}{c|}{$x_c$} & \multicolumn{3}{c|}{$\tilde{\sigma}$} & \multicolumn{1}{c|}{$N_{\text{data}}$} & \multicolumn{1}{c|}{$N^{\text{MC}}_{\text{bg}}$} \\
\multicolumn{3}{|c|}{($\gev^{2}$)} & \multicolumn{1}{c|}{($\gev^{2}$)} & \multicolumn{3}{|c|}{} & \multicolumn{1}{c|}{} & \multicolumn{3}{c|}{} & & \\ \hline \hline
530 & -- & 710 & 650 & 0.01 & -- & 0.017 & 0.013 & 0.87 & $\pm$ 0.02 & $^{+0.01} _{-0.01}$ & 2231 & 25.1 \\
& & & & 0.017 & -- & 0.025 & 0.021 & 0.76 & $\pm$ 0.02 & $^{+0.01} _{-0.01}$ & 1511 & 2.4 \\
& & & & 0.025 & -- & 0.037 & 0.032 & 0.60 & $\pm$ 0.02 & $^{+0.01} _{-0.01}$ & 1151 & 0.6 \\
& & & & 0.037 & -- & 0.06 & 0.05 & 0.50 & $\pm$ 0.01 & $^{+0.01} _{-0.00}$ & 1186 & 0.0 \\
& & & & 0.06 & -- & 0.1 & 0.08 & 0.41 & $\pm$ 0.01 & $^{+0.01} _{-0.01}$ & 988 & 0.0 \\
& & & & 0.1 & -- & 0.17 & 0.13 & 0.35 & $\pm$ 0.01 & $^{+0.01} _{-0.01}$ & 891 & 0.0 \\
& & & & 0.17 & -- & 0.3 & 0.25 & 0.25 & $\pm$ 0.01 & $^{+0.02} _{-0.02}$ & 780 & 0.0 \\
\hline
710 & -- & 900 & 800 & 0.009 & -- & 0.017 & 0.013 & 0.83 & $\pm$ 0.02 & $^{+0.01} _{-0.02}$ & 1347 & 20.7 \\
& & & & 0.017 & -- & 0.025 & 0.021 & 0.68 & $\pm$ 0.02 & $^{+0.01} _{-0.01}$ & 924 & 1.3 \\
& & & & 0.025 & -- & 0.037 & 0.032 & 0.62 & $\pm$ 0.02 & $^{+0.01} _{-0.01}$ & 1000 & 3.1 \\
& & & & 0.037 & -- & 0.06 & 0.05 & 0.49 & $\pm$ 0.02 & $^{+0.01} _{-0.01}$ & 1082 & 0.0 \\
& & & & 0.06 & -- & 0.1 & 0.08 & 0.45 & $\pm$ 0.01 & $^{+0.01} _{-0.01}$ & 974 & 0.0 \\
& & & & 0.1 & -- & 0.17 & 0.13 & 0.38 & $\pm$ 0.01 & $^{+0.00} _{-0.00}$ & 797 & 0.0 \\
& & & & 0.17 & -- & 0.3 & 0.25 & 0.24 & $\pm$ 0.01 & $^{+0.01} _{-0.01}$ & 552 & 0.0 \\
\hline
900 & -- & 1300 & 1200 & 0.01 & -- & 0.017 & 0.014 & 0.88 & $\pm$ 0.03 & $^{+0.02} _{-0.03}$ & 911 & 31.9 \\
& & & & 0.017 & -- & 0.025 & 0.021 & 0.74 & $\pm$ 0.02 & $^{+0.01} _{-0.01}$ & 1005 & 9.8 \\
& & & & 0.025 & -- & 0.037 & 0.032 & 0.59 & $\pm$ 0.02 & $^{+0.01} _{-0.01}$ & 1013 & 4.9 \\
& & & & 0.037 & -- & 0.06 & 0.05 & 0.56 & $\pm$ 0.02 & $^{+0.01} _{-0.01}$ & 1352 & 0.5 \\
& & & & 0.06 & -- & 0.1 & 0.08 & 0.44 & $\pm$ 0.01 & $^{+0.01} _{-0.01}$ & 1215 & 0.0 \\
& & & & 0.1 & -- & 0.17 & 0.13 & 0.36 & $\pm$ 0.01 & $^{+0.00} _{-0.00}$ & 995 & 0.0 \\
& & & & 0.17 & -- & 0.3 & 0.25 & 0.23 & $\pm$ 0.01 & $^{+0.01} _{-0.01}$ & 716 & 0.0 \\
& & & & 0.3 & -- & 0.53 & 0.4 & 0.12 & $\pm$ 0.01 & $^{+0.01} _{-0.01}$ & 257 & 0.0 \\
\hline
1300 & -- & 1800 & 1500 & 0.017 & -- & 0.025 & 0.021 & 0.76 & $\pm$ 0.03 & $^{+0.02} _{-0.02}$ & 530 & 24.5 \\
& & & & 0.025 & -- & 0.037 & 0.032 & 0.62 & $\pm$ 0.03 & $^{+0.00} _{-0.00}$ & 575 & 3.0 \\
& & & & 0.037 & -- & 0.06 & 0.05 & 0.54 & $\pm$ 0.02 & $^{+0.00} _{-0.00}$ & 756 & 0.6 \\
& & & & 0.06 & -- & 0.1 & 0.08 & 0.45 & $\pm$ 0.02 & $^{+0.00} _{-0.01}$ & 751 & 0.0 \\
& & & & 0.1 & -- & 0.15 & 0.13 & 0.36 & $\pm$ 0.02 & $^{+0.01} _{-0.01}$ & 487 & 0.0 \\
& & & & 0.15 & -- & 0.23 & 0.18 & 0.29 & $\pm$ 0.01 & $^{+0.01} _{-0.00}$ & 403 & 0.0 \\
& & & & 0.23 & -- & 0.35 & 0.25 & 0.26 & $\pm$ 0.02 & $^{+0.00} _{-0.01}$ & 304 & 0.0 \\
& & & & 0.35 & -- & 0.53 & 0.4 & 0.12 & $\pm$ 0.01 & $^{+0.01} _{-0.01}$ & 126 & 0.0 \\
\hline
1800 & -- & 2500 & 2000 & 0.023 & -- & 0.037 & 0.032 & 0.63 & $\pm$ 0.03 & $^{+0.02} _{-0.01}$ & 405 & 8.0 \\
& & & & 0.037 & -- & 0.06 & 0.05 & 0.53 & $\pm$ 0.02 & $^{+0.01} _{-0.00}$ & 476 & 0.0 \\
& & & & 0.06 & -- & 0.1 & 0.08 & 0.44 & $\pm$ 0.02 & $^{+0.01} _{-0.01}$ & 497 & 0.5 \\
& & & & 0.1 & -- & 0.15 & 0.13 & 0.42 & $\pm$ 0.02 & $^{+0.00} _{-0.00}$ & 412 & 0.0 \\
& & & & 0.15 & -- & 0.23 & 0.18 & 0.29 & $\pm$ 0.02 & $^{+0.00} _{-0.01}$ & 288 & 0.0 \\
& & & & 0.23 & -- & 0.35 & 0.25 & 0.23 & $\pm$ 0.02 & $^{+0.00} _{-0.00}$ & 197 & 0.0 \\
& & & & 0.35 & -- & 0.53 & 0.4 & 0.12 & $^{+0.01} _{-0.01}$ & $^{+0.00} _{-0.00}$ & 98 & 0.0 \\

\hline
\end{tabular}
\end{center}
\end{scriptsize}
\vspace{0.5cm} \hspace{5.4cm} {\bf Table 11:}~~~{\it Continuation 1.}
\label{tab:ds2dxdq2Rh_2}
\end{table}

\begin{table} \begin{scriptsize} \begin{center} \begin{tabular}[t]{|rcr|r|lcl|l|rll|r|r|} \hline
\multicolumn{3}{|c|}{$Q^2$ range} & \multicolumn{1}{c|}{$Q^2_c$} & \multicolumn{3}{|c|}{$x$ range} & \multicolumn{1}{c|}{$x_c$} & \multicolumn{3}{c|}{$\tilde{\sigma}$} & \multicolumn{1}{c|}{$N_{\text{data}}$} & \multicolumn{1}{c|}{$N^{\text{MC}}_{\text{bg}}$} \\
\multicolumn{3}{|c|}{($\gev^{2}$)} & \multicolumn{1}{c|}{($\gev^{2}$)} & \multicolumn{3}{|c|}{} & \multicolumn{1}{c|}{} & \multicolumn{3}{c|}{} & & \\ \hline \hline
2500 & -- & 3500 & 3000 & 0.037 & -- & 0.06 & 0.05 & 0.56 & $\pm$ 0.03 & $^{+0.01} _{-0.01}$ & 311 & 3.8 \\
& & & & 0.06 & -- & 0.1 & 0.08 & 0.51 & $\pm$ 0.03 & $^{+0.01} _{-0.01}$ & 377 & 0.6 \\
& & & & 0.1 & -- & 0.15 & 0.13 & 0.38 & $\pm$ 0.02 & $^{+0.00} _{-0.00}$ & 260 & 0.0 \\
& & & & 0.15 & -- & 0.23 & 0.18 & 0.29 & $\pm$ 0.02 & $^{+0.00} _{-0.00}$ & 201 & 0.0 \\
& & & & 0.23 & -- & 0.35 & 0.25 & 0.30 & $\pm$ 0.02 & $^{+0.02} _{-0.02}$ & 190 & 0.0 \\
& & & & 0.35 & -- & 0.53 & 0.4 & 0.11 & $^{+0.02} _{-0.01}$ & $^{+0.00} _{-0.00}$ & 61 & 0.0 \\
& & & & 0.53 & -- & 0.75 & 0.65 & 0.01 & $^{+0.00} _{-0.00}$ & $^{+0.00} _{-0.00}$ & 23 & 0.0 \\
\hline
3500 & -- & 5600 & 5000 & 0.04 & -- & 0.1 & 0.08 & 0.48 & $\pm$ 0.02 & $^{+0.01} _{-0.00}$ & 380 & 3.1 \\
& & & & 0.1 & -- & 0.15 & 0.13 & 0.43 & $\pm$ 0.03 & $^{+0.00} _{-0.00}$ & 236 & 0.6 \\
& & & & 0.15 & -- & 0.23 & 0.18 & 0.33 & $\pm$ 0.02 & $^{+0.00} _{-0.00}$ & 199 & 0.0 \\
& & & & 0.23 & -- & 0.35 & 0.25 & 0.26 & $\pm$ 0.02 & $^{+0.00} _{-0.00}$ & 142 & 0.0 \\
& & & & 0.35 & -- & 0.53 & 0.4 & 0.15 & $^{+0.02} _{-0.02}$ & $^{+0.01} _{-0.01}$ & 80 & 0.0 \\
\hline
5600 & -- & 9000 & 8000 & 0.07 & -- & 0.15 & 0.13 & 0.53 & $\pm$ 0.04 & $^{+0.01} _{-0.01}$ & 231 & 1.9 \\
& & & & 0.15 & -- & 0.23 & 0.18 & 0.43 & $\pm$ 0.04 & $^{+0.01} _{-0.01}$ & 143 & 0.0 \\
& & & & 0.23 & -- & 0.35 & 0.25 & 0.29 & $^{+0.03} _{-0.03}$ & $^{+0.01} _{-0.01}$ & 92 & 0.0 \\
& & & & 0.35 & -- & 0.53 & 0.4 & 0.11 & $^{+0.02} _{-0.02}$ & $^{+0.01} _{-0.01}$ & 36 & 0.0 \\
& & & & 0.53 & -- & 0.75 & 0.65 & 0.02 & $^{+0.01} _{-0.00}$ & $^{+0.00} _{-0.00}$ & 14 & 0.0 \\
\hline
9000 & -- & 15000 & 12000 & 0.09 & -- & 0.23 & 0.18 & 0.40 & $^{+0.04} _{-0.04}$ & $^{+0.01} _{-0.02}$ & 99 & 0.6 \\
& & & & 0.23 & -- & 0.35 & 0.25 & 0.36 & $^{+0.05} _{-0.05}$ & $^{+0.01} _{-0.01}$ & 63 & 0.0 \\
& & & & 0.35 & -- & 0.53 & 0.4 & 0.17 & $^{+0.04} _{-0.03}$ & $^{+0.01} _{-0.01}$ & 33 & 0.0 \\
\hline
15000 & -- & 25000 & 20000 & 0.15 & -- & 0.35 & 0.25 & 0.46 & $^{+0.07} _{-0.06}$ & $^{+0.03} _{-0.02}$ & 53 & 1.8 \\
& & & & 0.35 & -- & 0.75 & 0.4 & 0.17 & $^{+0.05} _{-0.04}$ & $^{+0.01} _{-0.01}$ & 19 & 0.0 \\
\hline
25000 & -- & 50000 & 30000 & 0.25 & -- & 0.75 & 0.4 & 0.19 & $^{+0.07} _{-0.05}$ & $^{+0.01} _{-0.01}$ & 14 & 0.0 \\

\hline
\end{tabular}
\end{center}
\end{scriptsize}
\vspace{0.5cm} \hspace{5.4cm} {\bf Table 11:}~~~{\it Continuation 2.}
\label{tab:ds2dxdq2Rh_3}
\end{table}

\begin{table} \begin{scriptsize} \begin{center} \begin{tabular}[t]{|r|l|c|c|c||c|c|c|c|c|c|c|c|} \hline
\multicolumn{1}{|c|}{$Q^2_c$} & \multicolumn{1}{c|}{$x_c$} & \multicolumn{1}{c|}{$\tilde{\sigma}$} & \multicolumn{1}{c|}{stat.} & \multicolumn{1}{c||}{sys.} & \multicolumn{1}{c|}{$\delta_{1}$} & \multicolumn{1}{c|}{$\delta_{2}$} & \multicolumn{1}{c|}{$\delta_{3}$} & \multicolumn{1}{c|}{$\delta_{4}$} & \multicolumn{1}{c|}{$\delta_{5}$} & \multicolumn{1}{c|}{$\delta_{6}$} & \multicolumn{1}{c|}{$\delta_{7}$} & \multicolumn{1}{c|}{$\delta_{8} - \delta_{13}$}\\
\multicolumn{1}{|c|}{($\gev^{2}$)} & & & \multicolumn{1}{c|}{(\%)} & \multicolumn{1}{c||}{(\%)} & \multicolumn{1}{c|}{(\%)} & \multicolumn{1}{c|}{(\%)} & \multicolumn{1}{c|}{(\%)} & \multicolumn{1}{c|}{(\%)} & \multicolumn{1}{c|}{(\%)} & \multicolumn{1}{c|}{(\%)} & \multicolumn{1}{c|}{(\%)} & \multicolumn{1}{c|}{(\%)} \\ \hline \hline
200 & 0.005 & 1.09 & $\pm$ 1.4 & $^{+1.3} _{-1.0}$ & $^{+0.7} _{-0.0}$ & $^{-0.4} _{+0.6}$ & $^{+0.4} _{-0.4}$ & $^{-0.4} _{+0.4}$ & $^{+0.0} _{+0.0}$ & $^{+0.1} _{-0.1}$ & $^{+0.3} _{-0.3}$ & $^{+0.7} _{-0.7}$ \\
& 0.008 & 0.93 & $\pm$ 1.2 & $^{+1.9} _{-1.9}$ & $^{+0.3} _{-0.0}$ & $^{-0.4} _{+0.5}$ & $^{+0.2} _{-0.2}$ & $^{-1.6} _{+1.6}$ & $^{+0.0} _{+0.0}$ & $^{+0.1} _{+0.4}$ & $^{+0.1} _{-0.1}$ & $^{+0.8} _{-1.0}$ \\
& 0.013 & 0.78 & $\pm$ 1.2 & $^{+1.7} _{-1.6}$ & $^{+0.2} _{+0.0}$ & $^{-0.5} _{+0.5}$ & $^{-0.2} _{+0.2}$ & $^{-1.3} _{+1.3}$ & $^{+0.0} _{+0.0}$ & $^{+0.0} _{+0.4}$ & $^{+0.0} _{-0.0}$ & $^{+0.8} _{-0.9}$ \\
& 0.021 & 0.67 & $\pm$ 1.4 & $^{+2.1} _{-2.1}$ & $^{+0.4} _{+0.0}$ & $^{-0.3} _{+0.5}$ & $^{-1.6} _{+1.6}$ & $^{+0.9} _{-0.9}$ & $^{+0.0} _{+0.0}$ & $^{-0.1} _{+0.1}$ & $^{+0.0} _{+0.0}$ & $^{+0.8} _{-0.9}$ \\
& 0.032 & 0.56 & $\pm$ 1.5 & $^{+4.0} _{-4.0}$ & $^{+0.1} _{-0.0}$ & $^{-0.3} _{+0.4}$ & $^{-1.5} _{+1.5}$ & $^{+3.6} _{-3.6}$ & $^{+0.0} _{+0.0}$ & $^{+0.2} _{+0.4}$ & $^{+0.0} _{+0.0}$ & $^{+0.8} _{-0.8}$ \\
& 0.05 & 0.50 & $\pm$ 1.5 & $^{+2.5} _{-2.4}$ & $^{+0.0} _{-0.3}$ & $^{-0.2} _{+0.3}$ & $^{-0.5} _{+0.5}$ & $^{+2.2} _{-2.2}$ & $^{+0.0} _{+0.0}$ & $^{+0.0} _{+0.3}$ & $^{+0.0} _{+0.0}$ & $^{+1.0} _{-0.9}$ \\
& 0.08 & 0.43 & $\pm$ 1.3 & $^{+5.5} _{-5.5}$ & $^{+0.0} _{-0.1}$ & $^{-0.2} _{+0.2}$ & $^{+5.0} _{-5.0}$ & $^{+2.0} _{-2.0}$ & $^{-0.0} _{+0.0}$ & $^{-0.0} _{+0.1}$ & $^{+0.0} _{+0.0}$ & $^{+0.8} _{-1.0}$ \\
& 0.18 & 0.33 & $\pm$ 1.9 & $^{+5.4} _{-5.4}$ & $^{+0.1} _{+0.0}$ & $^{-0.1} _{+0.2}$ & $^{+5.3} _{-5.3}$ & $^{+0.5} _{-0.5}$ & $^{+0.7} _{+0.2}$ & $^{-0.1} _{+0.1}$ & $^{+0.0} _{+0.0}$ & $^{+0.7} _{-0.9}$ \\
\hline
250 & 0.008 & 0.92 & $\pm$ 1.5 & $^{+1.5} _{-1.3}$ & $^{+0.6} _{-0.0}$ & $^{-0.3} _{+0.4}$ & $^{+0.5} _{-0.5}$ & $^{-0.9} _{+0.9}$ & $^{+0.0} _{+0.0}$ & $^{-0.2} _{+0.3}$ & $^{+0.1} _{-0.1}$ & $^{+0.7} _{-0.7}$ \\
& 0.013 & 0.79 & $\pm$ 1.4 & $^{+1.7} _{-1.5}$ & $^{+0.7} _{+0.0}$ & $^{-0.5} _{+0.6}$ & $^{+0.0} _{-0.0}$ & $^{-1.3} _{+1.3}$ & $^{+0.0} _{+0.0}$ & $^{-0.1} _{+0.2}$ & $^{+0.0} _{-0.0}$ & $^{+0.5} _{-0.7}$ \\
& 0.021 & 0.65 & $\pm$ 1.7 & $^{+1.6} _{-1.4}$ & $^{+0.7} _{-0.0}$ & $^{-0.3} _{+0.5}$ & $^{-0.1} _{+0.1}$ & $^{+1.1} _{-1.1}$ & $^{+0.0} _{+0.0}$ & $^{-0.2} _{+0.2}$ & $^{+0.0} _{-0.0}$ & $^{+0.8} _{-0.7}$ \\
& 0.032 & 0.56 & $\pm$ 1.7 & $^{+3.3} _{-3.3}$ & $^{+0.6} _{-0.0}$ & $^{-0.3} _{+0.4}$ & $^{-0.8} _{+0.8}$ & $^{+3.0} _{-3.0}$ & $^{+0.0} _{+0.0}$ & $^{+0.0} _{+0.1}$ & $^{+0.0} _{+0.0}$ & $^{+0.7} _{-0.9}$ \\
& 0.05 & 0.48 & $\pm$ 1.8 & $^{+3.4} _{-3.3}$ & $^{+0.4} _{+0.0}$ & $^{-0.1} _{+0.3}$ & $^{-2.4} _{+2.4}$ & $^{+2.2} _{-2.2}$ & $^{+0.0} _{+0.0}$ & $^{-0.0} _{-0.1}$ & $^{+0.0} _{+0.0}$ & $^{+0.7} _{-0.6}$ \\
& 0.08 & 0.43 & $\pm$ 1.6 & $^{+5.2} _{-5.1}$ & $^{+0.7} _{-0.0}$ & $^{-0.1} _{+0.3}$ & $^{+4.6} _{-4.6}$ & $^{+2.2} _{-2.2}$ & $^{+0.0} _{+0.0}$ & $^{-0.0} _{+0.1}$ & $^{+0.0} _{+0.0}$ & $^{+0.7} _{-0.7}$ \\
& 0.18 & 0.32 & $\pm$ 1.9 & $^{+4.8} _{-4.7}$ & $^{+0.2} _{+0.0}$ & $^{+0.0} _{+0.2}$ & $^{+4.1} _{-4.1}$ & $^{+2.0} _{-2.0}$ & $^{+0.3} _{+1.0}$ & $^{+0.0} _{+0.2}$ & $^{+0.0} _{+0.0}$ & $^{+0.8} _{-0.8}$ \\
\hline
350 & 0.008 & 0.97 & $\pm$ 1.9 & $^{+2.1} _{-1.5}$ & $^{+1.6} _{-0.1}$ & $^{-0.5} _{+0.6}$ & $^{+0.9} _{-0.9}$ & $^{-0.4} _{+0.4}$ & $^{+0.0} _{+0.0}$ & $^{-0.6} _{+0.4}$ & $^{+0.4} _{-0.4}$ & $^{+0.6} _{-0.8}$ \\
& 0.013 & 0.83 & $\pm$ 1.6 & $^{+1.9} _{-1.2}$ & $^{+1.4} _{-0.0}$ & $^{-0.2} _{+0.5}$ & $^{-0.2} _{+0.2}$ & $^{-1.0} _{+1.0}$ & $^{+0.0} _{+0.0}$ & $^{-0.2} _{+0.5}$ & $^{+0.0} _{-0.0}$ & $^{+0.7} _{-0.6}$ \\
& 0.021 & 0.69 & $\pm$ 1.9 & $^{+0.9} _{-0.6}$ & $^{+0.4} _{-0.1}$ & $^{-0.3} _{+0.4}$ & $^{+0.3} _{-0.3}$ & $^{+0.0} _{-0.0}$ & $^{+0.0} _{+0.0}$ & $^{-0.2} _{-0.0}$ & $^{+0.0} _{+0.0}$ & $^{+0.6} _{-0.5}$ \\
& 0.032 & 0.60 & $\pm$ 1.9 & $^{+1.9} _{-1.8}$ & $^{+0.8} _{-0.0}$ & $^{-0.2} _{+0.4}$ & $^{-1.6} _{+1.6}$ & $^{+0.2} _{-0.2}$ & $^{+0.0} _{+0.0}$ & $^{-0.3} _{+0.1}$ & $^{+0.0} _{+0.0}$ & $^{+0.6} _{-0.7}$ \\
& 0.05 & 0.50 & $\pm$ 1.9 & $^{+2.7} _{-2.6}$ & $^{+0.6} _{-0.0}$ & $^{-0.1} _{+0.2}$ & $^{-1.9} _{+1.9}$ & $^{+1.6} _{-1.6}$ & $^{+0.0} _{+0.0}$ & $^{-0.1} _{+0.2}$ & $^{+0.0} _{+0.0}$ & $^{+0.5} _{-0.7}$ \\
& 0.08 & 0.43 & $\pm$ 1.7 & $^{+2.0} _{-2.0}$ & $^{+0.3} _{+0.0}$ & $^{-0.1} _{+0.2}$ & $^{+1.1} _{-1.1}$ & $^{+1.6} _{-1.6}$ & $^{+0.0} _{+0.0}$ & $^{+0.0} _{+0.1}$ & $^{+0.0} _{+0.0}$ & $^{+0.7} _{-0.7}$ \\
& 0.18 & 0.31 & $\pm$ 2.0 & $^{+6.4} _{-6.3}$ & $^{+0.4} _{+0.0}$ & $^{-0.0} _{+0.2}$ & $^{+5.7} _{-5.7}$ & $^{+2.6} _{-2.6}$ & $^{-0.1} _{+1.2}$ & $^{+0.1} _{+0.2}$ & $^{+0.0} _{+0.0}$ & $^{+0.7} _{-0.6}$ \\
\hline
450 & 0.008 & 1.00 & $\pm$ 2.2 & $^{+2.4} _{-2.1}$ & $^{+0.7} _{-0.3}$ & $^{-1.4} _{+1.6}$ & $^{+1.3} _{-1.3}$ & $^{-0.5} _{+0.5}$ & $^{+0.0} _{+0.0}$ & $^{-0.2} _{+0.3}$ & $^{+0.6} _{-0.6}$ & $^{+0.5} _{-0.4}$ \\
& 0.013 & 0.84 & $\pm$ 2.3 & $^{+1.1} _{-1.1}$ & $^{+0.1} _{-0.1}$ & $^{-0.5} _{+0.6}$ & $^{-0.4} _{+0.4}$ & $^{+0.2} _{-0.2}$ & $^{+0.0} _{+0.0}$ & $^{-0.6} _{+0.6}$ & $^{+0.1} _{-0.1}$ & $^{+0.5} _{-0.6}$ \\
& 0.021 & 0.68 & $\pm$ 2.6 & $^{+1.8} _{-1.7}$ & $^{+0.2} _{-0.1}$ & $^{-0.4} _{+0.5}$ & $^{-0.1} _{+0.1}$ & $^{+1.3} _{-1.3}$ & $^{+0.0} _{+0.0}$ & $^{-0.7} _{+0.9}$ & $^{+0.0} _{-0.0}$ & $^{+0.8} _{-0.8}$ \\
& 0.032 & 0.57 & $\pm$ 2.5 & $^{+1.9} _{-1.7}$ & $^{+0.8} _{-0.1}$ & $^{-0.2} _{+0.4}$ & $^{+0.0} _{-0.0}$ & $^{+1.4} _{-1.4}$ & $^{+0.0} _{+0.0}$ & $^{-0.5} _{+0.4}$ & $^{+0.0} _{+0.0}$ & $^{+1.0} _{-0.9}$ \\
& 0.05 & 0.50 & $\pm$ 2.2 & $^{+2.5} _{-2.4}$ & $^{+0.7} _{-0.0}$ & $^{+0.0} _{+0.2}$ & $^{-1.6} _{+1.6}$ & $^{+1.6} _{-1.6}$ & $^{+0.0} _{+0.0}$ & $^{-0.4} _{+0.6}$ & $^{+0.0} _{+0.0}$ & $^{+0.7} _{-0.7}$ \\
& 0.08 & 0.41 & $\pm$ 2.5 & $^{+2.1} _{-1.9}$ & $^{+0.3} _{+0.0}$ & $^{+0.1} _{+0.2}$ & $^{-0.7} _{+0.7}$ & $^{+1.7} _{-1.7}$ & $^{+0.0} _{+0.0}$ & $^{+0.1} _{+0.6}$ & $^{+0.0} _{+0.0}$ & $^{+0.7} _{-0.5}$ \\
& 0.13 & 0.35 & $\pm$ 2.6 & $^{+4.9} _{-4.9}$ & $^{+0.6} _{+0.0}$ & $^{+0.1} _{+0.1}$ & $^{+4.4} _{-4.4}$ & $^{+1.9} _{-1.9}$ & $^{+0.0} _{+0.0}$ & $^{-0.4} _{+0.4}$ & $^{+0.0} _{+0.0}$ & $^{+0.5} _{-0.6}$ \\
& 0.25 & 0.25 & $\pm$ 3.0 & $^{+5.6} _{-5.5}$ & $^{+1.1} _{+0.0}$ & $^{+0.2} _{+0.1}$ & $^{+5.2} _{-5.2}$ & $^{+1.1} _{-1.1}$ & $^{-0.9} _{+0.8}$ & $^{-0.1} _{+0.6}$ & $^{+0.0} _{+0.0}$ & $^{+0.8} _{-0.8}$ \\

\hline
\end{tabular}
\end{center}
\end{scriptsize}
\caption[]
{Systematic uncertainties with bin-to-bin correlations
for the reduced cross section $\tilde{\sigma}$
for the reaction $e^{-}p \rightarrow e^{-}X$ ($\mathcal{L} = 71.2 \pbi, P_{e} = +0.29$).
The left five columns of the table contain
the bin centres, $Q^{2}_{c}$ and $x_{c}$, the measured cross section,
the statistical uncertainty and the total systematic uncertainty.
The right eight columns of the table list
the bin-to-bin correlated systematic uncertainties
for $\delta_{1} - \delta_{7}$,
and the systematic uncertainties
summed in quadrature for $\delta_{8} - \delta_{13}$,
as defined in the section \ref{sec-sys}.
The upper and lower correlated uncertainties correspond to
a positive or negative variation of a cut value for example.
However, if this is not possible for a particular systematic,
the uncertainty is symmetrised.
This table has two continuations.}
\label{tab:ds2dxdq2RhSys_1}
\end{table}

\begin{table} \begin{scriptsize} \begin{center} \begin{tabular}[t]{|r|l|c|c|c||c|c|c|c|c|c|c|c|} \hline
\multicolumn{1}{|c|}{$Q^2_c$} & \multicolumn{1}{c|}{$x_c$} & \multicolumn{1}{c|}{$\tilde{\sigma}$} & \multicolumn{1}{c|}{stat.} & \multicolumn{1}{c||}{sys.} & \multicolumn{1}{c|}{$\delta_{1}$} & \multicolumn{1}{c|}{$\delta_{2}$} & \multicolumn{1}{c|}{$\delta_{3}$} & \multicolumn{1}{c|}{$\delta_{4}$} & \multicolumn{1}{c|}{$\delta_{5}$} & \multicolumn{1}{c|}{$\delta_{6}$} & \multicolumn{1}{c|}{$\delta_{7}$} & \multicolumn{1}{c|}{$\delta_{8} - \delta_{13}$}\\
\multicolumn{1}{|c|}{($\gev^{2}$)} & & & \multicolumn{1}{c|}{(\%)} & \multicolumn{1}{c||}{(\%)} & \multicolumn{1}{c|}{(\%)} & \multicolumn{1}{c|}{(\%)} & \multicolumn{1}{c|}{(\%)} & \multicolumn{1}{c|}{(\%)} & \multicolumn{1}{c|}{(\%)} & \multicolumn{1}{c|}{(\%)} & \multicolumn{1}{c|}{(\%)} & \multicolumn{1}{c|}{(\%)} \\ \hline \hline
650 & 0.013 & 0.87 & $\pm$ 2.2 & $^{+1.5} _{-0.8}$ & $^{+1.2} _{-0.2}$ & $^{-0.5} _{+0.6}$ & $^{-0.0} _{+0.0}$ & $^{-0.1} _{+0.1}$ & $^{+0.0} _{+0.0}$ & $^{-0.2} _{+0.3}$ & $^{+0.5} _{-0.5}$ & $^{+0.4} _{-0.3}$ \\
& 0.021 & 0.76 & $\pm$ 2.6 & $^{+1.1} _{-1.1}$ & $^{+0.0} _{-0.1}$ & $^{-0.4} _{+0.5}$ & $^{-0.3} _{+0.3}$ & $^{-0.9} _{+0.9}$ & $^{+0.0} _{+0.0}$ & $^{-0.2} _{+0.0}$ & $^{+0.1} _{-0.1}$ & $^{+0.2} _{-0.2}$ \\
& 0.032 & 0.60 & $\pm$ 3.0 & $^{+1.2} _{-1.0}$ & $^{+0.5} _{-0.2}$ & $^{-0.4} _{+0.5}$ & $^{-0.5} _{+0.5}$ & $^{+0.7} _{-0.7}$ & $^{+0.0} _{+0.0}$ & $^{+0.5} _{+0.5}$ & $^{+0.0} _{-0.0}$ & $^{+0.2} _{-0.2}$ \\
& 0.05 & 0.50 & $\pm$ 2.9 & $^{+1.0} _{-0.9}$ & $^{+0.1} _{-0.0}$ & $^{-0.3} _{+0.4}$ & $^{-0.4} _{+0.4}$ & $^{+0.7} _{-0.7}$ & $^{+0.0} _{+0.0}$ & $^{+0.1} _{+0.0}$ & $^{+0.0} _{+0.0}$ & $^{+0.5} _{-0.4}$ \\
& 0.08 & 0.41 & $\pm$ 3.2 & $^{+2.4} _{-2.3}$ & $^{+0.7} _{-0.0}$ & $^{-0.1} _{+0.3}$ & $^{-1.8} _{+1.8}$ & $^{+1.0} _{-1.0}$ & $^{+0.0} _{+0.0}$ & $^{+0.1} _{-0.0}$ & $^{+0.0} _{+0.0}$ & $^{+0.9} _{-1.1}$ \\
& 0.13 & 0.35 & $\pm$ 3.4 & $^{+2.7} _{-2.4}$ & $^{+0.9} _{-0.0}$ & $^{-0.1} _{+0.2}$ & $^{+2.1} _{-2.1}$ & $^{-0.8} _{+0.8}$ & $^{+0.0} _{+0.0}$ & $^{+0.1} _{+0.4}$ & $^{+0.0} _{+0.0}$ & $^{+1.0} _{-0.9}$ \\
& 0.25 & 0.25 & $\pm$ 3.6 & $^{+6.6} _{-6.5}$ & $^{+1.0} _{-0.0}$ & $^{-0.1} _{+0.3}$ & $^{+6.3} _{-6.3}$ & $^{+1.0} _{-1.0}$ & $^{-0.8} _{+0.2}$ & $^{-0.1} _{+0.5}$ & $^{+0.0} _{+0.0}$ & $^{+1.0} _{-1.1}$ \\
\hline
800 & 0.013 & 0.83 & $\pm$ 2.8 & $^{+1.7} _{-1.9}$ & $^{+0.4} _{-1.0}$ & $^{-0.8} _{+1.0}$ & $^{+0.4} _{-0.4}$ & $^{-1.0} _{+1.0}$ & $^{+0.0} _{+0.0}$ & $^{-0.1} _{+0.0}$ & $^{+0.6} _{-0.6}$ & $^{+0.3} _{-0.4}$ \\
& 0.021 & 0.68 & $\pm$ 3.3 & $^{+1.1} _{-0.9}$ & $^{+0.5} _{-0.1}$ & $^{-0.4} _{+0.5}$ & $^{+0.6} _{-0.6}$ & $^{-0.2} _{+0.2}$ & $^{+0.0} _{+0.0}$ & $^{+0.0} _{+0.5}$ & $^{+0.1} _{-0.1}$ & $^{+0.4} _{-0.4}$ \\
& 0.032 & 0.62 & $\pm$ 3.2 & $^{+1.4} _{-1.4}$ & $^{+0.0} _{-0.4}$ & $^{-0.4} _{+0.5}$ & $^{-0.8} _{+0.8}$ & $^{-0.9} _{+0.9}$ & $^{+0.0} _{+0.0}$ & $^{+0.3} _{+0.0}$ & $^{+0.1} _{-0.1}$ & $^{+0.2} _{-0.4}$ \\
& 0.05 & 0.49 & $\pm$ 3.1 & $^{+2.2} _{-2.2}$ & $^{+0.3} _{+0.0}$ & $^{-0.3} _{+0.4}$ & $^{+0.7} _{-0.7}$ & $^{+2.0} _{-2.0}$ & $^{+0.0} _{+0.0}$ & $^{-0.4} _{+0.1}$ & $^{+0.0} _{+0.0}$ & $^{+0.3} _{-0.4}$ \\
& 0.08 & 0.45 & $\pm$ 3.2 & $^{+2.6} _{-2.6}$ & $^{+0.5} _{-0.0}$ & $^{-0.2} _{+0.3}$ & $^{-2.5} _{+2.5}$ & $^{+0.1} _{-0.1}$ & $^{+0.0} _{+0.0}$ & $^{+0.5} _{-0.7}$ & $^{+0.0} _{+0.0}$ & $^{+0.3} _{-0.4}$ \\
& 0.13 & 0.38 & $\pm$ 3.6 & $^{+1.1} _{-1.0}$ & $^{+0.5} _{-0.0}$ & $^{-0.2} _{+0.3}$ & $^{+0.7} _{-0.7}$ & $^{+0.6} _{-0.6}$ & $^{+0.0} _{+0.0}$ & $^{+0.1} _{-0.4}$ & $^{+0.0} _{+0.0}$ & $^{+0.2} _{-0.2}$ \\
& 0.25 & 0.24 & $\pm$ 4.3 & $^{+3.7} _{-3.4}$ & $^{+0.3} _{+0.0}$ & $^{-0.2} _{+0.2}$ & $^{+3.3} _{-3.3}$ & $^{+0.8} _{-0.8}$ & $^{+1.2} _{-0.1}$ & $^{+0.7} _{-0.5}$ & $^{+0.0} _{+0.0}$ & $^{+0.2} _{-0.3}$ \\
\hline
1200 & 0.014 & 0.88 & $\pm$ 3.4 & $^{+2.4} _{-3.2}$ & $^{+0.6} _{-2.2}$ & $^{-0.8} _{+1.0}$ & $^{-0.7} _{+0.7}$ & $^{-1.4} _{+1.4}$ & $^{+0.0} _{+0.0}$ & $^{-0.3} _{-0.8}$ & $^{+1.4} _{-1.4}$ & $^{+0.5} _{-0.2}$ \\
& 0.021 & 0.74 & $\pm$ 3.2 & $^{+1.9} _{-1.6}$ & $^{+0.9} _{-0.1}$ & $^{-0.5} _{+0.5}$ & $^{+1.5} _{-1.5}$ & $^{+0.0} _{-0.0}$ & $^{+0.0} _{+0.0}$ & $^{+0.1} _{+0.4}$ & $^{+0.4} _{-0.4}$ & $^{+0.3} _{-0.4}$ \\
& 0.032 & 0.59 & $\pm$ 3.2 & $^{+1.2} _{-1.1}$ & $^{+0.0} _{-0.1}$ & $^{-0.4} _{+0.5}$ & $^{-0.5} _{+0.5}$ & $^{-0.7} _{+0.7}$ & $^{+0.0} _{+0.0}$ & $^{+0.1} _{+0.1}$ & $^{+0.2} _{-0.2}$ & $^{+0.6} _{-0.6}$ \\
& 0.05 & 0.56 & $\pm$ 2.8 & $^{+1.3} _{-1.4}$ & $^{+0.1} _{-0.0}$ & $^{-0.3} _{+0.3}$ & $^{-0.1} _{+0.1}$ & $^{+1.2} _{-1.2}$ & $^{+0.0} _{+0.0}$ & $^{-0.3} _{-0.4}$ & $^{+0.0} _{-0.0}$ & $^{+0.4} _{-0.4}$ \\
& 0.08 & 0.44 & $\pm$ 2.9 & $^{+1.4} _{-1.3}$ & $^{+0.3} _{-0.0}$ & $^{-0.2} _{+0.4}$ & $^{-0.3} _{+0.3}$ & $^{+1.2} _{-1.2}$ & $^{+0.0} _{+0.0}$ & $^{+0.3} _{-0.3}$ & $^{+0.0} _{+0.0}$ & $^{+0.4} _{-0.3}$ \\
& 0.13 & 0.36 & $\pm$ 3.2 & $^{+1.1} _{-1.2}$ & $^{+0.1} _{-0.0}$ & $^{-0.1} _{+0.2}$ & $^{-0.9} _{+0.9}$ & $^{+0.4} _{-0.4}$ & $^{+0.0} _{+0.0}$ & $^{+0.1} _{-0.5}$ & $^{+0.0} _{+0.0}$ & $^{+0.3} _{-0.4}$ \\
& 0.25 & 0.23 & $\pm$ 3.8 & $^{+2.4} _{-2.3}$ & $^{+0.3} _{+0.0}$ & $^{-0.2} _{+0.3}$ & $^{+2.0} _{-2.0}$ & $^{+1.0} _{-1.0}$ & $^{+0.0} _{+0.0}$ & $^{-0.3} _{-0.1}$ & $^{+0.0} _{+0.0}$ & $^{+0.5} _{-0.4}$ \\
& 0.4 & 0.12 & $\pm$ 6.3 & $^{+8.1} _{-5.7}$ & $^{+0.0} _{-0.1}$ & $^{-0.4} _{+0.4}$ & $^{+5.6} _{-5.6}$ & $^{+0.6} _{-0.6}$ & $^{+5.8} _{+3.4}$ & $^{+0.5} _{-1.0}$ & $^{+0.0} _{+0.0}$ & $^{+0.5} _{-0.5}$ \\
\hline
1500 & 0.021 & 0.76 & $\pm$ 4.5 & $^{+2.6} _{-2.3}$ & $^{+0.9} _{-0.2}$ & $^{-0.6} _{+0.4}$ & $^{-0.4} _{+0.4}$ & $^{-0.1} _{+0.1}$ & $^{+0.0} _{+0.0}$ & $^{-1.0} _{+0.1}$ & $^{+2.3} _{-1.9}$ & $^{+0.4} _{-0.4}$ \\
& 0.032 & 0.62 & $\pm$ 4.2 & $^{+0.7} _{-0.6}$ & $^{+0.3} _{-0.1}$ & $^{-0.3} _{+0.3}$ & $^{-0.1} _{+0.1}$ & $^{-0.3} _{+0.3}$ & $^{+0.0} _{+0.0}$ & $^{-0.1} _{-0.1}$ & $^{+0.2} _{-0.2}$ & $^{+0.4} _{-0.3}$ \\
& 0.05 & 0.54 & $\pm$ 3.7 & $^{+0.9} _{-0.8}$ & $^{+0.2} _{-0.0}$ & $^{-0.3} _{+0.4}$ & $^{+0.2} _{-0.2}$ & $^{+0.4} _{-0.4}$ & $^{+0.0} _{+0.0}$ & $^{-0.4} _{+0.5}$ & $^{+0.0} _{-0.0}$ & $^{+0.3} _{-0.4}$ \\
& 0.08 & 0.45 & $\pm$ 3.7 & $^{+1.0} _{-1.1}$ & $^{+0.2} _{-0.0}$ & $^{-0.3} _{+0.4}$ & $^{-0.9} _{+0.9}$ & $^{-0.2} _{+0.2}$ & $^{+0.0} _{+0.0}$ & $^{-0.4} _{+0.1}$ & $^{+0.0} _{+0.0}$ & $^{+0.3} _{-0.5}$ \\
& 0.13 & 0.36 & $\pm$ 4.6 & $^{+2.1} _{-1.8}$ & $^{+0.8} _{-0.0}$ & $^{-0.1} _{+0.3}$ & $^{-1.3} _{+1.3}$ & $^{-1.1} _{+1.1}$ & $^{+0.0} _{+0.0}$ & $^{+0.8} _{-0.2}$ & $^{+0.0} _{+0.0}$ & $^{+0.3} _{-0.4}$ \\
& 0.18 & 0.29 & $\pm$ 5.0 & $^{+1.8} _{-1.7}$ & $^{+0.6} _{-0.1}$ & $^{-0.3} _{+0.2}$ & $^{+0.0} _{-0.0}$ & $^{-1.6} _{+1.6}$ & $^{+0.0} _{+0.0}$ & $^{-0.3} _{-0.2}$ & $^{+0.0} _{+0.0}$ & $^{+0.4} _{-0.4}$ \\
& 0.25 & 0.26 & $\pm$ 5.8 & $^{+1.9} _{-1.9}$ & $^{+0.0} _{-0.3}$ & $^{-0.3} _{+0.3}$ & $^{+1.7} _{-1.7}$ & $^{+0.6} _{-0.6}$ & $^{+0.1} _{+0.0}$ & $^{-0.1} _{-0.3}$ & $^{+0.0} _{+0.0}$ & $^{+0.3} _{-0.3}$ \\
& 0.4 & 0.12 & $\pm$ 9.0 & $^{+8.8} _{-4.3}$ & $^{+0.6} _{+0.0}$ & $^{-0.5} _{+0.3}$ & $^{+2.0} _{-2.0}$ & $^{-0.9} _{+0.9}$ & $^{+8.4} _{-3.6}$ & $^{+0.3} _{-0.4}$ & $^{+0.0} _{+0.0}$ & $^{+0.6} _{-0.5}$ \\
\hline
2000 & 0.032 & 0.63 & $\pm$ 5.1 & $^{+2.5} _{-1.4}$ & $^{+1.8} _{+0.0}$ & $^{-0.2} _{+0.2}$ & $^{+0.3} _{-0.3}$ & $^{+1.0} _{-1.0}$ & $^{+0.0} _{+0.0}$ & $^{+0.4} _{+0.7}$ & $^{+1.1} _{-0.8}$ & $^{+0.3} _{-0.4}$ \\
& 0.05 & 0.53 & $\pm$ 4.6 & $^{+1.0} _{-0.8}$ & $^{+0.5} _{-0.0}$ & $^{-0.3} _{+0.4}$ & $^{-0.3} _{+0.3}$ & $^{-0.6} _{+0.6}$ & $^{+0.0} _{+0.0}$ & $^{+0.1} _{+0.1}$ & $^{+0.0} _{+0.0}$ & $^{+0.3} _{-0.3}$ \\
& 0.08 & 0.44 & $\pm$ 4.5 & $^{+2.0} _{-2.0}$ & $^{+0.1} _{-0.1}$ & $^{-0.3} _{+0.4}$ & $^{-0.1} _{+0.1}$ & $^{+1.9} _{-1.9}$ & $^{+0.0} _{+0.0}$ & $^{+0.2} _{+0.0}$ & $^{+0.0} _{-0.0}$ & $^{+0.3} _{-0.3}$ \\
& 0.13 & 0.42 & $\pm$ 5.0 & $^{+0.8} _{-1.2}$ & $^{+0.1} _{-0.6}$ & $^{-0.2} _{+0.3}$ & $^{-0.5} _{+0.5}$ & $^{+0.5} _{-0.5}$ & $^{+0.0} _{+0.0}$ & $^{+0.1} _{-0.6}$ & $^{+0.0} _{+0.0}$ & $^{+0.3} _{-0.4}$ \\
& 0.18 & 0.29 & $\pm$ 5.9 & $^{+0.6} _{-1.9}$ & $^{+0.0} _{-0.7}$ & $^{-0.2} _{+0.2}$ & $^{-0.3} _{+0.3}$ & $^{+0.1} _{-0.1}$ & $^{+0.0} _{+0.0}$ & $^{-0.4} _{-1.6}$ & $^{+0.0} _{+0.0}$ & $^{+0.4} _{-0.6}$ \\
& 0.25 & 0.23 & $\pm$ 7.2 & $^{+1.7} _{-1.6}$ & $^{+0.3} _{-0.0}$ & $^{-0.2} _{+0.3}$ & $^{+1.5} _{-1.5}$ & $^{+0.5} _{-0.5}$ & $^{+0.0} _{+0.0}$ & $^{+0.4} _{+0.5}$ & $^{+0.0} _{+0.0}$ & $^{+0.5} _{-0.5}$ \\
& 0.4 & 0.12 & $^{+11.2} _{-10.1}$ & $^{+3.7} _{-3.9}$ & $^{+0.0} _{-0.2}$ & $^{-0.3} _{+0.4}$ & $^{+3.1} _{-3.1}$ & $^{+1.1} _{-1.1}$ & $^{-0.5} _{+1.0}$ & $^{+0.1} _{-1.3}$ & $^{+0.0} _{+0.0}$ & $^{+1.2} _{-1.4}$ \\

\hline
\end{tabular}
\end{center}
\end{scriptsize}
\vspace{0.5cm} \hspace{5.4cm} {\bf Table 12:}~~~{\it Continuation 1.}
\label{tab:ds2dxdq2RhSys_2}
\end{table}

\begin{table} \begin{scriptsize} \begin{center} \begin{tabular}[t]{|r|l|c|c|c||c|c|c|c|c|c|c|c|} \hline
\multicolumn{1}{|c|}{$Q^2_c$} & \multicolumn{1}{c|}{$x_c$} & \multicolumn{1}{c|}{$\tilde{\sigma}$} & \multicolumn{1}{c|}{stat.} & \multicolumn{1}{c||}{sys.} & \multicolumn{1}{c|}{$\delta_{1}$} & \multicolumn{1}{c|}{$\delta_{2}$} & \multicolumn{1}{c|}{$\delta_{3}$} & \multicolumn{1}{c|}{$\delta_{4}$} & \multicolumn{1}{c|}{$\delta_{5}$} & \multicolumn{1}{c|}{$\delta_{6}$} & \multicolumn{1}{c|}{$\delta_{7}$} & \multicolumn{1}{c|}{$\delta_{8} - \delta_{13}$}\\
\multicolumn{1}{|c|}{($\gev^{2}$)} & & & \multicolumn{1}{c|}{(\%)} & \multicolumn{1}{c||}{(\%)} & \multicolumn{1}{c|}{(\%)} & \multicolumn{1}{c|}{(\%)} & \multicolumn{1}{c|}{(\%)} & \multicolumn{1}{c|}{(\%)} & \multicolumn{1}{c|}{(\%)} & \multicolumn{1}{c|}{(\%)} & \multicolumn{1}{c|}{(\%)} & \multicolumn{1}{c|}{(\%)} \\ \hline \hline
3000 & 0.05 & 0.56 & $\pm$ 5.7 & $^{+2.0} _{-1.1}$ & $^{+1.6} _{-0.1}$ & $^{-0.3} _{+0.4}$ & $^{+0.9} _{-0.9}$ & $^{-0.0} _{+0.0}$ & $^{+0.0} _{+0.0}$ & $^{+0.4} _{+0.2}$ & $^{+0.5} _{-0.5}$ & $^{+0.4} _{-0.3}$ \\
& 0.08 & 0.51 & $\pm$ 5.2 & $^{+1.1} _{-1.1}$ & $^{+0.2} _{-0.4}$ & $^{-0.2} _{+0.3}$ & $^{-0.4} _{+0.4}$ & $^{+0.8} _{-0.8}$ & $^{+0.0} _{+0.0}$ & $^{+0.5} _{-0.1}$ & $^{+0.1} _{-0.1}$ & $^{+0.3} _{-0.5}$ \\
& 0.13 & 0.38 & $\pm$ 6.2 & $^{+0.9} _{-0.7}$ & $^{+0.1} _{-0.1}$ & $^{-0.2} _{+0.3}$ & $^{+0.6} _{-0.6}$ & $^{-0.1} _{+0.1}$ & $^{+0.0} _{+0.0}$ & $^{-0.3} _{-0.3}$ & $^{+0.0} _{+0.0}$ & $^{+0.7} _{-0.1}$ \\
& 0.18 & 0.29 & $\pm$ 7.1 & $^{+1.2} _{-1.3}$ & $^{+0.0} _{-0.3}$ & $^{-0.2} _{+0.2}$ & $^{-0.9} _{+0.9}$ & $^{-0.6} _{+0.6}$ & $^{+0.0} _{+0.0}$ & $^{-0.5} _{+0.2}$ & $^{+0.0} _{+0.0}$ & $^{+0.4} _{-0.3}$ \\
& 0.25 & 0.30 & $\pm$ 7.3 & $^{+7.4} _{-7.3}$ & $^{+0.6} _{-0.0}$ & $^{-0.2} _{+0.2}$ & $^{+7.3} _{-7.3}$ & $^{+0.2} _{-0.2}$ & $^{+0.0} _{+0.0}$ & $^{+1.1} _{+0.3}$ & $^{+0.0} _{+0.0}$ & $^{+0.8} _{-0.6}$ \\
& 0.4 & 0.11 & $^{+14.5} _{-12.8}$ & $^{+3.8} _{-3.5}$ & $^{+1.3} _{-0.0}$ & $^{-0.2} _{+0.4}$ & $^{-3.4} _{+3.4}$ & $^{+0.5} _{-0.5}$ & $^{+0.4} _{-0.0}$ & $^{+0.8} _{-0.3}$ & $^{+0.0} _{+0.0}$ & $^{+0.6} _{-0.6}$ \\
& 0.65 & 0.01 & $^{+25.5} _{-20.7}$ & $^{+10.2} _{-10.4}$ & $^{+0.0} _{-0.6}$ & $^{-0.8} _{+1.0}$ & $^{-8.7} _{+8.7}$ & $^{+4.6} _{-4.6}$ & $^{+1.8} _{+0.7}$ & $^{-2.7} _{+0.9}$ & $^{+0.0} _{+0.0}$ & $^{+1.7} _{-2.0}$ \\
\hline
5000 & 0.08 & 0.48 & $\pm$ 5.2 & $^{+1.8} _{-0.9}$ & $^{+1.6} _{-0.3}$ & $^{-0.2} _{+0.2}$ & $^{-0.3} _{+0.3}$ & $^{+0.2} _{-0.2}$ & $^{+0.0} _{+0.0}$ & $^{-0.6} _{-0.4}$ & $^{+0.3} _{-0.3}$ & $^{+0.5} _{-0.3}$ \\
& 0.13 & 0.43 & $\pm$ 6.5 & $^{+1.0} _{-0.8}$ & $^{+0.8} _{+0.0}$ & $^{-0.2} _{+0.2}$ & $^{-0.4} _{+0.4}$ & $^{+0.3} _{-0.3}$ & $^{+0.0} _{+0.0}$ & $^{+0.1} _{-0.3}$ & $^{+0.1} _{-0.1}$ & $^{+0.2} _{-0.5}$ \\
& 0.18 & 0.33 & $\pm$ 7.1 & $^{+1.0} _{-0.5}$ & $^{+0.2} _{-0.1}$ & $^{-0.3} _{+0.2}$ & $^{+0.2} _{-0.2}$ & $^{+0.3} _{-0.3}$ & $^{+0.0} _{+0.0}$ & $^{-0.1} _{+0.7}$ & $^{+0.0} _{+0.0}$ & $^{+0.5} _{-0.3}$ \\
& 0.25 & 0.26 & $\pm$ 8.4 & $^{+1.8} _{-1.2}$ & $^{+0.7} _{-0.0}$ & $^{-0.2} _{+0.3}$ & $^{+0.7} _{-0.7}$ & $^{-0.7} _{+0.7}$ & $^{+0.0} _{+0.0}$ & $^{+1.2} _{-0.6}$ & $^{+0.0} _{+0.0}$ & $^{+0.4} _{-0.3}$ \\
& 0.4 & 0.15 & $^{+12.5} _{-11.2}$ & $^{+6.5} _{-6.4}$ & $^{+0.0} _{-0.3}$ & $^{-0.2} _{+0.3}$ & $^{+6.0} _{-6.0}$ & $^{-2.1} _{+2.1}$ & $^{+0.0} _{+0.0}$ & $^{+0.4} _{+1.3}$ & $^{+0.0} _{+0.0}$ & $^{+1.1} _{-1.1}$ \\
\hline
8000 & 0.13 & 0.53 & $\pm$ 6.6 & $^{+1.0} _{-1.9}$ & $^{+0.0} _{-1.0}$ & $^{-0.3} _{+0.3}$ & $^{-0.5} _{+0.5}$ & $^{-0.1} _{+0.1}$ & $^{+0.0} _{+0.0}$ & $^{-1.3} _{+0.3}$ & $^{+0.5} _{-0.3}$ & $^{+0.5} _{-0.5}$ \\
& 0.18 & 0.43 & $\pm$ 8.4 & $^{+2.6} _{-2.9}$ & $^{+0.1} _{-0.3}$ & $^{-0.2} _{+0.2}$ & $^{-2.4} _{+2.4}$ & $^{+0.6} _{-0.6}$ & $^{+0.0} _{+0.0}$ & $^{+0.5} _{-0.3}$ & $^{+0.0} _{+0.0}$ & $^{+0.6} _{-1.3}$ \\
& 0.25 & 0.29 & $^{+11.6} _{-10.4}$ & $^{+3.8} _{-3.9}$ & $^{+0.7} _{-0.1}$ & $^{-0.3} _{+0.2}$ & $^{-3.6} _{+3.6}$ & $^{+0.4} _{-0.4}$ & $^{+0.0} _{+0.0}$ & $^{-1.1} _{-0.6}$ & $^{+0.0} _{+0.0}$ & $^{+0.4} _{-0.4}$ \\
& 0.4 & 0.11 & $^{+19.6} _{-16.6}$ & $^{+7.1} _{-7.3}$ & $^{+0.1} _{-0.0}$ & $^{-0.3} _{+0.4}$ & $^{+7.0} _{-7.0}$ & $^{-0.6} _{+0.6}$ & $^{+0.0} _{+0.0}$ & $^{-0.3} _{+0.5}$ & $^{+0.0} _{+0.0}$ & $^{+0.6} _{-2.0}$ \\
& 0.65 & 0.02 & $^{+34.5} _{-26.5}$ & $^{+4.1} _{-4.2}$ & $^{+0.2} _{-1.4}$ & $^{-0.9} _{+0.9}$ & $^{+1.5} _{-1.5}$ & $^{+2.3} _{-2.3}$ & $^{+0.0} _{+0.0}$ & $^{-2.6} _{+2.8}$ & $^{+0.0} _{+0.0}$ & $^{+0.8} _{-0.6}$ \\
\hline
12000 & 0.18 & 0.40 & $^{+11.1} _{-10.1}$ & $^{+1.3} _{-3.8}$ & $^{+1.0} _{-3.6}$ & $^{-0.3} _{+0.4}$ & $^{-0.4} _{+0.4}$ & $^{-0.2} _{+0.2}$ & $^{+0.0} _{+0.0}$ & $^{-0.3} _{-0.8}$ & $^{+0.2} _{-0.3}$ & $^{+0.5} _{-0.6}$ \\
& 0.25 & 0.36 & $^{+14.3} _{-12.6}$ & $^{+2.8} _{-2.3}$ & $^{+2.4} _{-0.6}$ & $^{-0.2} _{+0.2}$ & $^{+0.3} _{-0.3}$ & $^{+0.8} _{-0.8}$ & $^{+0.0} _{+0.0}$ & $^{+0.3} _{-1.6}$ & $^{+0.0} _{+0.0}$ & $^{+1.0} _{-1.2}$ \\
& 0.4 & 0.17 & $^{+20.6} _{-17.3}$ & $^{+4.6} _{-3.8}$ & $^{+2.7} _{-0.2}$ & $^{-0.3} _{+0.4}$ & $^{+3.2} _{-3.2}$ & $^{-0.8} _{+0.8}$ & $^{+0.0} _{+0.0}$ & $^{-1.2} _{+1.5}$ & $^{+0.0} _{+0.0}$ & $^{+0.6} _{-1.3}$ \\
\hline
20000 & 0.25 & 0.46 & $^{+16.0} _{-14.0}$ & $^{+6.1} _{-3.6}$ & $^{+4.9} _{-1.8}$ & $^{-1.5} _{+0.2}$ & $^{-2.2} _{+2.2}$ & $^{+0.1} _{-0.1}$ & $^{+0.0} _{+0.0}$ & $^{-0.7} _{-0.2}$ & $^{+2.7} _{-1.4}$ & $^{+0.6} _{-0.5}$ \\
& 0.4 & 0.17 & $^{+28.6} _{-22.7}$ & $^{+4.9} _{-3.8}$ & $^{+3.6} _{+0.0}$ & $^{-0.4} _{+0.4}$ & $^{+0.9} _{-0.9}$ & $^{+3.0} _{-3.0}$ & $^{+0.0} _{+0.0}$ & $^{-1.8} _{-0.3}$ & $^{+0.0} _{+0.0}$ & $^{+1.1} _{-1.2}$ \\
\hline
30000 & 0.4 & 0.19 & $^{+34.5} _{-26.4}$ & $^{+6.9} _{-5.9}$ & $^{+6.7} _{+0.0}$ & $^{-0.2} _{+0.3}$ & $^{-0.8} _{+0.8}$ & $^{+0.7} _{-0.7}$ & $^{+0.0} _{+0.0}$ & $^{+0.1} _{-4.9}$ & $^{+0.0} _{-2.9}$ & $^{+1.3} _{-1.3}$ \\

\hline
\end{tabular}
\end{center}
\end{scriptsize}
\vspace{0.5cm} \hspace{5.4cm} {\bf Table 12:}~~~{\it Continuation 2.}
\label{tab:ds2dxdq2RhSys_3}
\end{table}

\clearpage
\newpage
\begin{table} \begin{scriptsize} \begin{center} \begin{tabular}[t]{|rcr|r|lcl|l|rll|r|r|} \hline
\multicolumn{3}{|c|}{$Q^2$ range} & \multicolumn{1}{c|}{$Q^2_c$} & \multicolumn{3}{|c|}{$x$ range} & \multicolumn{1}{c|}{$x_c$} & \multicolumn{3}{c|}{$\tilde{\sigma}$} & \multicolumn{1}{c|}{$N_{\text{data}}$} & \multicolumn{1}{c|}{$N^{\text{MC}}_{\text{bg}}$} \\
\multicolumn{3}{|c|}{($\gev^{2}$)} & \multicolumn{1}{c|}{($\gev^{2}$)} & \multicolumn{3}{|c|}{} & \multicolumn{1}{c|}{} & \multicolumn{3}{c|}{} & & \\ \hline \hline
185 & -- & 240 & 200 & 0.0037 & -- & 0.006 & 0.005 & 1.09 & $\pm$ 0.01 & $^{+0.01} _{-0.01}$ & 8166 & 51.2 \\
& & & & 0.006 & -- & 0.01 & 0.008 & 0.92 & $\pm$ 0.01 & $^{+0.02} _{-0.02}$ & 9475 & 13.6 \\
& & & & 0.01 & -- & 0.017 & 0.013 & 0.78 & $\pm$ 0.01 & $^{+0.01} _{-0.01}$ & 10222 & 1.7 \\
& & & & 0.017 & -- & 0.025 & 0.021 & 0.66 & $\pm$ 0.01 & $^{+0.01} _{-0.01}$ & 7261 & 0.0 \\
& & & & 0.025 & -- & 0.037 & 0.032 & 0.56 & $\pm$ 0.01 & $^{+0.02} _{-0.02}$ & 6934 & 0.0 \\
& & & & 0.037 & -- & 0.06 & 0.05 & 0.50 & $\pm$ 0.01 & $^{+0.01} _{-0.01}$ & 6387 & 0.0 \\
& & & & 0.06 & -- & 0.12 & 0.08 & 0.45 & $\pm$ 0.01 & $^{+0.02} _{-0.02}$ & 8344 & 0.0 \\
& & & & 0.12 & -- & 0.25 & 0.18 & 0.34 & $\pm$ 0.01 & $^{+0.02} _{-0.02}$ & 4369 & 0.0 \\
\hline
240 & -- & 310 & 250 & 0.006 & -- & 0.01 & 0.008 & 0.95 & $\pm$ 0.01 & $^{+0.01} _{-0.01}$ & 6666 & 15.2 \\
& & & & 0.01 & -- & 0.017 & 0.013 & 0.80 & $\pm$ 0.01 & $^{+0.01} _{-0.01}$ & 7355 & 3.5 \\
& & & & 0.017 & -- & 0.025 & 0.021 & 0.66 & $\pm$ 0.01 & $^{+0.01} _{-0.01}$ & 5232 & 0.8 \\
& & & & 0.025 & -- & 0.037 & 0.032 & 0.57 & $\pm$ 0.01 & $^{+0.02} _{-0.02}$ & 5088 & 0.0 \\
& & & & 0.037 & -- & 0.06 & 0.05 & 0.50 & $\pm$ 0.01 & $^{+0.02} _{-0.02}$ & 5029 & 0.0 \\
& & & & 0.06 & -- & 0.12 & 0.08 & 0.43 & $\pm$ 0.01 & $^{+0.02} _{-0.02}$ & 5997 & 0.0 \\
& & & & 0.12 & -- & 0.25 & 0.18 & 0.34 & $\pm$ 0.01 & $^{+0.02} _{-0.02}$ & 4121 & 0.0 \\
\hline
310 & -- & 410 & 350 & 0.006 & -- & 0.01 & 0.008 & 0.97 & $\pm$ 0.02 & $^{+0.02} _{-0.01}$ & 4001 & 41.5 \\
& & & & 0.01 & -- & 0.017 & 0.013 & 0.82 & $\pm$ 0.01 & $^{+0.02} _{-0.01}$ & 5495 & 5.2 \\
& & & & 0.017 & -- & 0.025 & 0.021 & 0.69 & $\pm$ 0.01 & $^{+0.01} _{-0.00}$ & 4230 & 0.0 \\
& & & & 0.025 & -- & 0.037 & 0.032 & 0.61 & $\pm$ 0.01 & $^{+0.01} _{-0.01}$ & 4118 & 0.0 \\
& & & & 0.037 & -- & 0.06 & 0.05 & 0.51 & $\pm$ 0.01 & $^{+0.01} _{-0.01}$ & 4397 & 0.0 \\
& & & & 0.06 & -- & 0.12 & 0.08 & 0.44 & $\pm$ 0.01 & $^{+0.01} _{-0.01}$ & 4916 & 0.0 \\
& & & & 0.12 & -- & 0.25 & 0.18 & 0.32 & $\pm$ 0.01 & $^{+0.02} _{-0.02}$ & 3882 & 0.0 \\
\hline
410 & -- & 530 & 450 & 0.006 & -- & 0.01 & 0.008 & 0.99 & $\pm$ 0.02 & $^{+0.02} _{-0.02}$ & 3072 & 48.9 \\
& & & & 0.01 & -- & 0.017 & 0.013 & 0.87 & $\pm$ 0.02 & $^{+0.01} _{-0.01}$ & 2771 & 9.3 \\
& & & & 0.017 & -- & 0.025 & 0.021 & 0.69 & $\pm$ 0.02 & $^{+0.01} _{-0.01}$ & 2145 & 0.8 \\
& & & & 0.025 & -- & 0.037 & 0.032 & 0.61 & $\pm$ 0.01 & $^{+0.01} _{-0.01}$ & 2497 & 0.0 \\
& & & & 0.037 & -- & 0.06 & 0.05 & 0.53 & $\pm$ 0.01 & $^{+0.01} _{-0.01}$ & 3100 & 0.0 \\
& & & & 0.06 & -- & 0.1 & 0.08 & 0.45 & $\pm$ 0.01 & $^{+0.01} _{-0.01}$ & 2562 & 0.0 \\
& & & & 0.1 & -- & 0.17 & 0.13 & 0.36 & $\pm$ 0.01 & $^{+0.02} _{-0.02}$ & 2182 & 0.0 \\
& & & & 0.17 & -- & 0.3 & 0.25 & 0.27 & $\pm$ 0.01 & $^{+0.02} _{-0.01}$ & 1684 & 0.0 \\

\hline
\end{tabular}
\end{center}
\end{scriptsize}
\caption[]
{The reduced cross section $\tilde{\sigma}$ for the reaction
$e^{-}p \rightarrow e^{-}X$ ($\mathcal{L} = 98.7 \pbi, P_{e} = -0.27$).
The bin range, bin centre ($Q^2_c$ and $x_c$)
and measured cross section corrected to the electroweak Born level are shown.
The first (second) error on the cross section
corresponds to the statistical (systematic) uncertainties.
The number of observed data events ($N_{\text{data}}$)
and simulated background events ($N^{\text{MC}}_{\text{bg}}$) are also shown.
This table has two continuations.}
\label{tab:ds2dxdq2Lh_1}
\end{table}

\begin{table} \begin{scriptsize} \begin{center} \begin{tabular}[t]{|rcr|r|lcl|l|rll|r|r|} \hline
\multicolumn{3}{|c|}{$Q^2$ range} & \multicolumn{1}{c|}{$Q^2_c$} & \multicolumn{3}{|c|}{$x$ range} & \multicolumn{1}{c|}{$x_c$} & \multicolumn{3}{c|}{$\tilde{\sigma}$} & \multicolumn{1}{c|}{$N_{\text{data}}$} & \multicolumn{1}{c|}{$N^{\text{MC}}_{\text{bg}}$} \\
\multicolumn{3}{|c|}{($\gev^{2}$)} & \multicolumn{1}{c|}{($\gev^{2}$)} & \multicolumn{3}{|c|}{} & \multicolumn{1}{c|}{} & \multicolumn{3}{c|}{} & & \\ \hline \hline
530 & -- & 710 & 650 & 0.01 & -- & 0.017 & 0.013 & 0.88 & $\pm$ 0.02 & $^{+0.01} _{-0.01}$ & 3116 & 34.4 \\
& & & & 0.017 & -- & 0.025 & 0.021 & 0.75 & $\pm$ 0.02 & $^{+0.01} _{-0.01}$ & 2057 & 3.3 \\
& & & & 0.025 & -- & 0.037 & 0.032 & 0.60 & $\pm$ 0.02 & $^{+0.01} _{-0.01}$ & 1589 & 0.8 \\
& & & & 0.037 & -- & 0.06 & 0.05 & 0.52 & $\pm$ 0.01 & $^{+0.01} _{-0.00}$ & 1722 & 0.0 \\
& & & & 0.06 & -- & 0.1 & 0.08 & 0.43 & $\pm$ 0.01 & $^{+0.01} _{-0.01}$ & 1430 & 0.0 \\
& & & & 0.1 & -- & 0.17 & 0.13 & 0.37 & $\pm$ 0.01 & $^{+0.01} _{-0.01}$ & 1310 & 0.0 \\
& & & & 0.17 & -- & 0.3 & 0.25 & 0.26 & $\pm$ 0.01 & $^{+0.02} _{-0.02}$ & 1107 & 0.0 \\
\hline
710 & -- & 900 & 800 & 0.009 & -- & 0.017 & 0.013 & 0.89 & $\pm$ 0.02 & $^{+0.02} _{-0.02}$ & 1992 & 28.8 \\
& & & & 0.017 & -- & 0.025 & 0.021 & 0.77 & $\pm$ 0.02 & $^{+0.01} _{-0.01}$ & 1436 & 1.8 \\
& & & & 0.025 & -- & 0.037 & 0.032 & 0.65 & $\pm$ 0.02 & $^{+0.01} _{-0.01}$ & 1453 & 4.3 \\
& & & & 0.037 & -- & 0.06 & 0.05 & 0.54 & $\pm$ 0.01 & $^{+0.01} _{-0.01}$ & 1637 & 0.0 \\
& & & & 0.06 & -- & 0.1 & 0.08 & 0.45 & $\pm$ 0.01 & $^{+0.01} _{-0.01}$ & 1358 & 0.0 \\
& & & & 0.1 & -- & 0.17 & 0.13 & 0.38 & $\pm$ 0.01 & $^{+0.00} _{-0.00}$ & 1106 & 0.0 \\
& & & & 0.17 & -- & 0.3 & 0.25 & 0.27 & $\pm$ 0.01 & $^{+0.01} _{-0.01}$ & 853 & 0.0 \\
\hline
900 & -- & 1300 & 1200 & 0.01 & -- & 0.017 & 0.014 & 0.91 & $\pm$ 0.03 & $^{+0.02} _{-0.03}$ & 1306 & 43.8 \\
& & & & 0.017 & -- & 0.025 & 0.021 & 0.83 & $\pm$ 0.02 & $^{+0.02} _{-0.01}$ & 1548 & 13.4 \\
& & & & 0.025 & -- & 0.037 & 0.032 & 0.62 & $\pm$ 0.02 & $^{+0.01} _{-0.01}$ & 1465 & 6.8 \\
& & & & 0.037 & -- & 0.06 & 0.05 & 0.55 & $\pm$ 0.01 & $^{+0.01} _{-0.01}$ & 1845 & 0.8 \\
& & & & 0.06 & -- & 0.1 & 0.08 & 0.47 & $\pm$ 0.01 & $^{+0.01} _{-0.01}$ & 1824 & 0.0 \\
& & & & 0.1 & -- & 0.17 & 0.13 & 0.38 & $\pm$ 0.01 & $^{+0.00} _{-0.00}$ & 1456 & 0.0 \\
& & & & 0.17 & -- & 0.3 & 0.25 & 0.26 & $\pm$ 0.01 & $^{+0.01} _{-0.01}$ & 1156 & 0.0 \\
& & & & 0.3 & -- & 0.53 & 0.4 & 0.13 & $\pm$ 0.01 & $^{+0.01} _{-0.01}$ & 384 & 0.0 \\
\hline
1300 & -- & 1800 & 1500 & 0.017 & -- & 0.025 & 0.021 & 0.77 & $\pm$ 0.03 & $^{+0.02} _{-0.02}$ & 743 & 33.8 \\
& & & & 0.025 & -- & 0.037 & 0.032 & 0.64 & $\pm$ 0.02 & $^{+0.00} _{-0.00}$ & 827 & 4.1 \\
& & & & 0.037 & -- & 0.06 & 0.05 & 0.51 & $\pm$ 0.02 & $^{+0.00} _{-0.00}$ & 1000 & 0.8 \\
& & & & 0.06 & -- & 0.1 & 0.08 & 0.50 & $\pm$ 0.01 & $^{+0.01} _{-0.01}$ & 1136 & 0.0 \\
& & & & 0.1 & -- & 0.15 & 0.13 & 0.39 & $\pm$ 0.01 & $^{+0.01} _{-0.01}$ & 714 & 0.0 \\
& & & & 0.15 & -- & 0.23 & 0.18 & 0.33 & $\pm$ 0.01 & $^{+0.01} _{-0.01}$ & 625 & 0.0 \\
& & & & 0.23 & -- & 0.35 & 0.25 & 0.27 & $\pm$ 0.01 & $^{+0.01} _{-0.01}$ & 425 & 0.0 \\
& & & & 0.35 & -- & 0.53 & 0.4 & 0.13 & $\pm$ 0.01 & $^{+0.01} _{-0.01}$ & 192 & 0.0 \\
\hline
1800 & -- & 2500 & 2000 & 0.023 & -- & 0.037 & 0.032 & 0.64 & $\pm$ 0.03 & $^{+0.02} _{-0.01}$ & 572 & 11.1 \\
& & & & 0.037 & -- & 0.06 & 0.05 & 0.61 & $\pm$ 0.02 & $^{+0.01} _{-0.00}$ & 753 & 0.0 \\
& & & & 0.06 & -- & 0.1 & 0.08 & 0.48 & $\pm$ 0.02 & $^{+0.01} _{-0.01}$ & 749 & 0.8 \\
& & & & 0.1 & -- & 0.15 & 0.13 & 0.34 & $\pm$ 0.02 & $^{+0.00} _{-0.00}$ & 471 & 0.0 \\
& & & & 0.15 & -- & 0.23 & 0.18 & 0.32 & $\pm$ 0.02 & $^{+0.00} _{-0.01}$ & 435 & 0.0 \\
& & & & 0.23 & -- & 0.35 & 0.25 & 0.26 & $\pm$ 0.01 & $^{+0.00} _{-0.00}$ & 305 & 0.0 \\
& & & & 0.35 & -- & 0.53 & 0.4 & 0.12 & $\pm$ 0.01 & $^{+0.00} _{-0.00}$ & 141 & 0.0 \\

\hline
\end{tabular}
\end{center}
\end{scriptsize}
\vspace{0.5cm} \hspace{5.4cm} {\bf Table 13:}~~~{\it Continuation 1.}
\label{tab:ds2dxdq2Lh_2}
\end{table}

\begin{table} \begin{scriptsize} \begin{center} \begin{tabular}[t]{|rcr|r|lcl|l|rll|r|r|} \hline
\multicolumn{3}{|c|}{$Q^2$ range} & \multicolumn{1}{c|}{$Q^2_c$} & \multicolumn{3}{|c|}{$x$ range} & \multicolumn{1}{c|}{$x_c$} & \multicolumn{3}{c|}{$\tilde{\sigma}$} & \multicolumn{1}{c|}{$N_{\text{data}}$} & \multicolumn{1}{c|}{$N^{\text{MC}}_{\text{bg}}$} \\
\multicolumn{3}{|c|}{($\gev^{2}$)} & \multicolumn{1}{c|}{($\gev^{2}$)} & \multicolumn{3}{|c|}{} & \multicolumn{1}{c|}{} & \multicolumn{3}{c|}{} & & \\ \hline \hline
2500 & -- & 3500 & 3000 & 0.037 & -- & 0.06 & 0.05 & 0.60 & $\pm$ 0.03 & $^{+0.01} _{-0.01}$ & 466 & 5.2 \\
& & & & 0.06 & -- & 0.1 & 0.08 & 0.51 & $\pm$ 0.02 & $^{+0.01} _{-0.01}$ & 525 & 0.9 \\
& & & & 0.1 & -- & 0.15 & 0.13 & 0.38 & $\pm$ 0.02 & $^{+0.00} _{-0.00}$ & 363 & 0.0 \\
& & & & 0.15 & -- & 0.23 & 0.18 & 0.34 & $\pm$ 0.02 & $^{+0.00} _{-0.00}$ & 327 & 0.0 \\
& & & & 0.23 & -- & 0.35 & 0.25 & 0.27 & $\pm$ 0.02 & $^{+0.02} _{-0.02}$ & 237 & 0.0 \\
& & & & 0.35 & -- & 0.53 & 0.4 & 0.16 & $\pm$ 0.01 & $^{+0.01} _{-0.01}$ & 124 & 0.0 \\
& & & & 0.53 & -- & 0.75 & 0.65 & 0.02 & $^{+0.00} _{-0.00}$ & $^{+0.00} _{-0.00}$ & 39 & 0.0 \\
\hline
3500 & -- & 5600 & 5000 & 0.04 & -- & 0.1 & 0.08 & 0.57 & $\pm$ 0.02 & $^{+0.01} _{-0.01}$ & 621 & 4.2 \\
& & & & 0.1 & -- & 0.15 & 0.13 & 0.49 & $\pm$ 0.03 & $^{+0.01} _{-0.00}$ & 374 & 0.8 \\
& & & & 0.15 & -- & 0.23 & 0.18 & 0.35 & $\pm$ 0.02 & $^{+0.00} _{-0.00}$ & 300 & 0.0 \\
& & & & 0.23 & -- & 0.35 & 0.25 & 0.23 & $\pm$ 0.02 & $^{+0.00} _{-0.00}$ & 176 & 0.0 \\
& & & & 0.35 & -- & 0.53 & 0.4 & 0.13 & $^{+0.01} _{-0.01}$ & $^{+0.01} _{-0.01}$ & 96 & 0.0 \\
\hline
5600 & -- & 9000 & 8000 & 0.07 & -- & 0.15 & 0.13 & 0.59 & $\pm$ 0.03 & $^{+0.01} _{-0.01}$ & 351 & 2.6 \\
& & & & 0.15 & -- & 0.23 & 0.18 & 0.44 & $\pm$ 0.03 & $^{+0.01} _{-0.01}$ & 203 & 0.0 \\
& & & & 0.23 & -- & 0.35 & 0.25 & 0.32 & $\pm$ 0.03 & $^{+0.01} _{-0.01}$ & 140 & 0.0 \\
& & & & 0.35 & -- & 0.53 & 0.4 & 0.12 & $^{+0.02} _{-0.02}$ & $^{+0.01} _{-0.01}$ & 51 & 0.0 \\
& & & & 0.53 & -- & 0.75 & 0.65 & 0.02 & $^{+0.01} _{-0.00}$ & $^{+0.00} _{-0.00}$ & 21 & 0.0 \\
\hline
9000 & -- & 15000 & 12000 & 0.09 & -- & 0.23 & 0.18 & 0.52 & $\pm$ 0.04 & $^{+0.01} _{-0.02}$ & 176 & 0.8 \\
& & & & 0.23 & -- & 0.35 & 0.25 & 0.35 & $^{+0.04} _{-0.04}$ & $^{+0.01} _{-0.01}$ & 86 & 0.0 \\
& & & & 0.35 & -- & 0.53 & 0.4 & 0.15 & $^{+0.03} _{-0.02}$ & $^{+0.01} _{-0.01}$ & 39 & 0.0 \\
\hline
15000 & -- & 25000 & 20000 & 0.15 & -- & 0.35 & 0.25 & 0.47 & $^{+0.06} _{-0.06}$ & $^{+0.03} _{-0.02}$ & 76 & 2.6 \\
& & & & 0.35 & -- & 0.75 & 0.4 & 0.19 & $^{+0.04} _{-0.04}$ & $^{+0.01} _{-0.01}$ & 29 & 0.0 \\
\hline
25000 & -- & 50000 & 30000 & 0.25 & -- & 0.75 & 0.4 & 0.28 & $^{+0.06} _{-0.05}$ & $^{+0.02} _{-0.02}$ & 28 & 0.0 \\

\hline
\end{tabular}
\end{center}
\end{scriptsize}
\vspace{0.5cm} \hspace{5.4cm} {\bf Table 13:}~~~{\it Continuation 2.}
\label{tab:ds2dxdq2Lh_3}
\end{table}

\begin{table} \begin{scriptsize} \begin{center} \begin{tabular}[t]{|r|l|c|c|c||c|c|c|c|c|c|c|c|} \hline
\multicolumn{1}{|c|}{$Q^2_c$} & \multicolumn{1}{c|}{$x_c$} & \multicolumn{1}{c|}{$\tilde{\sigma}$} & \multicolumn{1}{c|}{stat.} & \multicolumn{1}{c||}{sys.} & \multicolumn{1}{c|}{$\delta_{1}$} & \multicolumn{1}{c|}{$\delta_{2}$} & \multicolumn{1}{c|}{$\delta_{3}$} & \multicolumn{1}{c|}{$\delta_{4}$} & \multicolumn{1}{c|}{$\delta_{5}$} & \multicolumn{1}{c|}{$\delta_{6}$} & \multicolumn{1}{c|}{$\delta_{7}$} & \multicolumn{1}{c|}{$\delta_{8} - \delta_{13}$}\\
\multicolumn{1}{|c|}{($\gev^{2}$)} & & & \multicolumn{1}{c|}{(\%)} & \multicolumn{1}{c||}{(\%)} & \multicolumn{1}{c|}{(\%)} & \multicolumn{1}{c|}{(\%)} & \multicolumn{1}{c|}{(\%)} & \multicolumn{1}{c|}{(\%)} & \multicolumn{1}{c|}{(\%)} & \multicolumn{1}{c|}{(\%)} & \multicolumn{1}{c|}{(\%)} & \multicolumn{1}{c|}{(\%)} \\ \hline \hline
200 & 0.005 & 1.09 & $\pm$ 1.2 & $^{+1.3} _{-1.0}$ & $^{+0.7} _{-0.0}$ & $^{-0.4} _{+0.6}$ & $^{+0.4} _{-0.4}$ & $^{-0.4} _{+0.4}$ & $^{+0.0} _{+0.0}$ & $^{+0.1} _{-0.1}$ & $^{+0.3} _{-0.3}$ & $^{+0.7} _{-0.7}$ \\
& 0.008 & 0.92 & $\pm$ 1.1 & $^{+1.9} _{-1.9}$ & $^{+0.3} _{-0.0}$ & $^{-0.4} _{+0.5}$ & $^{+0.2} _{-0.2}$ & $^{-1.6} _{+1.6}$ & $^{+0.0} _{+0.0}$ & $^{+0.1} _{+0.4}$ & $^{+0.1} _{-0.1}$ & $^{+0.8} _{-1.0}$ \\
& 0.013 & 0.78 & $\pm$ 1.0 & $^{+1.7} _{-1.6}$ & $^{+0.2} _{+0.0}$ & $^{-0.5} _{+0.5}$ & $^{-0.2} _{+0.2}$ & $^{-1.3} _{+1.3}$ & $^{+0.0} _{+0.0}$ & $^{+0.0} _{+0.4}$ & $^{+0.0} _{-0.0}$ & $^{+0.8} _{-0.9}$ \\
& 0.021 & 0.66 & $\pm$ 1.2 & $^{+2.1} _{-2.1}$ & $^{+0.4} _{+0.0}$ & $^{-0.3} _{+0.5}$ & $^{-1.6} _{+1.6}$ & $^{+0.9} _{-0.9}$ & $^{+0.0} _{+0.0}$ & $^{-0.1} _{+0.1}$ & $^{+0.0} _{+0.0}$ & $^{+0.8} _{-0.9}$ \\
& 0.032 & 0.56 & $\pm$ 1.3 & $^{+4.0} _{-4.0}$ & $^{+0.1} _{-0.0}$ & $^{-0.3} _{+0.4}$ & $^{-1.5} _{+1.5}$ & $^{+3.6} _{-3.6}$ & $^{+0.0} _{+0.0}$ & $^{+0.2} _{+0.4}$ & $^{+0.0} _{+0.0}$ & $^{+0.8} _{-0.8}$ \\
& 0.05 & 0.50 & $\pm$ 1.3 & $^{+2.5} _{-2.4}$ & $^{+0.0} _{-0.3}$ & $^{-0.2} _{+0.3}$ & $^{-0.5} _{+0.5}$ & $^{+2.2} _{-2.2}$ & $^{+0.0} _{+0.0}$ & $^{+0.0} _{+0.3}$ & $^{+0.0} _{+0.0}$ & $^{+1.0} _{-0.9}$ \\
& 0.08 & 0.45 & $\pm$ 1.1 & $^{+5.5} _{-5.5}$ & $^{+0.0} _{-0.1}$ & $^{-0.2} _{+0.2}$ & $^{+5.0} _{-5.0}$ & $^{+2.0} _{-2.0}$ & $^{-0.0} _{+0.0}$ & $^{-0.0} _{+0.1}$ & $^{+0.0} _{+0.0}$ & $^{+0.8} _{-1.0}$ \\
& 0.18 & 0.34 & $\pm$ 1.6 & $^{+5.4} _{-5.4}$ & $^{+0.1} _{+0.0}$ & $^{-0.1} _{+0.2}$ & $^{+5.3} _{-5.3}$ & $^{+0.5} _{-0.5}$ & $^{+0.7} _{+0.2}$ & $^{-0.1} _{+0.1}$ & $^{+0.0} _{+0.0}$ & $^{+0.7} _{-0.9}$ \\
\hline
250 & 0.008 & 0.95 & $\pm$ 1.3 & $^{+1.5} _{-1.3}$ & $^{+0.6} _{-0.0}$ & $^{-0.3} _{+0.4}$ & $^{+0.5} _{-0.5}$ & $^{-0.9} _{+0.9}$ & $^{+0.0} _{+0.0}$ & $^{-0.2} _{+0.3}$ & $^{+0.1} _{-0.1}$ & $^{+0.7} _{-0.7}$ \\
& 0.013 & 0.80 & $\pm$ 1.2 & $^{+1.7} _{-1.5}$ & $^{+0.7} _{+0.0}$ & $^{-0.5} _{+0.6}$ & $^{+0.0} _{-0.0}$ & $^{-1.3} _{+1.3}$ & $^{+0.0} _{+0.0}$ & $^{-0.1} _{+0.2}$ & $^{+0.0} _{-0.0}$ & $^{+0.5} _{-0.7}$ \\
& 0.021 & 0.66 & $\pm$ 1.4 & $^{+1.6} _{-1.4}$ & $^{+0.7} _{-0.0}$ & $^{-0.3} _{+0.5}$ & $^{-0.1} _{+0.1}$ & $^{+1.1} _{-1.1}$ & $^{+0.0} _{+0.0}$ & $^{-0.2} _{+0.2}$ & $^{+0.0} _{-0.0}$ & $^{+0.8} _{-0.7}$ \\
& 0.032 & 0.57 & $\pm$ 1.5 & $^{+3.3} _{-3.3}$ & $^{+0.6} _{-0.0}$ & $^{-0.3} _{+0.4}$ & $^{-0.8} _{+0.8}$ & $^{+3.0} _{-3.0}$ & $^{+0.0} _{+0.0}$ & $^{+0.0} _{+0.1}$ & $^{+0.0} _{+0.0}$ & $^{+0.7} _{-0.9}$ \\
& 0.05 & 0.50 & $\pm$ 1.5 & $^{+3.4} _{-3.3}$ & $^{+0.4} _{+0.0}$ & $^{-0.1} _{+0.3}$ & $^{-2.4} _{+2.4}$ & $^{+2.2} _{-2.2}$ & $^{+0.0} _{+0.0}$ & $^{-0.0} _{-0.1}$ & $^{+0.0} _{+0.0}$ & $^{+0.7} _{-0.6}$ \\
& 0.08 & 0.43 & $\pm$ 1.4 & $^{+5.2} _{-5.1}$ & $^{+0.7} _{-0.0}$ & $^{-0.1} _{+0.3}$ & $^{+4.6} _{-4.6}$ & $^{+2.2} _{-2.2}$ & $^{+0.0} _{+0.0}$ & $^{-0.0} _{+0.1}$ & $^{+0.0} _{+0.0}$ & $^{+0.7} _{-0.7}$ \\
& 0.18 & 0.34 & $\pm$ 1.6 & $^{+4.8} _{-4.7}$ & $^{+0.2} _{+0.0}$ & $^{+0.0} _{+0.2}$ & $^{+4.1} _{-4.1}$ & $^{+2.0} _{-2.0}$ & $^{+0.3} _{+1.0}$ & $^{+0.0} _{+0.2}$ & $^{+0.0} _{+0.0}$ & $^{+0.8} _{-0.8}$ \\
\hline
350 & 0.008 & 0.97 & $\pm$ 1.7 & $^{+2.1} _{-1.5}$ & $^{+1.6} _{-0.1}$ & $^{-0.5} _{+0.6}$ & $^{+0.9} _{-0.9}$ & $^{-0.4} _{+0.4}$ & $^{+0.0} _{+0.0}$ & $^{-0.6} _{+0.4}$ & $^{+0.4} _{-0.4}$ & $^{+0.6} _{-0.8}$ \\
& 0.013 & 0.82 & $\pm$ 1.4 & $^{+1.9} _{-1.2}$ & $^{+1.4} _{-0.0}$ & $^{-0.2} _{+0.5}$ & $^{-0.2} _{+0.2}$ & $^{-1.0} _{+1.0}$ & $^{+0.0} _{+0.0}$ & $^{-0.2} _{+0.5}$ & $^{+0.0} _{-0.0}$ & $^{+0.7} _{-0.6}$ \\
& 0.021 & 0.69 & $\pm$ 1.6 & $^{+0.9} _{-0.6}$ & $^{+0.4} _{-0.1}$ & $^{-0.3} _{+0.4}$ & $^{+0.3} _{-0.3}$ & $^{+0.0} _{-0.0}$ & $^{+0.0} _{+0.0}$ & $^{-0.2} _{-0.0}$ & $^{+0.0} _{+0.0}$ & $^{+0.6} _{-0.5}$ \\
& 0.032 & 0.61 & $\pm$ 1.6 & $^{+1.9} _{-1.8}$ & $^{+0.8} _{-0.0}$ & $^{-0.2} _{+0.4}$ & $^{-1.6} _{+1.6}$ & $^{+0.2} _{-0.2}$ & $^{+0.0} _{+0.0}$ & $^{-0.3} _{+0.1}$ & $^{+0.0} _{+0.0}$ & $^{+0.6} _{-0.7}$ \\
& 0.05 & 0.51 & $\pm$ 1.6 & $^{+2.7} _{-2.6}$ & $^{+0.6} _{-0.0}$ & $^{-0.1} _{+0.2}$ & $^{-1.9} _{+1.9}$ & $^{+1.6} _{-1.6}$ & $^{+0.0} _{+0.0}$ & $^{-0.1} _{+0.2}$ & $^{+0.0} _{+0.0}$ & $^{+0.5} _{-0.7}$ \\
& 0.08 & 0.44 & $\pm$ 1.5 & $^{+2.0} _{-2.0}$ & $^{+0.3} _{+0.0}$ & $^{-0.1} _{+0.2}$ & $^{+1.1} _{-1.1}$ & $^{+1.6} _{-1.6}$ & $^{+0.0} _{+0.0}$ & $^{+0.0} _{+0.1}$ & $^{+0.0} _{+0.0}$ & $^{+0.7} _{-0.7}$ \\
& 0.18 & 0.32 & $\pm$ 1.7 & $^{+6.4} _{-6.3}$ & $^{+0.4} _{+0.0}$ & $^{-0.0} _{+0.2}$ & $^{+5.7} _{-5.7}$ & $^{+2.6} _{-2.6}$ & $^{-0.1} _{+1.2}$ & $^{+0.1} _{+0.2}$ & $^{+0.0} _{+0.0}$ & $^{+0.7} _{-0.6}$ \\
\hline
450 & 0.008 & 0.99 & $\pm$ 1.9 & $^{+2.4} _{-2.1}$ & $^{+0.7} _{-0.3}$ & $^{-1.4} _{+1.6}$ & $^{+1.3} _{-1.3}$ & $^{-0.5} _{+0.5}$ & $^{+0.0} _{+0.0}$ & $^{-0.2} _{+0.3}$ & $^{+0.6} _{-0.6}$ & $^{+0.5} _{-0.4}$ \\
& 0.013 & 0.87 & $\pm$ 2.0 & $^{+1.1} _{-1.1}$ & $^{+0.1} _{-0.1}$ & $^{-0.5} _{+0.6}$ & $^{-0.4} _{+0.4}$ & $^{+0.2} _{-0.2}$ & $^{+0.0} _{+0.0}$ & $^{-0.6} _{+0.6}$ & $^{+0.1} _{-0.1}$ & $^{+0.5} _{-0.6}$ \\
& 0.021 & 0.69 & $\pm$ 2.2 & $^{+1.8} _{-1.7}$ & $^{+0.2} _{-0.1}$ & $^{-0.4} _{+0.5}$ & $^{-0.1} _{+0.1}$ & $^{+1.3} _{-1.3}$ & $^{+0.0} _{+0.0}$ & $^{-0.7} _{+0.9}$ & $^{+0.0} _{-0.0}$ & $^{+0.8} _{-0.8}$ \\
& 0.032 & 0.61 & $\pm$ 2.0 & $^{+1.9} _{-1.7}$ & $^{+0.8} _{-0.1}$ & $^{-0.2} _{+0.4}$ & $^{+0.0} _{-0.0}$ & $^{+1.4} _{-1.4}$ & $^{+0.0} _{+0.0}$ & $^{-0.5} _{+0.4}$ & $^{+0.0} _{+0.0}$ & $^{+1.0} _{-0.9}$ \\
& 0.05 & 0.53 & $\pm$ 1.8 & $^{+2.5} _{-2.4}$ & $^{+0.7} _{-0.0}$ & $^{+0.0} _{+0.2}$ & $^{-1.6} _{+1.6}$ & $^{+1.6} _{-1.6}$ & $^{+0.0} _{+0.0}$ & $^{-0.4} _{+0.6}$ & $^{+0.0} _{+0.0}$ & $^{+0.7} _{-0.7}$ \\
& 0.08 & 0.45 & $\pm$ 2.0 & $^{+2.1} _{-1.9}$ & $^{+0.3} _{+0.0}$ & $^{+0.1} _{+0.2}$ & $^{-0.7} _{+0.7}$ & $^{+1.7} _{-1.7}$ & $^{+0.0} _{+0.0}$ & $^{+0.1} _{+0.6}$ & $^{+0.0} _{+0.0}$ & $^{+0.7} _{-0.5}$ \\
& 0.13 & 0.36 & $\pm$ 2.2 & $^{+4.9} _{-4.9}$ & $^{+0.6} _{+0.0}$ & $^{+0.1} _{+0.1}$ & $^{+4.4} _{-4.4}$ & $^{+1.9} _{-1.9}$ & $^{+0.0} _{+0.0}$ & $^{-0.4} _{+0.4}$ & $^{+0.0} _{+0.0}$ & $^{+0.5} _{-0.6}$ \\
& 0.25 & 0.27 & $\pm$ 2.5 & $^{+5.6} _{-5.5}$ & $^{+1.1} _{+0.0}$ & $^{+0.2} _{+0.1}$ & $^{+5.2} _{-5.2}$ & $^{+1.1} _{-1.1}$ & $^{-0.9} _{+0.8}$ & $^{-0.1} _{+0.6}$ & $^{+0.0} _{+0.0}$ & $^{+0.8} _{-0.8}$ \\

\hline
\end{tabular}
\end{center}
\end{scriptsize}
\caption[]
{Systematic uncertainties with bin-to-bin correlations
for the reduced cross section $\tilde{\sigma}$
for the reaction $e^{-}p \rightarrow e^{-}X$ ($\mathcal{L} = 98.7 \pbi, P_{e} = -0.27$).
The left five columns of the table contain
the bin centres, $Q^{2}_{c}$ and $x_{c}$, the measured cross section,
the statistical uncertainty and the total systematic uncertainty.
The right eight columns of the table list
the bin-to-bin correlated systematic uncertainties
for $\delta_{1} - \delta_{7}$,
and the systematic uncertainties
summed in quadrature for $\delta_{8} - \delta_{13}$,
as defined in the section \ref{sec-sys}.
The upper and lower correlated uncertainties correspond to
a positive or negative variation of a cut value for example.
However, if this is not possible for a particular systematic,
the uncertainty is symmetrised.
This table has two continuations.}
\label{tab:ds2dxdq2LhSys_1}
\end{table}

\begin{table} \begin{scriptsize} \begin{center} \begin{tabular}[t]{|r|l|c|c|c||c|c|c|c|c|c|c|c|} \hline
\multicolumn{1}{|c|}{$Q^2_c$} & \multicolumn{1}{c|}{$x_c$} & \multicolumn{1}{c|}{$\tilde{\sigma}$} & \multicolumn{1}{c|}{stat.} & \multicolumn{1}{c||}{sys.} & \multicolumn{1}{c|}{$\delta_{1}$} & \multicolumn{1}{c|}{$\delta_{2}$} & \multicolumn{1}{c|}{$\delta_{3}$} & \multicolumn{1}{c|}{$\delta_{4}$} & \multicolumn{1}{c|}{$\delta_{5}$} & \multicolumn{1}{c|}{$\delta_{6}$} & \multicolumn{1}{c|}{$\delta_{7}$} & \multicolumn{1}{c|}{$\delta_{8} - \delta_{13}$}\\
\multicolumn{1}{|c|}{($\gev^{2}$)} & & & \multicolumn{1}{c|}{(\%)} & \multicolumn{1}{c||}{(\%)} & \multicolumn{1}{c|}{(\%)} & \multicolumn{1}{c|}{(\%)} & \multicolumn{1}{c|}{(\%)} & \multicolumn{1}{c|}{(\%)} & \multicolumn{1}{c|}{(\%)} & \multicolumn{1}{c|}{(\%)} & \multicolumn{1}{c|}{(\%)} & \multicolumn{1}{c|}{(\%)} \\ \hline \hline
650 & 0.013 & 0.88 & $\pm$ 1.8 & $^{+1.5} _{-0.8}$ & $^{+1.2} _{-0.2}$ & $^{-0.5} _{+0.6}$ & $^{-0.0} _{+0.0}$ & $^{-0.1} _{+0.1}$ & $^{+0.0} _{+0.0}$ & $^{-0.2} _{+0.3}$ & $^{+0.5} _{-0.5}$ & $^{+0.4} _{-0.3}$ \\
& 0.021 & 0.75 & $\pm$ 2.3 & $^{+1.1} _{-1.1}$ & $^{+0.0} _{-0.1}$ & $^{-0.4} _{+0.5}$ & $^{-0.3} _{+0.3}$ & $^{-0.9} _{+0.9}$ & $^{+0.0} _{+0.0}$ & $^{-0.2} _{+0.0}$ & $^{+0.1} _{-0.1}$ & $^{+0.2} _{-0.2}$ \\
& 0.032 & 0.60 & $\pm$ 2.6 & $^{+1.2} _{-1.0}$ & $^{+0.5} _{-0.2}$ & $^{-0.4} _{+0.5}$ & $^{-0.5} _{+0.5}$ & $^{+0.7} _{-0.7}$ & $^{+0.0} _{+0.0}$ & $^{+0.5} _{+0.5}$ & $^{+0.0} _{-0.0}$ & $^{+0.2} _{-0.2}$ \\
& 0.05 & 0.52 & $\pm$ 2.5 & $^{+1.0} _{-0.9}$ & $^{+0.1} _{-0.0}$ & $^{-0.3} _{+0.4}$ & $^{-0.4} _{+0.4}$ & $^{+0.7} _{-0.7}$ & $^{+0.0} _{+0.0}$ & $^{+0.1} _{+0.0}$ & $^{+0.0} _{+0.0}$ & $^{+0.5} _{-0.4}$ \\
& 0.08 & 0.43 & $\pm$ 2.7 & $^{+2.4} _{-2.3}$ & $^{+0.7} _{-0.0}$ & $^{-0.1} _{+0.3}$ & $^{-1.8} _{+1.8}$ & $^{+1.0} _{-1.0}$ & $^{+0.0} _{+0.0}$ & $^{+0.1} _{-0.0}$ & $^{+0.0} _{+0.0}$ & $^{+0.9} _{-1.1}$ \\
& 0.13 & 0.37 & $\pm$ 2.8 & $^{+2.7} _{-2.4}$ & $^{+0.9} _{-0.0}$ & $^{-0.1} _{+0.2}$ & $^{+2.1} _{-2.1}$ & $^{-0.8} _{+0.8}$ & $^{+0.0} _{+0.0}$ & $^{+0.1} _{+0.4}$ & $^{+0.0} _{+0.0}$ & $^{+1.0} _{-0.9}$ \\
& 0.25 & 0.26 & $\pm$ 3.1 & $^{+6.6} _{-6.5}$ & $^{+1.0} _{-0.0}$ & $^{-0.1} _{+0.3}$ & $^{+6.3} _{-6.3}$ & $^{+1.0} _{-1.0}$ & $^{-0.8} _{+0.2}$ & $^{-0.1} _{+0.5}$ & $^{+0.0} _{+0.0}$ & $^{+1.0} _{-1.1}$ \\
\hline
800 & 0.013 & 0.89 & $\pm$ 2.3 & $^{+1.7} _{-1.9}$ & $^{+0.4} _{-1.0}$ & $^{-0.8} _{+1.0}$ & $^{+0.4} _{-0.4}$ & $^{-1.0} _{+1.0}$ & $^{+0.0} _{+0.0}$ & $^{-0.1} _{+0.0}$ & $^{+0.6} _{-0.6}$ & $^{+0.3} _{-0.4}$ \\
& 0.021 & 0.77 & $\pm$ 2.7 & $^{+1.1} _{-0.9}$ & $^{+0.5} _{-0.1}$ & $^{-0.4} _{+0.5}$ & $^{+0.6} _{-0.6}$ & $^{-0.2} _{+0.2}$ & $^{+0.0} _{+0.0}$ & $^{+0.0} _{+0.5}$ & $^{+0.1} _{-0.1}$ & $^{+0.4} _{-0.4}$ \\
& 0.032 & 0.65 & $\pm$ 2.7 & $^{+1.4} _{-1.4}$ & $^{+0.0} _{-0.4}$ & $^{-0.4} _{+0.5}$ & $^{-0.8} _{+0.8}$ & $^{-0.9} _{+0.9}$ & $^{+0.0} _{+0.0}$ & $^{+0.3} _{+0.0}$ & $^{+0.1} _{-0.1}$ & $^{+0.2} _{-0.4}$ \\
& 0.05 & 0.54 & $\pm$ 2.5 & $^{+2.2} _{-2.2}$ & $^{+0.3} _{+0.0}$ & $^{-0.3} _{+0.4}$ & $^{+0.7} _{-0.7}$ & $^{+2.0} _{-2.0}$ & $^{+0.0} _{+0.0}$ & $^{-0.4} _{+0.1}$ & $^{+0.0} _{+0.0}$ & $^{+0.3} _{-0.4}$ \\
& 0.08 & 0.45 & $\pm$ 2.8 & $^{+2.6} _{-2.6}$ & $^{+0.5} _{-0.0}$ & $^{-0.2} _{+0.3}$ & $^{-2.5} _{+2.5}$ & $^{+0.1} _{-0.1}$ & $^{+0.0} _{+0.0}$ & $^{+0.5} _{-0.7}$ & $^{+0.0} _{+0.0}$ & $^{+0.3} _{-0.4}$ \\
& 0.13 & 0.38 & $\pm$ 3.1 & $^{+1.1} _{-1.0}$ & $^{+0.5} _{-0.0}$ & $^{-0.2} _{+0.3}$ & $^{+0.7} _{-0.7}$ & $^{+0.6} _{-0.6}$ & $^{+0.0} _{+0.0}$ & $^{+0.1} _{-0.4}$ & $^{+0.0} _{+0.0}$ & $^{+0.2} _{-0.2}$ \\
& 0.25 & 0.27 & $\pm$ 3.5 & $^{+3.7} _{-3.4}$ & $^{+0.3} _{+0.0}$ & $^{-0.2} _{+0.2}$ & $^{+3.3} _{-3.3}$ & $^{+0.8} _{-0.8}$ & $^{+1.2} _{-0.1}$ & $^{+0.7} _{-0.5}$ & $^{+0.0} _{+0.0}$ & $^{+0.2} _{-0.3}$ \\
\hline
1200 & 0.014 & 0.91 & $\pm$ 2.9 & $^{+2.4} _{-3.2}$ & $^{+0.6} _{-2.2}$ & $^{-0.8} _{+1.0}$ & $^{-0.7} _{+0.7}$ & $^{-1.4} _{+1.4}$ & $^{+0.0} _{+0.0}$ & $^{-0.3} _{-0.8}$ & $^{+1.4} _{-1.4}$ & $^{+0.5} _{-0.2}$ \\
& 0.021 & 0.83 & $\pm$ 2.6 & $^{+1.9} _{-1.6}$ & $^{+0.9} _{-0.1}$ & $^{-0.5} _{+0.5}$ & $^{+1.5} _{-1.5}$ & $^{+0.0} _{-0.0}$ & $^{+0.0} _{+0.0}$ & $^{+0.1} _{+0.4}$ & $^{+0.4} _{-0.4}$ & $^{+0.3} _{-0.4}$ \\
& 0.032 & 0.62 & $\pm$ 2.7 & $^{+1.2} _{-1.1}$ & $^{+0.0} _{-0.1}$ & $^{-0.4} _{+0.5}$ & $^{-0.5} _{+0.5}$ & $^{-0.7} _{+0.7}$ & $^{+0.0} _{+0.0}$ & $^{+0.1} _{+0.1}$ & $^{+0.2} _{-0.2}$ & $^{+0.6} _{-0.6}$ \\
& 0.05 & 0.55 & $\pm$ 2.4 & $^{+1.3} _{-1.4}$ & $^{+0.1} _{-0.0}$ & $^{-0.3} _{+0.3}$ & $^{-0.1} _{+0.1}$ & $^{+1.2} _{-1.2}$ & $^{+0.0} _{+0.0}$ & $^{-0.3} _{-0.4}$ & $^{+0.0} _{-0.0}$ & $^{+0.4} _{-0.4}$ \\
& 0.08 & 0.47 & $\pm$ 2.4 & $^{+1.4} _{-1.3}$ & $^{+0.3} _{-0.0}$ & $^{-0.2} _{+0.4}$ & $^{-0.3} _{+0.3}$ & $^{+1.2} _{-1.2}$ & $^{+0.0} _{+0.0}$ & $^{+0.3} _{-0.3}$ & $^{+0.0} _{+0.0}$ & $^{+0.4} _{-0.3}$ \\
& 0.13 & 0.38 & $\pm$ 2.7 & $^{+1.1} _{-1.2}$ & $^{+0.1} _{-0.0}$ & $^{-0.1} _{+0.2}$ & $^{-0.9} _{+0.9}$ & $^{+0.4} _{-0.4}$ & $^{+0.0} _{+0.0}$ & $^{+0.1} _{-0.5}$ & $^{+0.0} _{+0.0}$ & $^{+0.3} _{-0.4}$ \\
& 0.25 & 0.26 & $\pm$ 3.0 & $^{+2.4} _{-2.3}$ & $^{+0.3} _{+0.0}$ & $^{-0.2} _{+0.3}$ & $^{+2.0} _{-2.0}$ & $^{+1.0} _{-1.0}$ & $^{+0.0} _{+0.0}$ & $^{-0.3} _{-0.1}$ & $^{+0.0} _{+0.0}$ & $^{+0.5} _{-0.4}$ \\
& 0.4 & 0.13 & $\pm$ 5.2 & $^{+8.1} _{-5.7}$ & $^{+0.0} _{-0.1}$ & $^{-0.4} _{+0.4}$ & $^{+5.6} _{-5.6}$ & $^{+0.6} _{-0.6}$ & $^{+5.8} _{+3.4}$ & $^{+0.5} _{-1.0}$ & $^{+0.0} _{+0.0}$ & $^{+0.5} _{-0.5}$ \\
\hline
1500 & 0.021 & 0.77 & $\pm$ 3.8 & $^{+2.6} _{-2.3}$ & $^{+0.9} _{-0.2}$ & $^{-0.6} _{+0.4}$ & $^{-0.4} _{+0.4}$ & $^{-0.1} _{+0.1}$ & $^{+0.0} _{+0.0}$ & $^{-1.0} _{+0.1}$ & $^{+2.3} _{-1.9}$ & $^{+0.4} _{-0.4}$ \\
& 0.032 & 0.64 & $\pm$ 3.5 & $^{+0.7} _{-0.6}$ & $^{+0.3} _{-0.1}$ & $^{-0.3} _{+0.3}$ & $^{-0.1} _{+0.1}$ & $^{-0.3} _{+0.3}$ & $^{+0.0} _{+0.0}$ & $^{-0.1} _{-0.1}$ & $^{+0.2} _{-0.2}$ & $^{+0.4} _{-0.3}$ \\
& 0.05 & 0.51 & $\pm$ 3.2 & $^{+0.9} _{-0.8}$ & $^{+0.2} _{-0.0}$ & $^{-0.3} _{+0.4}$ & $^{+0.2} _{-0.2}$ & $^{+0.4} _{-0.4}$ & $^{+0.0} _{+0.0}$ & $^{-0.4} _{+0.5}$ & $^{+0.0} _{-0.0}$ & $^{+0.3} _{-0.4}$ \\
& 0.08 & 0.50 & $\pm$ 3.0 & $^{+1.0} _{-1.1}$ & $^{+0.2} _{-0.0}$ & $^{-0.3} _{+0.4}$ & $^{-0.9} _{+0.9}$ & $^{-0.2} _{+0.2}$ & $^{+0.0} _{+0.0}$ & $^{-0.4} _{+0.1}$ & $^{+0.0} _{+0.0}$ & $^{+0.3} _{-0.5}$ \\
& 0.13 & 0.39 & $\pm$ 3.8 & $^{+2.1} _{-1.8}$ & $^{+0.8} _{-0.0}$ & $^{-0.1} _{+0.3}$ & $^{-1.3} _{+1.3}$ & $^{-1.1} _{+1.1}$ & $^{+0.0} _{+0.0}$ & $^{+0.8} _{-0.2}$ & $^{+0.0} _{+0.0}$ & $^{+0.3} _{-0.4}$ \\
& 0.18 & 0.33 & $\pm$ 4.0 & $^{+1.8} _{-1.7}$ & $^{+0.6} _{-0.1}$ & $^{-0.3} _{+0.2}$ & $^{+0.0} _{-0.0}$ & $^{-1.6} _{+1.6}$ & $^{+0.0} _{+0.0}$ & $^{-0.3} _{-0.2}$ & $^{+0.0} _{+0.0}$ & $^{+0.4} _{-0.4}$ \\
& 0.25 & 0.27 & $\pm$ 4.9 & $^{+1.9} _{-1.9}$ & $^{+0.0} _{-0.3}$ & $^{-0.3} _{+0.3}$ & $^{+1.7} _{-1.7}$ & $^{+0.6} _{-0.6}$ & $^{+0.1} _{+0.0}$ & $^{-0.1} _{-0.3}$ & $^{+0.0} _{+0.0}$ & $^{+0.3} _{-0.3}$ \\
& 0.4 & 0.13 & $\pm$ 7.3 & $^{+8.8} _{-4.3}$ & $^{+0.6} _{+0.0}$ & $^{-0.5} _{+0.3}$ & $^{+2.0} _{-2.0}$ & $^{-0.9} _{+0.9}$ & $^{+8.4} _{-3.6}$ & $^{+0.3} _{-0.4}$ & $^{+0.0} _{+0.0}$ & $^{+0.6} _{-0.5}$ \\
\hline
2000 & 0.032 & 0.64 & $\pm$ 4.3 & $^{+2.5} _{-1.4}$ & $^{+1.8} _{+0.0}$ & $^{-0.2} _{+0.2}$ & $^{+0.3} _{-0.3}$ & $^{+1.0} _{-1.0}$ & $^{+0.0} _{+0.0}$ & $^{+0.4} _{+0.7}$ & $^{+1.1} _{-0.8}$ & $^{+0.3} _{-0.4}$ \\
& 0.05 & 0.61 & $\pm$ 3.7 & $^{+1.0} _{-0.8}$ & $^{+0.5} _{-0.0}$ & $^{-0.3} _{+0.4}$ & $^{-0.3} _{+0.3}$ & $^{-0.6} _{+0.6}$ & $^{+0.0} _{+0.0}$ & $^{+0.1} _{+0.1}$ & $^{+0.0} _{+0.0}$ & $^{+0.3} _{-0.3}$ \\
& 0.08 & 0.48 & $\pm$ 3.7 & $^{+2.0} _{-2.0}$ & $^{+0.1} _{-0.1}$ & $^{-0.3} _{+0.4}$ & $^{-0.1} _{+0.1}$ & $^{+1.9} _{-1.9}$ & $^{+0.0} _{+0.0}$ & $^{+0.2} _{+0.0}$ & $^{+0.0} _{-0.0}$ & $^{+0.3} _{-0.3}$ \\
& 0.13 & 0.34 & $\pm$ 4.6 & $^{+0.8} _{-1.2}$ & $^{+0.1} _{-0.6}$ & $^{-0.2} _{+0.3}$ & $^{-0.5} _{+0.5}$ & $^{+0.5} _{-0.5}$ & $^{+0.0} _{+0.0}$ & $^{+0.1} _{-0.6}$ & $^{+0.0} _{+0.0}$ & $^{+0.3} _{-0.4}$ \\
& 0.18 & 0.32 & $\pm$ 4.8 & $^{+0.6} _{-1.9}$ & $^{+0.0} _{-0.7}$ & $^{-0.2} _{+0.2}$ & $^{-0.3} _{+0.3}$ & $^{+0.1} _{-0.1}$ & $^{+0.0} _{+0.0}$ & $^{-0.4} _{-1.6}$ & $^{+0.0} _{+0.0}$ & $^{+0.4} _{-0.6}$ \\
& 0.25 & 0.26 & $\pm$ 5.8 & $^{+1.7} _{-1.6}$ & $^{+0.3} _{-0.0}$ & $^{-0.2} _{+0.3}$ & $^{+1.5} _{-1.5}$ & $^{+0.5} _{-0.5}$ & $^{+0.0} _{+0.0}$ & $^{+0.4} _{+0.5}$ & $^{+0.0} _{+0.0}$ & $^{+0.5} _{-0.5}$ \\
& 0.4 & 0.12 & $\pm$ 8.5 & $^{+3.7} _{-3.9}$ & $^{+0.0} _{-0.2}$ & $^{-0.3} _{+0.4}$ & $^{+3.1} _{-3.1}$ & $^{+1.1} _{-1.1}$ & $^{-0.5} _{+1.0}$ & $^{+0.1} _{-1.3}$ & $^{+0.0} _{+0.0}$ & $^{+1.2} _{-1.4}$ \\

\hline
\end{tabular}
\end{center}
\end{scriptsize}
\vspace{0.5cm} \hspace{5.4cm} {\bf Table 14:}~~~{\it Continuation 1.}
\label{tab:ds2dxdq2LhSys_2}
\end{table}

\begin{table} \begin{scriptsize} \begin{center} \begin{tabular}[t]{|r|l|c|c|c||c|c|c|c|c|c|c|c|} \hline
\multicolumn{1}{|c|}{$Q^2_c$} & \multicolumn{1}{c|}{$x_c$} & \multicolumn{1}{c|}{$\tilde{\sigma}$} & \multicolumn{1}{c|}{stat.} & \multicolumn{1}{c||}{sys.} & \multicolumn{1}{c|}{$\delta_{1}$} & \multicolumn{1}{c|}{$\delta_{2}$} & \multicolumn{1}{c|}{$\delta_{3}$} & \multicolumn{1}{c|}{$\delta_{4}$} & \multicolumn{1}{c|}{$\delta_{5}$} & \multicolumn{1}{c|}{$\delta_{6}$} & \multicolumn{1}{c|}{$\delta_{7}$} & \multicolumn{1}{c|}{$\delta_{8} - \delta_{13}$}\\
\multicolumn{1}{|c|}{($\gev^{2}$)} & & & \multicolumn{1}{c|}{(\%)} & \multicolumn{1}{c||}{(\%)} & \multicolumn{1}{c|}{(\%)} & \multicolumn{1}{c|}{(\%)} & \multicolumn{1}{c|}{(\%)} & \multicolumn{1}{c|}{(\%)} & \multicolumn{1}{c|}{(\%)} & \multicolumn{1}{c|}{(\%)} & \multicolumn{1}{c|}{(\%)} & \multicolumn{1}{c|}{(\%)} \\ \hline \hline
3000 & 0.05 & 0.60 & $\pm$ 4.7 & $^{+2.0} _{-1.1}$ & $^{+1.6} _{-0.1}$ & $^{-0.3} _{+0.4}$ & $^{+0.9} _{-0.9}$ & $^{-0.0} _{+0.0}$ & $^{+0.0} _{+0.0}$ & $^{+0.4} _{+0.2}$ & $^{+0.5} _{-0.5}$ & $^{+0.4} _{-0.3}$ \\
& 0.08 & 0.51 & $\pm$ 4.4 & $^{+1.1} _{-1.1}$ & $^{+0.2} _{-0.4}$ & $^{-0.2} _{+0.3}$ & $^{-0.4} _{+0.4}$ & $^{+0.8} _{-0.8}$ & $^{+0.0} _{+0.0}$ & $^{+0.5} _{-0.1}$ & $^{+0.1} _{-0.1}$ & $^{+0.3} _{-0.5}$ \\
& 0.13 & 0.38 & $\pm$ 5.3 & $^{+0.9} _{-0.7}$ & $^{+0.1} _{-0.1}$ & $^{-0.2} _{+0.3}$ & $^{+0.6} _{-0.6}$ & $^{-0.1} _{+0.1}$ & $^{+0.0} _{+0.0}$ & $^{-0.3} _{-0.3}$ & $^{+0.0} _{+0.0}$ & $^{+0.7} _{-0.1}$ \\
& 0.18 & 0.34 & $\pm$ 5.6 & $^{+1.2} _{-1.3}$ & $^{+0.0} _{-0.3}$ & $^{-0.2} _{+0.2}$ & $^{-0.9} _{+0.9}$ & $^{-0.6} _{+0.6}$ & $^{+0.0} _{+0.0}$ & $^{-0.5} _{+0.2}$ & $^{+0.0} _{+0.0}$ & $^{+0.4} _{-0.3}$ \\
& 0.25 & 0.27 & $\pm$ 6.5 & $^{+7.4} _{-7.3}$ & $^{+0.6} _{-0.0}$ & $^{-0.2} _{+0.2}$ & $^{+7.3} _{-7.3}$ & $^{+0.2} _{-0.2}$ & $^{+0.0} _{+0.0}$ & $^{+1.1} _{+0.3}$ & $^{+0.0} _{+0.0}$ & $^{+0.8} _{-0.6}$ \\
& 0.4 & 0.16 & $\pm$ 9.0 & $^{+3.8} _{-3.5}$ & $^{+1.3} _{-0.0}$ & $^{-0.2} _{+0.4}$ & $^{-3.4} _{+3.4}$ & $^{+0.5} _{-0.5}$ & $^{+0.4} _{-0.0}$ & $^{+0.8} _{-0.3}$ & $^{+0.0} _{+0.0}$ & $^{+0.6} _{-0.6}$ \\
& 0.65 & 0.02 & $^{+18.8} _{-16.0}$ & $^{+10.2} _{-10.4}$ & $^{+0.0} _{-0.6}$ & $^{-0.8} _{+1.0}$ & $^{-8.7} _{+8.7}$ & $^{+4.6} _{-4.6}$ & $^{+1.8} _{+0.7}$ & $^{-2.7} _{+0.9}$ & $^{+0.0} _{+0.0}$ & $^{+1.7} _{-2.0}$ \\
\hline
5000 & 0.08 & 0.57 & $\pm$ 4.0 & $^{+1.8} _{-0.9}$ & $^{+1.6} _{-0.3}$ & $^{-0.2} _{+0.2}$ & $^{-0.3} _{+0.3}$ & $^{+0.2} _{-0.2}$ & $^{+0.0} _{+0.0}$ & $^{-0.6} _{-0.4}$ & $^{+0.3} _{-0.3}$ & $^{+0.5} _{-0.3}$ \\
& 0.13 & 0.49 & $\pm$ 5.2 & $^{+1.0} _{-0.8}$ & $^{+0.8} _{+0.0}$ & $^{-0.2} _{+0.2}$ & $^{-0.4} _{+0.4}$ & $^{+0.3} _{-0.3}$ & $^{+0.0} _{+0.0}$ & $^{+0.1} _{-0.3}$ & $^{+0.1} _{-0.1}$ & $^{+0.2} _{-0.5}$ \\
& 0.18 & 0.35 & $\pm$ 5.8 & $^{+1.0} _{-0.5}$ & $^{+0.2} _{-0.1}$ & $^{-0.3} _{+0.2}$ & $^{+0.2} _{-0.2}$ & $^{+0.3} _{-0.3}$ & $^{+0.0} _{+0.0}$ & $^{-0.1} _{+0.7}$ & $^{+0.0} _{+0.0}$ & $^{+0.5} _{-0.3}$ \\
& 0.25 & 0.23 & $\pm$ 7.6 & $^{+1.8} _{-1.2}$ & $^{+0.7} _{-0.0}$ & $^{-0.2} _{+0.3}$ & $^{+0.7} _{-0.7}$ & $^{-0.7} _{+0.7}$ & $^{+0.0} _{+0.0}$ & $^{+1.2} _{-0.6}$ & $^{+0.0} _{+0.0}$ & $^{+0.4} _{-0.3}$ \\
& 0.4 & 0.13 & $^{+11.3} _{-10.2}$ & $^{+6.5} _{-6.4}$ & $^{+0.0} _{-0.3}$ & $^{-0.2} _{+0.3}$ & $^{+6.0} _{-6.0}$ & $^{-2.1} _{+2.1}$ & $^{+0.0} _{+0.0}$ & $^{+0.4} _{+1.3}$ & $^{+0.0} _{+0.0}$ & $^{+1.1} _{-1.1}$ \\
\hline
8000 & 0.13 & 0.59 & $\pm$ 5.4 & $^{+1.0} _{-1.9}$ & $^{+0.0} _{-1.0}$ & $^{-0.3} _{+0.3}$ & $^{-0.5} _{+0.5}$ & $^{-0.1} _{+0.1}$ & $^{+0.0} _{+0.0}$ & $^{-1.3} _{+0.3}$ & $^{+0.5} _{-0.3}$ & $^{+0.5} _{-0.5}$ \\
& 0.18 & 0.44 & $\pm$ 7.0 & $^{+2.6} _{-2.9}$ & $^{+0.1} _{-0.3}$ & $^{-0.2} _{+0.2}$ & $^{-2.4} _{+2.4}$ & $^{+0.6} _{-0.6}$ & $^{+0.0} _{+0.0}$ & $^{+0.5} _{-0.3}$ & $^{+0.0} _{+0.0}$ & $^{+0.6} _{-1.3}$ \\
& 0.25 & 0.32 & $\pm$ 8.5 & $^{+3.8} _{-3.9}$ & $^{+0.7} _{-0.1}$ & $^{-0.3} _{+0.2}$ & $^{-3.6} _{+3.6}$ & $^{+0.4} _{-0.4}$ & $^{+0.0} _{+0.0}$ & $^{-1.1} _{-0.6}$ & $^{+0.0} _{+0.0}$ & $^{+0.4} _{-0.4}$ \\
& 0.4 & 0.12 & $^{+16.1} _{-14.0}$ & $^{+7.1} _{-7.3}$ & $^{+0.1} _{-0.0}$ & $^{-0.3} _{+0.4}$ & $^{+7.0} _{-7.0}$ & $^{-0.6} _{+0.6}$ & $^{+0.0} _{+0.0}$ & $^{-0.3} _{+0.5}$ & $^{+0.0} _{+0.0}$ & $^{+0.6} _{-2.0}$ \\
& 0.65 & 0.02 & $^{+26.9} _{-21.6}$ & $^{+4.1} _{-4.2}$ & $^{+0.2} _{-1.4}$ & $^{-0.9} _{+0.9}$ & $^{+1.5} _{-1.5}$ & $^{+2.3} _{-2.3}$ & $^{+0.0} _{+0.0}$ & $^{-2.6} _{+2.8}$ & $^{+0.0} _{+0.0}$ & $^{+0.8} _{-0.6}$ \\
\hline
12000 & 0.18 & 0.52 & $\pm$ 7.6 & $^{+1.3} _{-3.8}$ & $^{+1.0} _{-3.6}$ & $^{-0.3} _{+0.4}$ & $^{-0.4} _{+0.4}$ & $^{-0.2} _{+0.2}$ & $^{+0.0} _{+0.0}$ & $^{-0.3} _{-0.8}$ & $^{+0.2} _{-0.3}$ & $^{+0.5} _{-0.6}$ \\
& 0.25 & 0.35 & $^{+12.0} _{-10.8}$ & $^{+2.8} _{-2.3}$ & $^{+2.4} _{-0.6}$ & $^{-0.2} _{+0.2}$ & $^{+0.3} _{-0.3}$ & $^{+0.8} _{-0.8}$ & $^{+0.0} _{+0.0}$ & $^{+0.3} _{-1.6}$ & $^{+0.0} _{+0.0}$ & $^{+1.0} _{-1.2}$ \\
& 0.4 & 0.15 & $^{+18.7} _{-16.0}$ & $^{+4.6} _{-3.8}$ & $^{+2.7} _{-0.2}$ & $^{-0.3} _{+0.4}$ & $^{+3.2} _{-3.2}$ & $^{-0.8} _{+0.8}$ & $^{+0.0} _{+0.0}$ & $^{-1.2} _{+1.5}$ & $^{+0.0} _{+0.0}$ & $^{+0.6} _{-1.3}$ \\
\hline
20000 & 0.25 & 0.47 & $^{+13.1} _{-11.8}$ & $^{+6.1} _{-3.6}$ & $^{+4.9} _{-1.8}$ & $^{-1.5} _{+0.2}$ & $^{-2.2} _{+2.2}$ & $^{+0.1} _{-0.1}$ & $^{+0.0} _{+0.0}$ & $^{-0.7} _{-0.2}$ & $^{+2.7} _{-1.4}$ & $^{+0.6} _{-0.5}$ \\
& 0.4 & 0.19 & $^{+22.2} _{-18.5}$ & $^{+4.9} _{-3.8}$ & $^{+3.6} _{+0.0}$ & $^{-0.4} _{+0.4}$ & $^{+0.9} _{-0.9}$ & $^{+3.0} _{-3.0}$ & $^{+0.0} _{+0.0}$ & $^{-1.8} _{-0.3}$ & $^{+0.0} _{+0.0}$ & $^{+1.1} _{-1.2}$ \\
\hline
30000 & 0.4 & 0.28 & $^{+22.7} _{-18.8}$ & $^{+6.9} _{-5.9}$ & $^{+6.7} _{+0.0}$ & $^{-0.2} _{+0.3}$ & $^{-0.8} _{+0.8}$ & $^{+0.7} _{-0.7}$ & $^{+0.0} _{+0.0}$ & $^{+0.1} _{-4.9}$ & $^{+0.0} _{-2.9}$ & $^{+1.3} _{-1.3}$ \\

\hline
\end{tabular}
\end{center}
\end{scriptsize}
\vspace{0.5cm} \hspace{5.4cm} {\bf Table 14:}~~~{\it Continuation 2.}
\label{tab:ds2dxdq2LhSys_3}
\end{table}

\clearpage
\newpage
\begin{table}
\begin{scriptsize}
\begin{center} \begin{tabular}[t]{|rcr|r|rll|r|r|} \hline
\multicolumn{3}{|c|}{$Q^2$ range} & \multicolumn{1}{c|}{$Q^2_c$} & \multicolumn{3}{c|}{$d\sigma / dQ^{2}$} & \multicolumn{1}{c|}{$N_{\text{data}}$} & \multicolumn{1}{c|}{$N^{\text{MC}}_{\text{bg}}$}\\
\multicolumn{3}{|c|}{($\gev^{2}$)} & \multicolumn{1}{c|}{($\gev^{2}$)} & \multicolumn{3}{c|}{($\pb / \gev^{2}$)} & & \\ \hline \hline
185.0 & -- & 300.0 & 250 & (1.08 & $\pm$ 0.00 & $^{+0.01} _{-0.01}$ )$\cdot 10^{1}$ & 74098 & 105.5 \\
300.0 & -- & 400.0 & 350 & 4.77 & $\pm$ 0.03 & $^{+0.07} _{-0.05}$ & 25380 & 57.6 \\
400.0 & -- & 475.7 & 440 & 2.73 & $\pm$ 0.03 & $^{+0.05} _{-0.05}$ & 10769 & 34.0 \\
475.7 & -- & 565.7 & 520 & 1.78 & $\pm$ 0.02 & $^{+0.02} _{-0.02}$ & 7685 & 30.7 \\
565.7 & -- & 672.7 & 620 & 1.19 & $\pm$ 0.02 & $^{+0.02} _{-0.01}$ & 5715 & 32.6 \\
672.7 & -- & 800.0 & 730 & (7.77 & $\pm$ 0.11 & $^{+0.06} _{-0.06}$ )$\cdot 10^{-1}$ & 5242 & 17.8 \\
800.0 & -- & 1050.0 & 900 & (4.60 & $\pm$ 0.06 & $^{+0.04} _{-0.04}$ )$\cdot 10^{-1}$ & 6873 & 36.8 \\
1050.0 & -- & 1460.0 & 1230 & (2.13 & $\pm$ 0.03 & $^{+0.01} _{-0.02}$ )$\cdot 10^{-1}$ & 5558 & 38.9 \\
1460.0 & -- & 2080.0 & 1730 & (8.81 & $\pm$ 0.15 & $^{+0.05} _{-0.05}$ )$\cdot 10^{-2}$ & 3551 & 23.7 \\
2080.0 & -- & 3120.0 & 2500 & (3.42 & $\pm$ 0.07 & $^{+0.03} _{-0.02}$ )$\cdot 10^{-2}$ & 2269 & 14.2 \\
3120.0 & -- & 5220.0 & 3900 & (1.06 & $\pm$ 0.03 & $^{+0.01} _{-0.01}$ )$\cdot 10^{-2}$ & 1363 & 4.4 \\
5220.0 & -- & 12500.0 & 7000 & (2.34 & $\pm$ 0.08 & $^{+0.05} _{-0.01}$ )$\cdot 10^{-3}$ & 778 & 3.7 \\
12500.0 & -- & 51200.0 & 22400 & (6.24 & $\pm$ 0.52 & $^{+0.23} _{-0.14}$ )$\cdot 10^{-5}$ & 144 & 1.8 \\

 \hline
\end{tabular}
\end{center}
\caption[]
{The single differential cross section $d\sigma / dQ^{2}$ ($y < 0.9$)
for the reaction $e^{-}p \rightarrow e^{-}X$ ($\mathcal{L} = 71.2 \pbi, P_{e} = +0.29$).
The bin range, bin centre ($Q^2_c$) and measured cross section
corrected to the electroweak Born level are shown.
The first (second) error on the cross section
corresponds to the statistical (systematic) uncertainties.
The number of observed data events ($N_{\text{data}}$)
and simulated background events ($N^{\text{MC}}_{\text{bg}}$) are also shown.}
\label{tab:dsdq2Rh}
\end{scriptsize}
\end{table}

\begin{table}\begin{scriptsize}
\begin{center} \begin{tabular}[t]{|r|rl|c|c||c|c|c|c|c|c|c|c|} \hline
\multicolumn{1}{|c|}{$Q^2_c$} & \multicolumn{2}{c|}{$d\sigma / dQ^{2}$} & \multicolumn{1}{c|}{stat.} & \multicolumn{1}{c||}{sys.} & \multicolumn{1}{c|}{$\delta_{1}$} & \multicolumn{1}{c|}{$\delta_{2}$} & \multicolumn{1}{c|}{$\delta_{3}$} & \multicolumn{1}{c|}{$\delta_{4}$} & \multicolumn{1}{c|}{$\delta_{5}$} & \multicolumn{1}{c|}{$\delta_{6}$} & \multicolumn{1}{c|}{$\delta_{7}$} & \multicolumn{1}{c|}{$\delta_{8} - \delta_{13}$} \\
\multicolumn{1}{|c|}{($\gev^{2}$)} & \multicolumn{2}{c|}{($\pb / \gev^{2}$)} & \multicolumn{1}{c|}{(\%)} & \multicolumn{1}{c||}{(\%)} & \multicolumn{1}{c|}{(\%)} & \multicolumn{1}{c|}{(\%)} & \multicolumn{1}{c|}{(\%)} & \multicolumn{1}{c|}{(\%)} & \multicolumn{1}{c|}{(\%)} & \multicolumn{1}{c|}{(\%)} & \multicolumn{1}{c|}{(\%)} & \multicolumn{1}{c|}{(\%)} \\ \hline \hline
250 & 1.08 & $\cdot 10^{1}$ & $\pm$ 0.4 & $^{+1.2} _{-1.1}$ & $^{+0.3} _{-0.0}$ & $^{-0.5} _{+0.6}$ & $^{+0.5} _{-0.5}$ & $^{+0.4} _{-0.4}$ & $^{-0.1} _{+0.2}$ & $^{-0.0} _{+0.2}$ & $^{+0.1} _{-0.1}$ & $^{+0.7} _{-0.8}$ \\
350 & 4.77 & & $\pm$ 0.6 & $^{+1.5} _{-1.1}$ & $^{+0.8} _{-0.0}$ & $^{-0.5} _{+0.7}$ & $^{+0.5} _{-0.5}$ & $^{+0.6} _{-0.6}$ & $^{-0.0} _{+0.2}$ & $^{-0.2} _{+0.2}$ & $^{+0.1} _{-0.1}$ & $^{+0.6} _{-0.6}$ \\
440 & 2.73 & & $\pm$ 1.0 & $^{+1.8} _{-1.7}$ & $^{+0.3} _{-0.0}$ & $^{-0.5} _{+0.6}$ & $^{+0.5} _{-0.5}$ & $^{+1.3} _{-1.3}$ & $^{-0.2} _{+0.2}$ & $^{-0.4} _{+0.7}$ & $^{+0.1} _{-0.1}$ & $^{+0.6} _{-0.6}$ \\
520 & 1.78 & & $\pm$ 1.2 & $^{+1.3} _{-0.9}$ & $^{+0.7} _{-0.1}$ & $^{-0.5} _{+0.7}$ & $^{+0.0} _{-0.0}$ & $^{+0.1} _{-0.1}$ & $^{-0.1} _{+0.2}$ & $^{-0.2} _{+0.3}$ & $^{+0.2} _{-0.2}$ & $^{+0.7} _{-0.7}$ \\
620 & 1.19 & & $\pm$ 1.3 & $^{+1.4} _{-1.0}$ & $^{+0.9} _{-0.1}$ & $^{-0.6} _{+0.7}$ & $^{+0.2} _{-0.2}$ & $^{-0.6} _{+0.6}$ & $^{-0.1} _{-0.0}$ & $^{+0.1} _{+0.2}$ & $^{+0.2} _{-0.2}$ & $^{+0.4} _{-0.5}$ \\
730 & 7.77 & $\cdot 10^{-1}$ & $\pm$ 1.4 & $^{+0.8} _{-0.7}$ & $^{+0.0} _{-0.2}$ & $^{-0.5} _{+0.6}$ & $^{+0.3} _{-0.3}$ & $^{+0.1} _{-0.1}$ & $^{+0.2} _{+0.1}$ & $^{+0.1} _{-0.1}$ & $^{+0.1} _{-0.1}$ & $^{+0.2} _{-0.3}$ \\
900 & 4.60 & $\cdot 10^{-1}$ & $\pm$ 1.2 & $^{+0.9} _{-0.8}$ & $^{+0.1} _{-0.0}$ & $^{-0.4} _{+0.5}$ & $^{-0.2} _{+0.2}$ & $^{+0.6} _{-0.6}$ & $^{+0.1} _{-0.1}$ & $^{-0.0} _{-0.0}$ & $^{+0.2} _{-0.2}$ & $^{+0.3} _{-0.3}$ \\
1230 & 2.13 & $\cdot 10^{-1}$ & $\pm$ 1.4 & $^{+0.7} _{-0.7}$ & $^{+0.1} _{-0.0}$ & $^{-0.3} _{+0.4}$ & $^{+0.3} _{-0.3}$ & $^{-0.1} _{+0.1}$ & $^{+0.0} _{+0.1}$ & $^{-0.0} _{-0.3}$ & $^{+0.3} _{-0.3}$ & $^{+0.3} _{-0.4}$ \\
1730 & 8.81 & $\cdot 10^{-2}$ & $\pm$ 1.7 & $^{+0.6} _{-0.5}$ & $^{+0.2} _{-0.0}$ & $^{-0.3} _{+0.3}$ & $^{+0.1} _{-0.1}$ & $^{+0.0} _{-0.0}$ & $^{+0.2} _{-0.0}$ & $^{-0.2} _{-0.1}$ & $^{+0.3} _{-0.3}$ & $^{+0.3} _{-0.3}$ \\
2500 & 3.42 & $\cdot 10^{-2}$ & $\pm$ 2.1 & $^{+0.9} _{-0.6}$ & $^{+0.4} _{-0.0}$ & $^{-0.2} _{+0.3}$ & $^{+0.3} _{-0.3}$ & $^{+0.2} _{-0.2}$ & $^{-0.1} _{-0.0}$ & $^{+0.3} _{+0.1}$ & $^{+0.3} _{-0.2}$ & $^{+0.3} _{-0.3}$ \\
3900 & 1.06 & $\cdot 10^{-2}$ & $\pm$ 2.7 & $^{+1.3} _{-0.6}$ & $^{+1.2} _{-0.1}$ & $^{-0.2} _{+0.3}$ & $^{-0.4} _{+0.4}$ & $^{-0.1} _{+0.1}$ & $^{-0.0} _{-0.0}$ & $^{-0.2} _{-0.1}$ & $^{+0.1} _{-0.1}$ & $^{+0.3} _{-0.3}$ \\
7000 & 2.34 & $\cdot 10^{-3}$ & $\pm$ 3.6 & $^{+2.3} _{-0.6}$ & $^{+2.2} _{-0.0}$ & $^{-0.3} _{+0.3}$ & $^{-0.0} _{+0.0}$ & $^{+0.1} _{-0.1}$ & $^{+0.0} _{+0.0}$ & $^{-0.3} _{+0.1}$ & $^{+0.5} _{-0.2}$ & $^{+0.3} _{-0.4}$ \\
22400 & 6.24 & $\cdot 10^{-5}$ & $\pm$ 8.4 & $^{+3.7} _{-2.3}$ & $^{+3.4} _{-0.4}$ & $^{-0.7} _{+0.3}$ & $^{-1.0} _{+1.0}$ & $^{+0.7} _{-0.7}$ & $^{+0.0} _{+0.0}$ & $^{-1.0} _{-1.6}$ & $^{+0.5} _{-0.5}$ & $^{+0.6} _{-0.6}$ \\

\hline
\end{tabular}
\end{center}
\caption[]
{Systematic uncertainties with bin-to-bin correlations for $d\sigma / dQ^{2}$ ($y < 0.9$)
for the reaction $e^{-}p \rightarrow e^{-}X$ ($\mathcal{L} = 71.2 \pbi, P_{e} = +0.29$).
The left four columns of the table
contain the bin centre ($Q^2_c$), the measured cross section,
the statistical uncertainty and the total systematic uncertainty.
The right eight columns of the table list
the bin-to-bin correlated systematic uncertainties
for $\delta_{1} - \delta_{7}$,
and the total systematic uncertainties
summed in quadrature for $\delta_{8} - \delta_{13}$,
as defined in the section \ref{sec-sys}.
The upper and lower correlated uncertainties correspond to
a positive or negative variation of a cut value for example.
However, if this is not possible for a particular systematic,
the uncertainty is symmetrised.}
\label{tab:dsdq2RhSys}
\end{scriptsize}\end{table}

\newpage
\begin{table}
\begin{scriptsize}
\begin{center} \begin{tabular}[t]{|rcr|r|rll|r|r|} \hline
\multicolumn{3}{|c|}{$Q^2$ range} & \multicolumn{1}{c|}{$Q^2_c$} & \multicolumn{3}{c|}{$d\sigma / dQ^{2}$} & \multicolumn{1}{c|}{$N_{\text{data}}$} & \multicolumn{1}{c|}{$N^{\text{MC}}_{\text{bg}}$}\\
\multicolumn{3}{|c|}{($\gev^{2}$)} & \multicolumn{1}{c|}{($\gev^{2}$)} & \multicolumn{3}{c|}{($\pb / \gev^{2}$)} & & \\ \hline \hline
185.0 & -- & 300.0 & 250 & (1.08 & $\pm$ 0.00 & $^{+0.01} _{-0.01}$ )$\cdot 10^{1}$ & 103254 & 146.9 \\
300.0 & -- & 400.0 & 350 & 4.83 & $\pm$ 0.03 & $^{+0.07} _{-0.06}$ & 35547 & 80.0 \\
400.0 & -- & 475.7 & 440 & 2.80 & $\pm$ 0.02 & $^{+0.05} _{-0.05}$ & 15291 & 47.0 \\
475.7 & -- & 565.7 & 520 & 1.87 & $\pm$ 0.02 & $^{+0.02} _{-0.02}$ & 11190 & 42.4 \\
565.7 & -- & 672.7 & 620 & 1.19 & $\pm$ 0.01 & $^{+0.02} _{-0.01}$ & 7926 & 45.0 \\
672.7 & -- & 800.0 & 730 & (8.30 & $\pm$ 0.10 & $^{+0.07} _{-0.06}$ )$\cdot 10^{-1}$ & 7719 & 24.7 \\
800.0 & -- & 1050.0 & 900 & (4.89 & $\pm$ 0.05 & $^{+0.05} _{-0.04}$ )$\cdot 10^{-1}$ & 10094 & 51.1 \\
1050.0 & -- & 1460.0 & 1230 & (2.25 & $\pm$ 0.03 & $^{+0.02} _{-0.02}$ )$\cdot 10^{-1}$ & 8111 & 53.7 \\
1460.0 & -- & 2080.0 & 1730 & (9.25 & $\pm$ 0.13 & $^{+0.05} _{-0.05}$ )$\cdot 10^{-2}$ & 5158 & 32.7 \\
2080.0 & -- & 3120.0 & 2500 & (3.61 & $\pm$ 0.06 & $^{+0.03} _{-0.02}$ )$\cdot 10^{-2}$ & 3314 & 19.4 \\
3120.0 & -- & 5220.0 & 3900 & (1.16 & $\pm$ 0.03 & $^{+0.02} _{-0.01}$ )$\cdot 10^{-2}$ & 2065 & 6.0 \\
5220.0 & -- & 12500.0 & 7000 & (2.48 & $\pm$ 0.07 & $^{+0.06} _{-0.02}$ )$\cdot 10^{-3}$ & 1143 & 5.0 \\
12500.0 & -- & 51200.0 & 22400 & (6.26 & $\pm$ 0.45 & $^{+0.23} _{-0.14}$ )$\cdot 10^{-5}$ & 200 & 2.6 \\

 \hline
\end{tabular}
\end{center}
\caption[]
{The single differential cross section $d\sigma / dQ^{2}$ ($y < 0.9$)
for the reaction $e^{-}p \rightarrow e^{-}X$ ($\mathcal{L} = 98.7 \pbi, P_{e} = -0.27$).
The bin range, bin centre ($Q^2_c$) and measured cross section
corrected to the electroweak Born level are shown.
The first (second) error on the cross section
corresponds to the statistical (systematic) uncertainties.
The number of observed data events ($N_{\text{data}}$)
and simulated background events ($N^{\text{MC}}_{\text{bg}}$) are also shown.}
\label{tab:dsdq2Lh}
\end{scriptsize}
\end{table}

\begin{table}\begin{scriptsize}
\begin{center} \begin{tabular}[t]{|r|rl|c|c||c|c|c|c|c|c|c|c|} \hline
\multicolumn{1}{|c|}{$Q^2_c$} & \multicolumn{2}{c|}{$d\sigma / dQ^{2}$} & \multicolumn{1}{c|}{stat.} & \multicolumn{1}{c||}{sys.} & \multicolumn{1}{c|}{$\delta_{1}$} & \multicolumn{1}{c|}{$\delta_{2}$} & \multicolumn{1}{c|}{$\delta_{3}$} & \multicolumn{1}{c|}{$\delta_{4}$} & \multicolumn{1}{c|}{$\delta_{5}$} & \multicolumn{1}{c|}{$\delta_{6}$} & \multicolumn{1}{c|}{$\delta_{7}$} & \multicolumn{1}{c|}{$\delta_{8} - \delta_{13}$} \\
\multicolumn{1}{|c|}{($\gev^{2}$)} & \multicolumn{2}{c|}{($\pb / \gev^{2}$)} & \multicolumn{1}{c|}{(\%)} & \multicolumn{1}{c||}{(\%)} & \multicolumn{1}{c|}{(\%)} & \multicolumn{1}{c|}{(\%)} & \multicolumn{1}{c|}{(\%)} & \multicolumn{1}{c|}{(\%)} & \multicolumn{1}{c|}{(\%)} & \multicolumn{1}{c|}{(\%)} & \multicolumn{1}{c|}{(\%)} & \multicolumn{1}{c|}{(\%)} \\ \hline \hline
250 & 1.08 & $\cdot 10^{1}$ & $\pm$ 0.3 & $^{+1.2} _{-1.1}$ & $^{+0.3} _{-0.0}$ & $^{-0.5} _{+0.6}$ & $^{+0.5} _{-0.5}$ & $^{+0.4} _{-0.4}$ & $^{-0.1} _{+0.2}$ & $^{-0.0} _{+0.2}$ & $^{+0.1} _{-0.1}$ & $^{+0.7} _{-0.8}$ \\
350 & 4.83 & & $\pm$ 0.6 & $^{+1.5} _{-1.1}$ & $^{+0.8} _{-0.0}$ & $^{-0.5} _{+0.7}$ & $^{+0.5} _{-0.5}$ & $^{+0.6} _{-0.6}$ & $^{-0.0} _{+0.2}$ & $^{-0.2} _{+0.2}$ & $^{+0.1} _{-0.1}$ & $^{+0.6} _{-0.6}$ \\
440 & 2.80 & & $\pm$ 0.8 & $^{+1.8} _{-1.7}$ & $^{+0.3} _{-0.0}$ & $^{-0.5} _{+0.6}$ & $^{+0.5} _{-0.5}$ & $^{+1.3} _{-1.3}$ & $^{-0.2} _{+0.2}$ & $^{-0.4} _{+0.7}$ & $^{+0.1} _{-0.1}$ & $^{+0.6} _{-0.6}$ \\
520 & 1.87 & & $\pm$ 1.0 & $^{+1.3} _{-0.9}$ & $^{+0.7} _{-0.1}$ & $^{-0.5} _{+0.7}$ & $^{+0.0} _{-0.0}$ & $^{+0.1} _{-0.1}$ & $^{-0.1} _{+0.2}$ & $^{-0.2} _{+0.3}$ & $^{+0.2} _{-0.2}$ & $^{+0.7} _{-0.7}$ \\
620 & 1.19 & & $\pm$ 1.1 & $^{+1.4} _{-1.0}$ & $^{+0.9} _{-0.1}$ & $^{-0.6} _{+0.7}$ & $^{+0.2} _{-0.2}$ & $^{-0.6} _{+0.6}$ & $^{-0.1} _{-0.0}$ & $^{+0.1} _{+0.2}$ & $^{+0.2} _{-0.2}$ & $^{+0.4} _{-0.5}$ \\
730 & 8.30 & $\cdot 10^{-1}$ & $\pm$ 1.2 & $^{+0.8} _{-0.7}$ & $^{+0.0} _{-0.2}$ & $^{-0.5} _{+0.6}$ & $^{+0.3} _{-0.3}$ & $^{+0.1} _{-0.1}$ & $^{+0.2} _{+0.1}$ & $^{+0.1} _{-0.1}$ & $^{+0.1} _{-0.1}$ & $^{+0.2} _{-0.3}$ \\
900 & 4.89 & $\cdot 10^{-1}$ & $\pm$ 1.0 & $^{+0.9} _{-0.8}$ & $^{+0.1} _{-0.0}$ & $^{-0.4} _{+0.5}$ & $^{-0.2} _{+0.2}$ & $^{+0.6} _{-0.6}$ & $^{+0.1} _{-0.1}$ & $^{-0.0} _{-0.0}$ & $^{+0.2} _{-0.2}$ & $^{+0.3} _{-0.3}$ \\
1230 & 2.25 & $\cdot 10^{-1}$ & $\pm$ 1.1 & $^{+0.7} _{-0.7}$ & $^{+0.1} _{-0.0}$ & $^{-0.3} _{+0.4}$ & $^{+0.3} _{-0.3}$ & $^{-0.1} _{+0.1}$ & $^{+0.0} _{+0.1}$ & $^{-0.0} _{-0.3}$ & $^{+0.3} _{-0.3}$ & $^{+0.3} _{-0.4}$ \\
1730 & 9.25 & $\cdot 10^{-2}$ & $\pm$ 1.4 & $^{+0.6} _{-0.5}$ & $^{+0.2} _{-0.0}$ & $^{-0.3} _{+0.3}$ & $^{+0.1} _{-0.1}$ & $^{+0.0} _{-0.0}$ & $^{+0.2} _{-0.0}$ & $^{-0.2} _{-0.1}$ & $^{+0.3} _{-0.3}$ & $^{+0.3} _{-0.3}$ \\
2500 & 3.61 & $\cdot 10^{-2}$ & $\pm$ 1.8 & $^{+0.9} _{-0.6}$ & $^{+0.4} _{-0.0}$ & $^{-0.2} _{+0.3}$ & $^{+0.3} _{-0.3}$ & $^{+0.2} _{-0.2}$ & $^{-0.1} _{-0.0}$ & $^{+0.3} _{+0.1}$ & $^{+0.3} _{-0.2}$ & $^{+0.3} _{-0.3}$ \\
3900 & 1.16 & $\cdot 10^{-2}$ & $\pm$ 2.2 & $^{+1.3} _{-0.6}$ & $^{+1.2} _{-0.1}$ & $^{-0.2} _{+0.3}$ & $^{-0.4} _{+0.4}$ & $^{-0.1} _{+0.1}$ & $^{-0.0} _{-0.0}$ & $^{-0.2} _{-0.1}$ & $^{+0.1} _{-0.1}$ & $^{+0.3} _{-0.3}$ \\
7000 & 2.48 & $\cdot 10^{-3}$ & $\pm$ 3.0 & $^{+2.3} _{-0.6}$ & $^{+2.2} _{-0.0}$ & $^{-0.3} _{+0.3}$ & $^{-0.0} _{+0.0}$ & $^{+0.1} _{-0.1}$ & $^{+0.0} _{+0.0}$ & $^{-0.3} _{+0.1}$ & $^{+0.5} _{-0.2}$ & $^{+0.3} _{-0.4}$ \\
22400 & 6.26 & $\cdot 10^{-5}$ & $\pm$ 7.2 & $^{+3.7} _{-2.3}$ & $^{+3.4} _{-0.4}$ & $^{-0.7} _{+0.3}$ & $^{-1.0} _{+1.0}$ & $^{+0.7} _{-0.7}$ & $^{+0.0} _{+0.0}$ & $^{-1.0} _{-1.6}$ & $^{+0.5} _{-0.5}$ & $^{+0.6} _{-0.6}$ \\

\hline
\end{tabular}
\end{center}
\caption[]
{Systematic uncertainties with bin-to-bin correlations for $d\sigma / dQ^{2}$ ($y < 0.9$)
for the reaction $e^{-}p \rightarrow e^{-}X$ ($\mathcal{L} = 98.7 \pbi, P_{e} = -0.27$).
The left four columns of the table contain
the bin centre ($Q^2_c$), the measured cross section,
the statistical uncertainty and the total systematic uncertainty.
The right eight columns of the table list
the bin-to-bin correlated systematic uncertainties
for $\delta_{1} - \delta_{7}$,
and the total systematic uncertainties
summed in quadrature for $\delta_{8} - \delta_{13}$,
as defined in the section \ref{sec-sys}.
The upper and lower correlated uncertainties correspond to
a positive or negative variation of a cut value for example.
However, if this is not possible for a particular systematic,
the uncertainty is symmetrised.}
\label{tab:dsdq2LhSys}
\end{scriptsize}\end{table}

\newpage
\begin{table}
\begin{center}
\begin{tabular}[t]{|rcr|r|rl|} \hline
\multicolumn{3}{|c|}{$Q^{2}$ range} & \multicolumn{1}{c|}{$Q^{2}_c$} & \multicolumn{2}{c|}{Asymmetry $A^{-}$} \\
\multicolumn{3}{|c|}{($\gev^{2}$)} & \multicolumn{1}{c|}{($\gev^{2}$)} & \multicolumn{2}{c|}{$\times 10$} \\ \hline \hline
185.0 & -- & 300.0 & 250 & -0.11 & $\pm$0.09 \\
300.0 & -- & 400.0 & 350 & -0.21 & $\pm$0.15 \\
400.0 & -- & 475.7 & 440 & -0.45 & $\pm$0.23 \\
475.7 & -- & 565.7 & 520 & -0.89 & $\pm$0.27 \\
565.7 & -- & 672.7 & 620 & -0.03 & $\pm$0.32 \\
672.7 & -- & 800.0 & 730 & -1.19 & $\pm$0.32 \\
800.0 & -- & 1050.0 & 900 & -1.09 & $\pm$0.28 \\
1050.0 & -- & 1460.0 & 1230 & -0.95 & $\pm$0.32 \\
1460.0 & -- & 2080.0 & 1730 & -0.86 & $\pm$0.39 \\
2080.0 & -- & 3120.0 & 2500 & -0.96 & $\pm$0.49 \\
3120.0 & -- & 5220.0 & 3900 & -1.61 & $\pm$0.63 \\
5220.0 & -- & 12500.0 & 7000 & -1.06 & $\pm$0.83 \\
12500.0 & -- & 51200.0 & 22400 & -0.06 & $\pm$1.97 \\

\hline
\end{tabular}
\end{center}
\caption[]
{The polarisation asymmetry measured using
positively and negatively polarised $e^{-}p$ beams
($\mathcal{L} = 71.2 \pbi, P_{e} = +0.29$ and
 $\mathcal{L} = 98.7 \pbi, P_{e} = -0.27$, respectively).
The bin range, bin centre ($Q^2_c$) and measured Asymmetry $A^-$ are shown.
Only the statistical uncertainties on the measurement are shown
as systematic uncertainties are assumed to cancel.}
\label{tab:asym}
\end{table}

%
%
\end{document}